\documentclass{article}

\usepackage{arxiv}

\usepackage[utf8]{inputenc} 
\usepackage[T1]{fontenc}    
\usepackage{graphicx}       
\usepackage{xcolor}
\usepackage{natbib} 
\usepackage{amsfonts}       
\usepackage{amsmath}
\usepackage{amsthm}
\usepackage{booktabs}
\usepackage{algorithm}
\usepackage{algpseudocode}
\usepackage{adjustbox}
\PassOptionsToPackage{export}{adjustbox}
\usepackage{multirow} 
\usepackage{array}
\usepackage{eurosym}
\usepackage[shortlabels]{enumitem}

\newtheorem{theorem}{Theorem}
\newtheorem{lemma}[theorem]{Lemma}
\newtheorem{proposition}[theorem]{Proposition}%

\usepackage{comment}
\usepackage{standalone}

\def\showRoundOne{1} 
\def\showRoundTwo{1}

\ifnum\showRoundOne=1
    \newcommand{\revone}[1]{{\color{black}#1}}
\else
    \newcommand{\revone}[1]{#1}
\fi

\ifnum\showRoundTwo=1
    
\else
    
\fi

\title{A monotone single-index modal regression powered by deep neural networks for non-Gaussian periodontal data}

\newif\ifuniqueAffiliation
\uniqueAffiliationtrue

\ifuniqueAffiliation 
\author{ \href{https://orcid.org/0000-0003-3265-6330}{\includegraphics[scale=0.06]{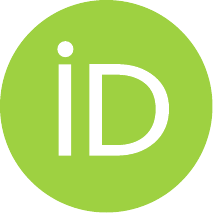}\hspace{1mm}Qingyang Liu} \\
	Department of Statistical Sciences\\
	Wake Forest University\\
	Winston--Salem, NC 27109 \\
	\texttt{liuqi@wfu.edu} \\
	\And
	\href{https://orcid.org/0009-0004-8522-6363}{\includegraphics[scale=0.06]{orcid.pdf}\hspace{1mm}Shijie Wang} \\
	Gauss Labs\\
	Palo Alto, CA 94301 \\
	\texttt{shijiew.usc@gmail.com} \\
	\And
	\href{https://orcid.org/0000-0002-7190-7844}{\includegraphics[scale=0.06]{orcid.pdf}\hspace{1mm}Ray Bai} \\
	Department of Statistics\\
	George Mason University\\
	Fairfax, VA 22030 \\
	\texttt{rbai2@gmu.edu} \\
	\And
	\href{https://orcid.org/0000-0001-5421-1725}{\includegraphics[scale=0.06]{orcid.pdf}\hspace{1mm}Dipankar Bandyopadhyay} \\
	Department of Biostatistics\\
	Virginia Commonwealth University\\
	Richmond, VA 23219 \\
	\texttt{dbandyop@vcu.edu} \\
}

\renewcommand{\shorttitle}{Robust Deep Monotonic Single-Index Model}

\usepackage{hyperref}       
\hypersetup{
pdftitle={Robust Deep Monotonic Single-Index Model},
pdfsubject={q-bio.NC, q-bio.QM},
pdfauthor={Qingyang Liu, Shijie Wang, Ray Bai, Dipankar Bandyopadhyay},
pdfkeywords={Single-Index Model, Deep Neural Network, Robust, Modal Regression},
}

\begin{document}
\maketitle

\begin{abstract}
Pocket depth (PD) is a widely used biomarker for diagnosing risk of periodontal disease (PrD). However, PD typically exhibits skewness and heavy-tailedness, and its relationship with clinical risk factors is often nonlinear. Motivated by PrD studies, this paper develops a robust single-index modal regression framework for analyzing skewed and heavy-tailed data. Our method has the following novel features: (a) a flexible two-piece scale Student-$t$ error distribution that generalizes both normal and two-piece scale normal distributions; (b) a neural network with guaranteed monotonicity constraints to estimate the unknown single-index function; and (c) theoretical \revone{support}, including model identifiability and a universal approximation theorem. Our single-index model combines the flexibility of neural networks and the two-piece scaled Student-$t$ distribution, delivering robust mode-based estimation that is resistant to outliers, while retaining clinical interpretability through parametric index coefficients. We demonstrate the performance of our method through simulation studies, and an application to PrD electronic health records obtained from the HealthPartners Institute of Minnesota. The proposed methodology is implemented in the \texttt{R} package \href{https://doi.org/10.32614/CRAN.package.DNNSIM}{\texttt{DNNSIM}}.
\end{abstract}


\section{Introduction} \label{sec:Introduction}

\subsection{Background and Our Contributions}

Despite significant recent advances in preventive strategies, such as water fluoridation and dental sealants, periodontal disease (PrD) remains a major public health problem worldwide, with a combined prevalence of nearly 62\% among dentate adults \citep{Villoria2024}. If left untreated, it can lead to progressive bone loss around the tooth, resulting in loosening and eventual tooth loss. This incurs a significant economic burden on individuals and healthcare systems. In 2018, the estimated direct and indirect costs of PrD were \$3.49 billion in the United States and \euro2.52 billion in Europe \citep{BotelhoPeriodontology2022}. As a complex chronic disease, the progression of PrD is also multifactorial, influenced by age, gender, race, tobacco use, and many other risk factors. The significant burden and complex nature of PrD underscore the need to develop evaluation tools for assessing the risk of PrD.

Periodontal pocket depth (PD) is a widely used biomarker for evaluating PrD, and the clinical success of periodontal therapy is often measured by PD reduction \citep{Donos2017}. However, there are three main challenges in developing statistical models to assess PrD risk. First, the data distributions for PD are typically skewed and heavy-tailed \citep{Bandyopadhyay2010}. This reflects the fact that most individuals have healthy (shallow) PD, but a few individuals exhibit extreme values for PD. This characteristic of PrD data makes conventional statistical tools assuming Gaussian errors particularly unsuitable. Transformations to normality also face practical challenges, such as the lack of a universally accepted class of transformations, and difficulty in interpreting the results on the original scale of PD \citep{Bandyopadhyay2010}. Second, the relationship between PrD and covariates is often nonlinear, necessitating the development of statistical tools without stringent linear assumptions. Finally, for ease of use by clinicians, an ideal risk assessment tool should balance interpretability with sufficient flexibility. 

To simultaneously address these three challenges, we propose robust single-index modal regression \citep{feng2020statistical} for PrD risk assessment. Our proposal models the conditional mode (rather than the conditional mean) of PD given covariates, ensuring its robustness to potential skewness and heavy-tailedness. Since we do not specify the functional form of the single-index function, we also avoid the pitfalls of restrictive linear assumptions. However, to balance flexibility with interpretability, we impose a monotonicity constraint on the single-index function, thereby facilitating clinically actionable findings. For example, clinicians can rank subjects' PrD risk directly from their index values, with higher scores indicating greater risk. Our model has the following novel features:

\begin{enumerate}
    \item We model the PD response using the two-piece scale Student-$t$ (ST) distribution \citep{Rubio2015}. The ST distribution generalizes the normal distribution, enabling robust modeling across a wide range of data types, including those with the characteristics commonly observed in PrD measurements. The conditional mode of the response is then linked to covariates through a monotonic single-index function. This modal regression formulation naturally accommodates skewed and heavy-tailed data, and is pleasingly interpretable as the ``most probable'' value of the response given a set of covariates. In contrast, mean regression is very sensitive to outliers and data asymmetry\revone{, and this} can lead to erroneous statistical inferences under violations of the Gaussian errors assumption \citep{Yao2014,feng2020statistical}. 
    \item For flexible modeling of the unknown monotone single-index function, we use a neural network \citep[NN;][]{caterini2018deep}. As we discuss in Section \ref{sec:related-work}, several existing methods for modeling the single-index function are highly sensitive to the choice of smoothing parameters. In contrast, our proposed NN achieves superior approximation accuracy using a default network architecture of \revone{one hidden layer with 256 nodes}. This obviates the need for manual tuning and offers a computationally efficient and user-friendly alternative to other methods. To the best of our knowledge, our work is the first NN-based monotonic single-index model in the context of modal regression.

\item We equip our model with the following theoretical guarantees. First, we prove that the proposed model is identifiable, thus enabling us to uniquely estimate both the single-index function and the regression coefficients. \revone{As a consequence, we establish the Fisher consistency of our estimator.} Second, we prove that the proposed NN architecture enforces monotonicity, a critical property for the interpretability of single-index models. Finally, we establish a universal approximation theorem for our constrained architecture, demonstrating its ability to approximate any continuous monotonic single-index function with arbitrary accuracy. \revone{Our identifiability result synthesizes standard single-index model assumptions with properties of the ST distribution, while the monotonicity and universal approximation results are adaptations of classical NN theory to our specific architecture with positive weights and hyperbolic tangent activations. We acknowledge, however, that we have not established sample consistency or formal asymptotic properties of the proposed estimator.}

\end{enumerate}

We demonstrate our method's practical utility through an application to real PD data and comprehensive simulation studies. We further provide clinicians with a practical framework for using the trained model to calculate risk indices and rank subjects' PD severity. 
	
%


\subsection{Motivating Data} \label{Sec:Motivation}

\begin{figure}
	\centering
	\includegraphics[width=1.0\textwidth]{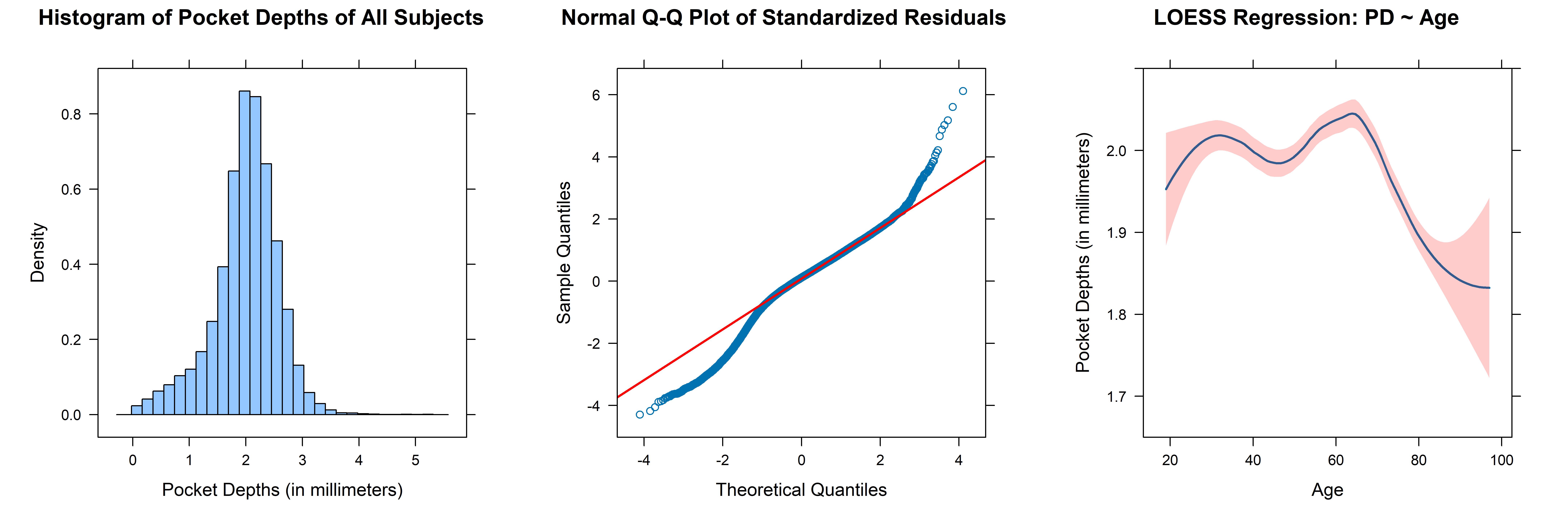}
	\caption{Exploratory data analysis of the HP data. Left panel: Histogram of PD (in millimeters) for all subjects. Middle panel: normal Q-Q plot of the residuals obtained from fitting a linear regression to the PD response. Right panel: plot of the predicted LOESS regression function for PD vs. age, along with its 95\% confidence band.} \label{fig:Exploratory Data Analysis: Histogram, quantile quantile plot and LOESS regression}
\end{figure}

To motivate our methodology, we first present an exploratory data analysis of a PrD dataset generated from the HealthPartners Institute of Minnesota electronic health records, henceforth HP data  \citep{guan2020bayesian, michalowicz2023periodontal}, collected between 2007-2014. This dataset recorded the longitudinal PD (in millimeter, mm), at each of the six probing sites for all available tooth per subject, for $n=24{,}871$ subjects. Our focus in this paper is to model the subject-level PD responses, averaged across all probed sites, at the baseline, and explore its association with eight (important) clinical risk factors, including race, gender, and age. In the left panel of Figure~\ref{fig:Exploratory Data Analysis: Histogram, quantile quantile plot and LOESS regression}, we plot the histogram of PD for all $n$ subjects. The histogram reveals an overall left-skewed distribution, with the right tail tapering off rapidly and the data centered around 2 mm. Upon closer inspection, a few large outliers near the 5 mm mark can also be identified, further highlighting the heavy-tailed and skewed characteristics of the PD measurements.

Next, we fit a traditional linear regression model under the normality assumption, regressing PD against all eight covariates. The normal quantile-quantile (Q-Q) plot of the standardized residuals, shown in the middle panel of Figure~\ref{fig:Exploratory Data Analysis: Histogram, quantile quantile plot and LOESS regression}, reveals a clear departure from normality. Although the residuals align with the theoretical quantiles near the center, significant deviations occur in both tails. The presence of extreme values and the spread of the points in the tail regions suggest that the response follows a heavy-tailed distribution. 

Finally, we employed a locally estimated scatterplot smoothing (LOESS) regression model to examine the relationship between PD and age. The right panel of Figure~\ref{fig:Exploratory Data Analysis: Histogram, quantile quantile plot and LOESS regression} displays the LOESS regression curve along with its 95\% confidence interval. The curve reveals a distinct nonlinear relationship between PD and age that is characterized by multiple fluctuations. This nonlinearity underscores the complexity of the association between age and PD and highlights the need for flexible modeling approaches in analyzing PD data.

\subsection{Related Work} \label{sec:related-work}

The single-index model is a well-established statistical framework that provides a pragmatic compromise between the restrictive fully parametric regressions requiring restrictive and often unrealistic assumptions, and hard-to-interpret fully nonparametric tools. Common methods for estimating the unknown single-index function include kernel-based and spline-based approaches. \citet{Ichimura1993} proposed semiparametric least squares estimation using kernel-based methods, while \citet{Carroll1997} later extended this framework to generalized linear models. \citet{Yu2002} introduced penalized spline estimation for partially linear single-index models, offering a computationally efficient alternative to kernel-based methods. \citet{wang2009spline} further advanced the theoretical foundations of spline estimation for single-index models. For skewed and heavy-tailed data, single-index models have been extended to focus on quantile regression and tail behavior. \citet{Wu2010} proposed kernel-based single-index quantile regression, \citet{Zhu2012} developed semiparametric quantile regression for high-dimensional covariates using kernel-based methods, and \citet{Ma2016} introduced profile optimization for single-index quantile regression with B-spline approximations. Furthermore, \citet{Gardes2017} focused on tail dimension reduction for extreme quantile estimation, while \citet{Xu2022} proposed a tail single-index model using kernel-based methods, explicitly modeling tail dependence and extreme quantiles.

Compared to general single-index models, monotonic single-index models offer greater interpretability since the indexes can be used to rank subjects, albeit at the cost of more restrictive shape constraints. The literature on estimating shape-constrained functions is extensive. \citet{Groeneboom2018} developed estimation methods for monotone single-index models, focusing on consistency and asymptotic properties. \citet{Balabdaoui2019} proposed least squares estimation techniques for monotonic single-index models. A key theoretical foundation for these methods is the Bernstein's theorem, which states that every real-valued, fully monotone function on the half-line $[0, +\infty)$ can be represented as a mixture of exponential functions. Building on this, \citet{hupf2020methods} introduced Bayesian and frequentist methodologies using the Bernstein polynomial basis, and \citet{Acharyya2023} developed semiparametric beta regression models using the same approach. Although Bernstein polynomials are a popular modeling choice for monotonic single-index models, they require the strong assumption of infinite differentiability \citep{Schilling2009} and can face practical challenges in optimal polynomial degree selection \citep{deMelloeSilva2024}.


Neural networks (NNs) have also been applied to monotonic function estimation. Early work by \citet{archer1993application} and \citet{sill1997monotonic} introduced methods for enforcing monotonicity, such as constrained back-propagation and weight-constrained architectures. \citet{daniels2010monotone} and \citet{WangJCGS2024} later proposed partially monotone networks, which enforce monotonicity only on a subset of inputs. \citet{runje2023constrained} also developed constrained monotonic NNs that can approximate any continuous monotonic function, including nonconvex ones, by incorporating additional activation functions. Recently, \citet{Hosseini2023ICLR} showed that shallow rectified linear unit (ReLU) networks trained with stochastic gradient descent can learn monotonic single-index functions with linear sample complexity (up to logarithmic factors).   

Despite this impressive array of methodological advances for monotonic single-index models, there are, to our knowledge, no neural network-based methods which are \emph{specifically} tailored for skewed and heavy-tailed data. Existing methods have focused predominantly on estimating the conditional mean single-index function given covariates. However, when data are heavily skewed and heavy-tailed, as is typically case for PD data (see Section \ref{Sec:Motivation}), the conditional mode offers a much more representative summary of central tendency. The conditional mode is also highly interpretable as the ``most probable'' value of the response given the covariates, and by nature, it is extremely robust to outliers which may obscure the inherent covariate effects suggested by the majority of the data \citep{Liu2024}. Finally, for unimodal and asymmetric distributions, intervals around the conditional mode tend to have higher coverage probability than intervals of the same length around the conditional mean or median \citep{Yao2014, xiang2022modal}. All of these features make modal regression a worthy alternative to quantile regression or extreme value tail modeling.

Modal regression has gained popularity in recent years due to its robustness to outliers. \citet{Yao2014} proposed kernel-based modal linear regression, while \citet{Chen2016} generalized it by removing the linearity assumption. Recently, \citet{feng2020statistical} proposed an alternative approach using classical empirical risk minimization. For a comprehensive review of kernel-based modal regression developments, see \citet{chen2018modal}. However, we are not aware of any modal regression models designed for \emph{monotonic single-index models}. 
Thus, we endeavor to combine the robustness of modal regression with the flexibility of DNNs to construct a monotonic single-index modal regression framework. 

In our framework, we model the response using the ST distribution employed by \citet{Rubio2015} and \citet{Liu2024}. The conditional mode of the response is then linked to the covariates through a monotonic single-index function, which we model using a NN. It should be noted that other distributions are also suitable for modeling skewed and heavy-tailed data. We refer readers to \citet{Azzalini2013} for a comprehensive review of such distributions. However, the ST distribution distinguishes itself through several unique characteristics. First, the ST distribution features a single location parameter that fully controls the mode. This simplifies the estimation, and more crucially, ensures the identifiability of our model. Secondly, the ST distribution contains the normal distribution as a special limiting case, making it flexible enough to handle not only skewed and/or heavy-tailed data but \emph{also} symmetric and thin-tailed data. 


The remainder of this paper is organized as follows. Section~\ref{sec:Methodology} formally presents our monotonic single-index modal regression framework. We describe the procedures for parameter estimation and uncertainty quantification and provide key theoretical results of our model, including model identifiability, monotonicity guarantees, and universal approximation properties. We also provide practical model selection guidelines. Section~\ref{sec:real data application} demonstrates the model's practical utility through an application to the HP data, while Section~\ref{sec:simulation studies} evaluates the finite-sample performance of our method through simulation studies. Finally, Section~\ref{sec:discussion} discusses broader implications, limitations, and potential extensions of this work.

\section{Methodology}\label{sec:Methodology}

Our monotonic single-index modal regression model relates a scalar response variable $y$ to covariates $\mathbf{x}$ through an unknown monotonic function $g(\cdot)$. Throughout this work, we maintain the fundamental assumption that $g(\cdot)$ is monotonic increasing, as this directly reflects the clinical relationship where higher values of the index $u = \boldsymbol{\beta}^\top \mathbf{x}$ correspond to greater severity of PD (or larger PDs). We also emphasize that the proposed methodology can be straightforwardly adapted to estimate monotonic decreasing relationships, although we exclusively consider the increasing case to maintain focus on PD applications. 

Suppose we observe $n$ pairs of data $\{ (y_i, \mathbf{x}_i) \}, i = 1, \ldots, n$. Our model takes the form,
\begin{equation}
	y_i = g(u_i) + e_i, \quad \text{for } i = 1, \dots, n,
	\label{eq:DNNSIM}
\end{equation}
where, $u_i = \boldsymbol{\beta}^\top \mathbf{x}_i$ is the index formed by the $p$-dimensional vector of covariates $\mathbf{x}_i = (x_{i1}, \ldots, x_{ip})^\top$ and a $p$-dimensional coefficient vector $\boldsymbol{\beta}$. For identifiability, we constrain $\boldsymbol{\beta}$ to have unit $L_2$ norm, i.e., $\boldsymbol{\beta}^{\top} \boldsymbol{\beta} = 1$. We further assume that the error terms $e_i$ in \eqref{eq:DNNSIM} independently follow a ST distribution, denoted as $e_i \sim \operatorname{ST}(w,\theta = 0, \sigma, \delta)$, where $w \in [0,1]$, $\sigma > 0$, and $\delta > 2$ are parameters governing the shape and tail behavior of the errors. The probability density function (PDF) for the ST distribution is defined as
\begin{equation}
f_{\text{ST}}(x \mid w, \theta, \sigma, \delta) = w f_{\text{LT}}\left(x~\bigg|~\theta, \sigma \sqrt{\frac{w}{1-w}}, \delta\right) + (1-w) f_{\text{RT}}\left(x~\bigg|~\theta, \sigma \sqrt{\frac{1-w}{w}}, \delta\right),
	\label{eq:ST_Distribution_Density}
\end{equation}
where
\begin{align*}
f_{\text{LT}}(x \mid \theta, \sigma, \delta) = \frac{2}{\sigma} f_{\text{Student-}t}\left(\frac{x-\theta}{\sigma} ~\bigg|~ \delta\right) \mathbb{I}(x < \theta) ~~~\text{and}~~~
f_{\text{RT}}(x \mid \theta, \sigma, \delta) = \frac{2}{\sigma} f_{\text{Student-}t}\left(\frac{x-\theta}{\sigma} ~\bigg|~ \delta\right) \mathbb{I}(x \ge \theta).
\end{align*}
Here, LT and RT denote left-truncated and right-truncated Student-$t$ distributions respectively, $f_{\text{Student-}t}(x \mid \delta)$ is the PDF of a Student-$t$ distribution with $\delta > 2$ degrees of freedom (mode 0, variance $\delta/(\delta-2)$), and $\theta$ serves as both the location parameter and the global mode of the ST distribution. The ST distribution \eqref{eq:ST_Distribution_Density}, originally introduced by \citet{Rubio2015}, generalizes the normal distribution and the two-piece scale normal (SN) distribution, defined below in \eqref{eq:density function of the SN distribution}, to accommodate symmetric, asymmetric, heavy-tailed, \emph{and} light-tailed distributions. Recently, \citet{Liu2024} introduced a Bayesian modal linear regression framework, which includes the ST distribution as a special case.

The skewness parameter $ w \in [0,1] $ in \eqref{eq:ST_Distribution_Density} controls the direction and magnitude of skewness of the ST distribution. When $ w > 0.5 $, the distribution is left-skewed; when $ w < 0.5 $, it is right-skewed; and when $ w = 0.5 $, it is symmetric; S-Figure 1 (in the Supplementary Material) demonstrates the role of $w$ in controlling the direction of skewness in the ST distribution. Meanwhile, the degrees of freedom parameter $\delta$ controls the heaviness of the tails, with smaller $\delta$ leading to heavier tails. Finally, the scale parameter $\sigma$ controls the spread of the ST distribution. The ST distribution with $w \neq 0.5$ and small $\delta$ is especially suitable for modeling PD responses, and other responses exhibiting both (pronounced) asymmetry and heavy tails. However, the ST distribution can also capture symmetric or thin-tailed data. For example, if $w = 0.5$ and $\sigma = 1$, the ST distribution coincides with a standard Student-$t$ distribution. Notably, as $ \delta \rightarrow +\infty $, the ST distribution converges to an SN distribution whose PDF is given by 
\begin{equation}
	f_{\text{SN}}(x \mid w, \theta, \sigma, \delta)=w f_{\mathrm{LT}}\left(x ~\bigg|~ \theta, \sigma \sqrt{\frac{w}{1-w}}~\right)+(1-w) f_{\mathrm{RT}}\left(x ~\bigg|~ \theta, \sigma \sqrt{\frac{1-w}{w}}~\right),
	\label{eq:density function of the SN distribution}
\end{equation}
where $f_{\mathrm{LT}}(x \mid \theta, \sigma)=\frac{2}{\sigma} \phi\left(\frac{x-\theta}{\sigma}\right) \mathbb{I}(x<\theta)$,
$f_{\mathrm{RT}}(x \mid \theta, \sigma)=\frac{2}{\sigma} \phi\left(\frac{x-\theta}{\sigma}\right) \mathbb{I}(x \geq \theta)
$, and $\phi(\cdot)$ represents the PDF of the standard normal distribution. If $ w = 0.5, \sigma = 1$, and $\delta \rightarrow +\infty$ in \eqref{eq:ST_Distribution_Density}, then \eqref{eq:density function of the SN distribution} converges to the standard normal distribution. Thus, the normal distribution, Student-$t$ distribution, and SN distribution can all be viewed as special (limiting) cases of the ST distribution. This flexibility makes the ST distribution a very appealing choice for modeling (continuous) responses that exhibit non-Gaussianity (via any combination of asymmetry and thick tails).


Apart from its flexibility, a crucial feature of the ST distribution \eqref{eq:ST_Distribution_Density} is that its location parameter $\theta$ equals its mode. This enables modal regression by linking the conditional mode of the response to our single-index predictor. Under \eqref{eq:DNNSIM}-\eqref{eq:ST_Distribution_Density}, $$\text{Mode}(y_i \mid \mathbf{x}_i ) = g(\boldsymbol{\beta}^\top \mathbf{x}_i),$$ establishing a direct relationship between the conditional mode of the response and the single-index function. Thus, our framework provides greater robustness against skewed and heavy-tailed data, compared to traditional mean single-index regression models where $\mathbb{E}(y_i \mid \mathbf{x}_i) = g( \boldsymbol{\beta}^\top \mathbf{x}_i)$. The fact that the mode is controlled by a single location parameter is also critical in ensuring model identifiability. This is formally proven in Section \ref{sec: Theoretical Results}. 

\revone{Finally, we discuss the interpretability of $\boldsymbol{\beta}$ within our proposed single-index framework. While the linear combination $\boldsymbol{\beta}^{\top} \mathbf{x}$ provides more transparency than a fully ``black box'' model, the nonlinear link function $g(\cdot)$ inherently complicates a direct marginal interpretation.  Specifically, the marginal effect of a continuous covariate $X_j$ on the conditional mode of the response is given by the partial derivative $\frac{\partial}{\partial x_j} g(\boldsymbol{\beta}^{\top} \mathbf{x}) = g'(\boldsymbol{\beta}^{\top} \mathbf{x}) \beta_j$. Because the NN architecture structurally constrains $g(\cdot)$ to be strictly monotonically increasing, we are mathematically guaranteed that $g'(\cdot) > 0$. Consequently, the sign of $\beta_j$ reliably indicates the direction of the covariate's association with the response. Furthermore, due to the identifiability constraint $\|\boldsymbol{\beta}\| = 1$, the relative magnitudes of the coefficients (for instance, the ratio $\beta_j / \beta_k$) can be meaningfully interpreted as the relative importance of the corresponding covariates within the composite risk index. However, researchers must be careful to note that the absolute values of $\boldsymbol{\beta}$ do not represent constant marginal effects. Thus, our model provides directional and relative interpretability rather than absolute marginal interpretability.}

\subsection{\texorpdfstring{\revone{Neural Network Architecture and Parameter Estimation}}{Neural Network Architecture and Parameter Estimation}} \label{sec: Parameter Estimation and Deep Neural Network Architecture}

Under \eqref{eq:DNNSIM}-\eqref{eq:ST_Distribution_Density}, we observe $n$ pairs of data $\left\{ (y_i, \mathbf{x}_i ) \right\}, i = 1, \dots, n$, and the unknown parameters of interest are $\boldsymbol{\varphi} = \left\{g, \boldsymbol{\beta}, w, \sigma, \delta\right\}$. In order to flexibly estimate the unknown monotonic single-index function $g(\cdot)$, we employ a NN, which we denote by a mapping $G: \mathbb{R}^{1} \mapsto \mathbb{R}^{1}$.  The NN architecture consists of an input layer (i.e. layer 0), \revone{$K \geq 1$} hidden layers (i.e. layers $1, \ldots, K$), and an output layer (i.e. layer $K+1$). For $k = 1, \ldots, K$, let $h_k$ denote the number of neurons in the $k$th hidden layer, and let $h_0 = 1$ and $h_{K+1} = 1$. For each $k$th layer, $k = 1, \ldots, K+1$, the NN first performs an affine transformation by premultiplying the output from the previous layer by a weights matrix $\boldsymbol{A}^{(k)} \in \mathbb{R}^{h_{k} \times h_{k-1}}$ and adding a bias term $\boldsymbol{b}^{(k)} \in \mathbb{R}^{h_{k}}$. Then, we apply a possibly non-affine \textit{activation} function elementwise to these transformed inputs to produce the final outputs of the $k$th layer. The ReLU activation function, $\text{ReLU}(x) = \max \{ 0, x \}$, is arguably the most popular activation function for the hidden layers of a NN \citep{nair2010rectified}. However, when \revone{all of} the NN weights are constrained to be nonnegative -- a typical requirement for \revone{NNs} to estimate monotonically increasing functions \citep{archer1993application, sill1997monotonic}, \revone{a ReLU network is always convex. This may be restrictive in practice.} We only assume that the single index function $g(\cdot)$ in \eqref{eq:DNNSIM} is monotonically increasing, not that it is necessarily convex. Therefore, instead of using ReLU, we choose the hyperbolic tangent function $\tanh(x) = (e^x - e^{-x}) / (e^x + e^{-x})$ as the activation function for the hidden layers of our NN. This does \emph{not} restrict $g(\cdot)$ to be convex, and our research convey that it demonstrates satisfactory empirical performance. Finally, since the responses are continuous, we simply use the identity function as the activation in the output layer. To sum up, the NN model $G(u):\mathbb{R}^{1} \mapsto \mathbb{R}^{1}$ is defined by the composition of $K+1$ functions,\begin{equation}G(u) = \sigma_{K+1} \circ \sigma_K \circ \cdots \circ \sigma_1 (u),\label{eq:definition of G}\end{equation}where, for inputs $\boldsymbol{v} \in \mathbb{R}^{h_{k-1}}$ to the $k$th layer, we have $\sigma_{k}(\boldsymbol{v}) = \tanh(\boldsymbol{A}^{(k)} \boldsymbol{v} + \boldsymbol{b}^{(k)}), \text{ for } k = 1, \ldots, K$, and 
$\sigma_{K+1}(\boldsymbol{v}) = \boldsymbol{A}^{(K+1)} \boldsymbol{v} + b^{(K+1)}.$

\revone{In all of our synthetic and real data examples, we select $K=1$ hidden layer with $h_1 = 256$ nodes to ensure sufficient flexibility. We found that this default setting for the NN architecture strikes a good balance, providing ample capacity to reliably approximate complex nonlinear index functions while computationally avoiding the destructive overfitting and divergence risks frequently associated with excessively deep networks. To rigorously justify this specific architecture in advance of our formal simulation studies, we conducted an extensive hyperparameter sensitivity analysis. This analysis evaluated various network depths, widths, and learning rates under a controlled Monte Carlo simulation setting with a sample size of $n=1000$.  We refer readers to Section 5 of the Supplementary Material for the complete details and visual diagnostics (S-Figures 15--18) supporting this model choice.} This flexibility is one of the key advantages of using a NN over the Bernstein polynomial bases, the latter of which can be highly sensitive to the choice of polynomial degree \citep{deMelloeSilva2024}. In particular, if the true single-index function $g(\cdot)$ is non-smooth or discontinuous, a very high degree polynomial may be required to approximate it sufficiently well. In all of our subsequent real data analyses and simulations presented in Sections \ref{sec:real data application} and \ref{sec:simulation studies} respectively, our NN with the exact same architecture consistently outperformed the competing methods based on Bernstein polynomials. 

To further ensure that $G(u)$ in \eqref{eq:definition of G} is monotonically increasing with respect to $u$, we restrict all the weights, or entries in $\boldsymbol{A}^{(k)}, k= 1, \ldots, K+1$, to be positive. For completeness, we prove in Theorem~\ref{thm: monotocity} that $G(u)$ is monotonically increasing in $u$. The monotonic decreasing case can be easily handled by restricting all the weights in $\boldsymbol{A}^{(k)}, k = 1, \ldots, K+1$, to be negative, and the methodology can be adapted accordingly, without loss of generality. The associated loss function is the negative log-likelihood function of the ST distribution \eqref{eq:ST_Distribution_Density}. Recall that $f_{\text{ST}}\left(y \mid w, \theta, \sigma, \delta\right)$ denotes the PDF of the ST distribution with location/mode parameter $\theta$, skewness parameter $w$, scale parameter $\sigma$, and degrees of freedom $\delta$ controlling the heaviness of the tails. Given observed data $\{ (y_i, \mathbf{x}_i) \}, i = 1, \ldots, n$, our single-index modal regression model minimizes the following loss function with respect to the NN $G$ (i.e. the weights and biases of the NN) and the additional model parameters $\{ \boldsymbol{\beta}, w, \sigma, \delta \}$:\begin{equation}\hat{\boldsymbol{\varphi}} = \underset{G,\boldsymbol{\beta}, w, \sigma, \delta}{\arg\min } \sum_{i = 1} ^{n} -\log \left[f_{\text{ST}}\left( y_i \mid w, \theta = G ( \boldsymbol{\beta}^\top \mathbf{x}_{i} ), \sigma, \delta\right)\right]. \label{eq:loss-function}\end{equation}In particular, $\hat{G}(\hat{\boldsymbol{\beta}}^\top \mathbf{x})$ is the predicted \emph{modal} value of the response $Y$ given $\mathbf{x}$, where $\hat{G}$ denotes the optimized NN and $\hat{\boldsymbol{\beta}}$ is the optimized $\boldsymbol{\beta}$ under \eqref{eq:loss-function}. In the context of our PrD application, $\hat{G}(\hat{\boldsymbol{\beta}}^\top \mathbf{x})$ gives the estimated most probable PD for a subject with observed covariates $\mathbf{x}$. To solve for a local minimum $\hat{\boldsymbol{\varphi}}$ in \eqref{eq:loss-function}, we employ the backpropagation algorithm \citep{Rumelhart1986} combined with the \revone{Adam optimization algorithm \citep{ADAM}}.  \revone{To address the numerical instability frequently encountered when directly optimizing the remaining parameters $\boldsymbol{\beta}, w, \sigma,$ and $\delta$, we employ specific overparameterization and random initialization strategies, with full implementation details provided in Section 4 of the Supplementary Material.} In all of our simulations and real data analyses, we chose 1000 epochs (i.e. the number of passes through the complete dataset) for \revone{Adam}. To assess convergence, we recommend that researchers plot the loss value at the end of each epoch against the epoch number. Examples of these learning curves for the real data application (Section \ref{sec:real data application}) are presented in S-Figure~1 of the Supplementary Material. \revone{Furthermore, because the optimization of NNs is generally non-convex, the estimation of the model parameters $\{\boldsymbol{\beta}, w, \sigma, \delta\}$ may occasionally converge to a local optimum. If local convergence remains a practical concern, we highly recommend that practitioners experiment with multiple different initial values or random seeds. By comparing the resulting models and selecting the configuration that achieves the highest maximized log-likelihood, one can systematically identify the optimal parameter estimates.} Our method was implemented using \texttt{PyTorch} and is publicly available in the \texttt{R} package \href{https://doi.org/10.32614/CRAN.package.DNNSIM}{\texttt{DNNSIM}} on the Comprehensive \texttt{R} Archive Network \texttt{CRAN}.

\subsection{Uncertainty Quantification} \label{sec:uncertainty}

The solution to \eqref{eq:loss-function} returns point estimates for both the single-index function and other model parameters. While this provides valuable insights into the model, it does not convey the reliability or precision of our estimates. Uncertainty quantification is a critical component for providing a measure of confidence for the estimates produced by our model. To leverage the parametric assumption on the residual errors, we adopt the well-established parametric bootstrap procedure \citep{Efron2012} for uncertainty quantification. 

To begin the parametric bootstrap procedure, we require the training dataset $\{(y_{i}, \mathbf{x}_{i}) \}, i = 1,\dots, n$, and initial estimates $\{ \hat{G}^{(0)}(\cdot), \hat{\boldsymbol{\beta}}^{(0)}, \hat{w}^{(0)},  \hat{\sigma}^{(0)},\hat{\delta}^{(0)} \}$ obtained from solving the loss function \eqref{eq:loss-function}. Then, for the $b$th iteration, we randomly sample $n$ realizations $y_i^{(b)}, i = 1, \ldots, n$, from the ST distribution with parameters $\hat{w}^{(b-1)}$, $\hat{\theta}^{(b-1)} = \hat{G}^{(b-1)}(\hat{\boldsymbol{\beta}}^{(b-1)\top} \mathbf{x}_i)$, $\hat{\sigma}^{(b-1)}$, and $\hat{\delta}^{(b-1)}$, and subsequently solve \eqref{eq:loss-function} using the $y_i^{(b)}$'s to produce a bootstrap estimate $\hat{\boldsymbol{\varphi}}^{(b)}$. Repeating this process $B$ times results in $B$ bootstrap estimates.  The $2.5\%$ and $97.5\%$ sample quantiles of these $B$ estimates can then be used to construct the associated 95\% confidence intervals. The complete parametric bootstrap procedure is given in Algorithm~\ref{alg:parametric bootstrap}.

\renewcommand{\algorithmicrequire}{\textbf{Input}}
\renewcommand{\algorithmicensure}{\textbf{Output}}
\begin{algorithm}[t!]
	\caption{Parametric Bootstrap for our Single-Index Modal Regression Model}\label{alg:parametric bootstrap}
	\begin{algorithmic}
		\Require: Training dataset: $\{(y_{i}, \mathbf{x}_{i}) \}, i = 1,\dots, n$, initial estimates of parameters $\big\{  \hat{G}^{(0)} (\cdot),\hat{\boldsymbol{\beta}}^{(0)},\hat{w}^{(0)},\hat{\sigma}^{(0)},\hat{\delta}^{(0)} \big\}$, \\
        \hspace{.85cm} bootstrap size $B$
		\Ensure: $B$ bootstrap estimates $\hat{\boldsymbol{\varphi}}^{(b)} = \big\{ \hat{G}^{(b)}(\cdot),\hat{\boldsymbol{\beta}}^{(b)},\hat{w}^{(b)},\hat{\sigma}^{(b)},\hat{\delta}^{(b)} \big\},~~b = 1, \dots, B$
		\State $b \gets 1$ \Comment{The beginning of the bootstrap procedure}
		\While{$b \leq B$}     
		\State \text{For } $i = 1, \ldots, n$, \text{draw a random sample } $y^{(b)}_{i} \text{ from ST} \big( \hat{w}^{(b-1)}, \hat{\theta}^{(b-1)} = \hat{G}^{(b-1)}(\hat{\boldsymbol{\beta}}^{(b-1)\top}\mathbf{x}_{i}), \hat{\sigma}^{(b-1)}, \hat{\delta}^{(b-1)} \big)$
		\State $\big\{ \hat{G}^{(b)}(\cdot),\hat{\boldsymbol{\beta}}^{(b)},\hat{w}^{(b)},\hat{\sigma}^{(b)},\hat{\delta}^{(b)} \big\} \gets$ Solution to \eqref{eq:loss-function} using $\{ (y^{(b)}_{i}, \mathbf{x}_{i}) \}, i = 1,\dots, n$
        \State $b \gets b + 1$
		\EndWhile \Comment{The end of the bootstrap procedure}
	\end{algorithmic}
\end{algorithm}

\subsection{Theoretical Results} \label{sec: Theoretical Results}

In this section, we provide several theoretical results that justify the use of our NN-based monotonic single-index modal regression model. The detailed proofs for these results are provided in Section~2 of the Supplementary Material.  Initially, we demonstrate that the model \eqref{eq:DNNSIM} with ST-distributed noise \eqref{eq:ST_Distribution_Density} is fully identifiable.  Without identifiability, the model parameters cannot be meaningfully estimated, and the model training process will tend to be unstable. We first state a lemma which establishes that all the parameters $\{w, \theta, \sigma, \delta \}$ in the ST distribution \eqref{eq:ST_Distribution_Density} can be uniquely estimated from data that arise from an ST$(w, \theta, \sigma, \delta)$ distribution.

\begin{lemma}[Identifiability of the ST Distribution]
	The ST distribution in \eqref{eq:ST_Distribution_Density} is identifiable.
	\label{lemma:identifiable}
\end{lemma}

In our single-index model \eqref{eq:DNNSIM}, the location/mode parameter for the $i$th observation is $\theta_i = g(\boldsymbol{\beta}^\top \mathbf{x}_i)$. Therefore, to ensure identifiability of $g(\cdot)$ and $\boldsymbol{\beta}$, we require some additional assumptions on these parameters. These assumptions are presented in Conditions (C1)-(C3) of Theorem \ref{theorem:identifiable}. As a result of Lemma \ref{lemma:identifiable} and these conditions, our model in \eqref{eq:DNNSIM} can also be identified from the data. This is formalized in the following theorem.

\begin{theorem}[Identifiability of the Single-Index Modal Regression]
Let $ m(\mathbf{x}) = g(\boldsymbol{\beta}^\top \mathbf{x}) $ be a function with a vector input $ \mathbf{x}$ and a scalar-valued output. Suppose the following conditions hold:
\begin{enumerate}[(C{\arabic*})]
		\item The support of $ m(\cdot) $, denoted as $ S $, is a bounded convex set with at least one interior point.
		\item The single-index function $ g(\cdot) $ is a monotonic increasing function on its support.
		\item The $ L_2 $ norm of $ \boldsymbol{\beta} $ is one, i.e., $ \boldsymbol{\beta}^{\top} \boldsymbol{\beta} = 1 $.
	\end{enumerate}
	Then the model in \eqref{eq:DNNSIM} is identifiable. \label{theorem:identifiable}
\end{theorem}
It is important to note that Theorem \ref{theorem:identifiable} does not imply that the NN $ G(\cdot) $ is identifiable. NNs are well-known to lack identifiability \citep{Sussmann1992}, since different networks with different weights and biases can lead to the same output. Instead, Theorem \ref{theorem:identifiable} establishes that all the parameters of interest $ \boldsymbol{\varphi} = \left\{g, \boldsymbol{\beta}, w, \sigma, \delta\right\} $ are identifiable, providing a theoretical foundation for accurate estimation of \eqref{eq:DNNSIM}. Without identifiability of \eqref{eq:DNNSIM}, no method, NN or otherwise, can meaningfully estimate the parameters.

\revone{A natural implication of this identifiability result is the Fisher consistency of our estimator, which we formalize in the following proposition. Fisher consistency guarantees that, at the population level, the true model parameters uniquely minimize the expected loss function. While Theorem~\ref{theorem:identifiable} ensures that the model parameters are uniquely determined by the data distribution, the Fisher consistency result confirms that our optimization objective targets the correct minimizer. Fisher consistency is often a desirable  property for an estimator. In particular, if an estimator lacks Fisher consistency, then the estimation procedure would fail to converge to the true parameter vector even if we had access to an infinite amount of training data. It is important to emphasize, however, that Fisher consistency does not guarantee that stochastic gradient-based optimization procedures will converge to this global minimizer, as the loss surface may contain multiple local minima and the optimization algorithm may become trapped in one of them.}

\revone{\begin{proposition}[Fisher Consistency]
\label{prop:fisher_consistency}
Assume the data are generated according to~\eqref{eq:DNNSIM}--\eqref{eq:ST_Distribution_Density} with true parameters
$\boldsymbol{\varphi}_0 = (g_0,\boldsymbol{\beta}_0,w_0,\sigma_0,\delta_0)$
satisfying conditions (C1)--(C3). Additionally, assume that $\mathbf{x}$ is a random vector with distribution $p_{\mathbf{x}}$ whose support satisfies Condition (C1) for both $\boldsymbol{\beta}_0$ and any candidate $\boldsymbol{\beta}$ under consideration (C4). Let $\boldsymbol{\varphi} = (g,\boldsymbol{\beta},w,\sigma,\delta)$ be any candidate parameters satisfying the same regularity conditions, and define the population negative log-likelihood
$$
\Psi(\boldsymbol{\varphi}) = \mathbb{E}_{(\mathbf{x},Y)}\Bigl[-\log f_{\mathrm{ST}}\bigl(Y \mid w,\; \theta = g(\boldsymbol{\beta}^\top \mathbf{x}),\; \sigma,\; \delta\bigr)\Bigr],
$$
where the expectation is taken with respect to the true joint distribution of $(\mathbf{x},Y)$.
Then, $\Psi(\boldsymbol{\varphi}) \geq \Psi(\boldsymbol{\varphi}_0)$, 
for all $\boldsymbol{\varphi}$, with equality if and only if $\boldsymbol{\varphi} = \boldsymbol{\varphi}_0$. Consequently, the true parameter vector $\boldsymbol{\varphi}_0$ is the unique minimizer of the population loss functional $\Psi(\boldsymbol{\varphi})$.
\end{proposition}}

Next, we prove that the NN $ G(u)$ in \eqref{eq:definition of G} is monotonic increasing in its index value $u$. Theorem \ref{thm: monotocity} ensures that our NN \eqref{eq:definition of G} with hyperbolic tangent activation functions and positive weights can be used to approximate the unknown monotonic single-index function $ g(\cdot) $ in \eqref{eq:DNNSIM}.

\begin{theorem}[Monotonicity of the NN Model]
	If $\boldsymbol{A}^{(k)} > 0$ for $k = 1,\dots, K+1$, then the NN $G(u)$ in \eqref{eq:definition of G} is monotonically increasing. Here, $\boldsymbol{A}^{(k)} > 0$ denotes that all elements of $\boldsymbol{A}^{(k)}$ are greater than zero.
	\label{thm: monotocity}
\end{theorem}

Finally, from the fact that $G(\cdot)$ in \eqref{eq:definition of G} is guaranteed to be a monotonic network, we establish a new universal approximation theorem. Theorem \ref{thm:universal approximation theory} states that with sufficiently many hidden layers, $G(\cdot)$ can approximate any univariate, continuous, monotonic increasing single-index function $ g(\cdot) $. This provides theoretical support for the excellent empirical performance of our NN-based single-index modal regression model.

\begin{theorem}[Universal Approximation Theorem for the Monotonic NN Model]
For any univariate continuous, monotonic increasing function $m(\cdot): \Omega \mapsto \mathbb{R}^{1}$, where $\Omega$ is a compact subset of $\mathbb{R}^{1}$, there exists a NN $G(\cdot)$ in \eqref{eq:definition of G}  with at most $k$ hidden layers and strictly positive weights such that, for any $u \in \Omega$ and $\epsilon > 0$, $|m(u) - G(u)| < \epsilon.$ \label{thm:universal approximation theory}
\end{theorem}

\revone{We briefly clarify the scope and limitations of our theoretical contributions. The identifiability result in Theorem~\ref{theorem:identifiable} combines standard single-index assumptions with properties of the ST distribution to justify estimation within our modal regression framework. Although the proof parallels existing single-index arguments, the result is necessary for our novel model formulation. The monotonicity guarantee in Theorem~\ref{thm: monotocity} and the universal approximation result in Theorem~\ref{thm:universal approximation theory} adapt classical feedforward neural-network theory to our constrained architecture with positive weights and hyperbolic tangent activations. We do not present these as standalone theoretical advances, but as guarantees that the proposed architecture remains sufficiently expressive to approximate any continuous monotone function. Several limitations remain. We do not establish convergence guarantees for the optimization algorithm; the universal approximation result provides no approximation rates and therefore does not characterize the trade-off between network complexity and estimation accuracy; and Fisher consistency in Proposition~\ref{prop:fisher_consistency} does not imply asymptotic consistency of the estimator. Full statistical consistency, especially contraction rates under the ST loss, remains open and is discussed as future work in Section~\ref{sec:discussion}.}

\subsection{Model Selection Guidelines}

As discussed in Section \ref{sec:Introduction}, the ST distribution \eqref{eq:ST_Distribution_Density} is very flexible in its ability to capture both skewness and symmetry, as well as both heavy and thin tails, depending on the model parameters. Nevertheless, since we assume ST errors in our single-index model \eqref{eq:DNNSIM}, practitioners may wish to assess the adequacy of this parametric assumption. Our proposed framework offers multiple model selection approaches grounded in its theoretical structure.  

First, researchers can visually assess the adequacy of the ST assumption through a residual diagnostic plot. Specifically, we can plot a histogram of the residuals $\hat{e}_i = y_i - \hat{G}(\hat{\boldsymbol{\beta}}^\top \mathbf{x}_i), i = 1, \ldots, n$, and compare the empirical distribution of the $\hat{e}_i$'s against the theoretical ST density \eqref{eq:ST_Distribution_Density} evaluated with the estimated parameters $\{ \hat{w}, \hat{\theta}_i = \hat{G}(\hat{\boldsymbol{\beta}}^\top \mathbf{x}_i),  \hat{\sigma}, \hat{\delta} \}$ from \eqref{eq:loss-function}. We demonstrate this diagnostic plot approach in our real data application in Section \ref{sec:real data application} (top left panel of Figure~\ref{fig:Real Data Application}). 

Second, the hierarchical structure of the ST distribution \eqref{eq:ST_Distribution_Density} provides heuristic selection criteria for the distribution of $e_i$ in \eqref{eq:DNNSIM}. Recall that in \eqref{eq:ST_Distribution_Density}, the parameters $w$ and $\delta$ control the skewness and the tail heaviness, respectively. We can obtain the bootstrap confidence intervals (CIs) of $w$ and $\delta$ using the parametric bootstrap procedure in Section \ref{sec:uncertainty}. If the 95\% CI for $w$ does \textit{not} contain 0.5 and the 95\% CI for $\delta$ has an upper endpoint less than 30, then we can conclude that the ST distribution is indeed the most appropriate error distribution for our data. On the other hand, if the 95\% CI for $w$ contains 0.5 and the 95\% CI for $\delta$ either contains 30 or has a lower endpoint greater than 30, then the normal distribution may have been an adequate choice. If the 95\% CI for $w$ excludes 0.5 but the 95\% CI for $\delta$ either contains 30 or has both endpoints greater than 30, then the SN distribution may be the most appropriate. Finally, if the 95\% CI for $w$ contains 0.5 but the upper endpoint of the 95\% CI for $\delta$ is less than 30, then we could use the symmetric Student-$t$ distribution for $e_i$ in \eqref{eq:DNNSIM}. In Section \ref{sec:real data application}, we describe how to fit these alternative single-index models.  

Finally, if the primary focus is on prediction, then comparative evaluation through $K$-fold cross-validation is a suitable way to select an appropriate error distribution in our single-index model \eqref{eq:DNNSIM}. In this case, we can fit several models with varying error distributions and select the one with the lowest cross-validated mean squared error (MSE) for the single-index function $g(u)$. We demonstrate the utility of this approach in our real data analysis in Section \ref{sec:real data application} (see, bottom panels of Figure~\ref{fig:Real Data Application}).

\section{Application: HealthPartners Data} \label{sec:real data application}

In this section, we illustrate our proposed methodology via application to the aforementioned HP dataset consisting of $n=24{,}871$ subjects. In addition to the average PD (measured in millimeters) at baseline for each subject, the other covariates available include gender (male or female), race (White, Black, or Other), age, diabetes status (yes or no), tobacco usage (user or non-user), flossing frequency (daily or less than daily), and insurance status (insured or uninsured). We regress PD on the other covariates. Note, age is the only continuous explanatory variable in this study, and it is standardized to have a mean of zero and a standard deviation of one.

\revone{We denote our proposed NN-based single-index model with noise $e_i$ from the ST distribution \eqref{eq:ST_Distribution_Density} as ST-GX-D. We also considered four other competing NN-based models: SN-GX-D assuming noise from the SN distribution \eqref{eq:density function of the SN distribution}, N-GX-D assuming noise from the normal distribution $\mathcal{N}(0, \sigma^2)$, MED-GX-D for median regression, and T-GX-D assuming noise from the symmetric Student-$t$ distribution.  For fitting the SN-GX-D and T-GX-D models, we replaced $f_{\text{ST}} ( y_i \mid \cdot)$ in \eqref{eq:loss-function} with the corresponding PDFs of the SN and symmetric Student-$t$ distributions, respectively. For the N-GX-D model, we replaced $f_{\text{ST}}( y_i \mid \cdot )$ in \eqref{eq:loss-function} with $\phi( (y_i - G(\boldsymbol{\beta}^\top \mathbf{x}_i))/\sigma )$, where $\phi(\cdot)$ is the standard normal PDF. For the MED-GX-D model, we directly utilized the mean absolute error (MAE) as the objective loss function instead of a likelihood-based formulation.  All five of these NN-based single-index models employed the exact same network architecture described in Section \ref{sec: Parameter Estimation and Deep Neural Network Architecture}.}

\revone{We also compared the empirical performance of the NN models to methods based on Bernstein polynomials \citep{hupf2020methods, Acharyya2023}. For these approaches, we used Bernstein polynomial bases to estimate $g(\cdot)$ in \eqref{eq:DNNSIM}. We use the notations ST-GX-B, SN-GX-B, N-GX-B, and MED-GX-B to denote Bernstein polynomial methods fitted utilizing the ST distribution, the SN distribution, the normal distribution, and the MAE loss, respectively. Additionally, we evaluated T-GX-PB, which designates the Student-$t$ single-index model where the unknown single-index function is estimated using a Bernstein polynomial augmented with an $L_2$ penalty applied directly to the polynomial coefficients to regularize the estimation process.  For all these polynomial-based baseline models, we set the dimension of the Bernstein coefficients to be 51, which is sufficiently large for most practical purposes \citep{McKayCurtis2011}.}

\revone{In addition to the ten single-index models (ST-GX-D, SN-GX-D, N-GX-D, MED-GX-D, T-GX-D, ST-GX-B, SN-GX-B, N-GX-B, MED-GX-B, and T-GX-PB), we also considered models that directly modeled the central location using a NN. These models, denoted as ST-FX, SN-FX, and N-FX, take the form,
\begin{equation}
	y_i = f\left(\mathbf{x}_i\right) + e_i, \quad \text{for } i = 1, \dots, n,
	\label{eq:model FX}
\end{equation}
where, a NN is used to approximate the unknown $ f(\mathbf{x}_i)$ directly, and $e_i$ follows the ST, SN and normal distributions respectively. Note that there is no vector of regression coefficients $\boldsymbol{\beta}$ in \eqref{eq:model FX}. This comparison aims to demonstrate that single-index models with a parametric component $\boldsymbol{\beta}^\top \mathbf{x}$ can strike a balance between flexibility and interpretability, offering a robust framework for analyzing complex data structures. The NN architectures for these fully flexible FX models consist of two hidden layers with a size of 512 nodes each.} For all models, we set the number of epochs to 1000 for model training. To verify the convergence of these methods, we plotted the learning curves of the loss value at the end of each epoch vs. the epoch number in S-Figures 2--4 of the Supplementary Material. The figure reveals that training became relatively stable around 100 epochs, and that 1000 epochs was adequate for all models.

\begin{figure}
\centering
\includegraphics[width=0.495\textwidth]{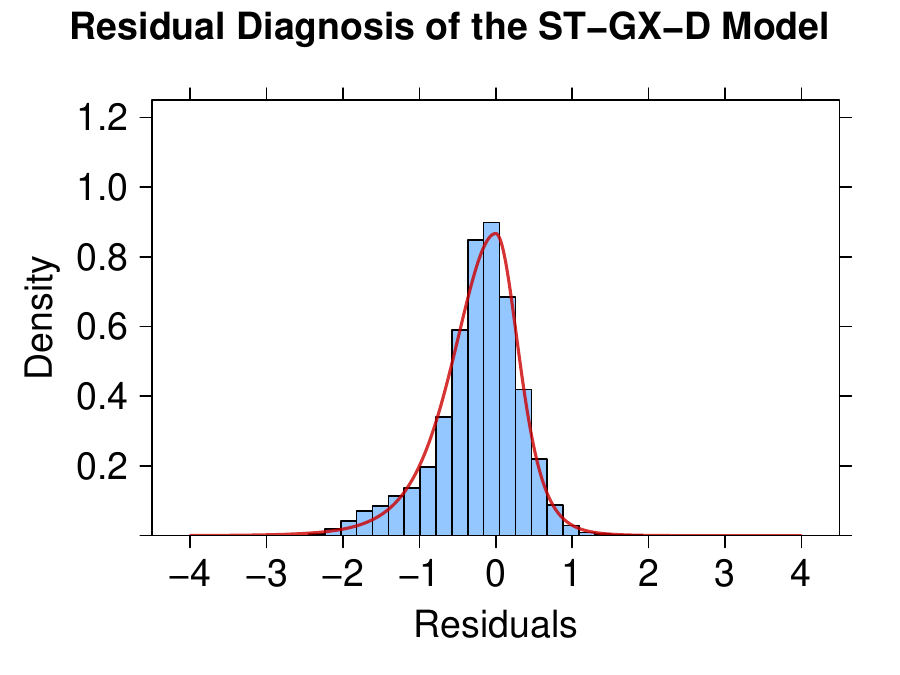}
\includegraphics[width=0.495\textwidth]{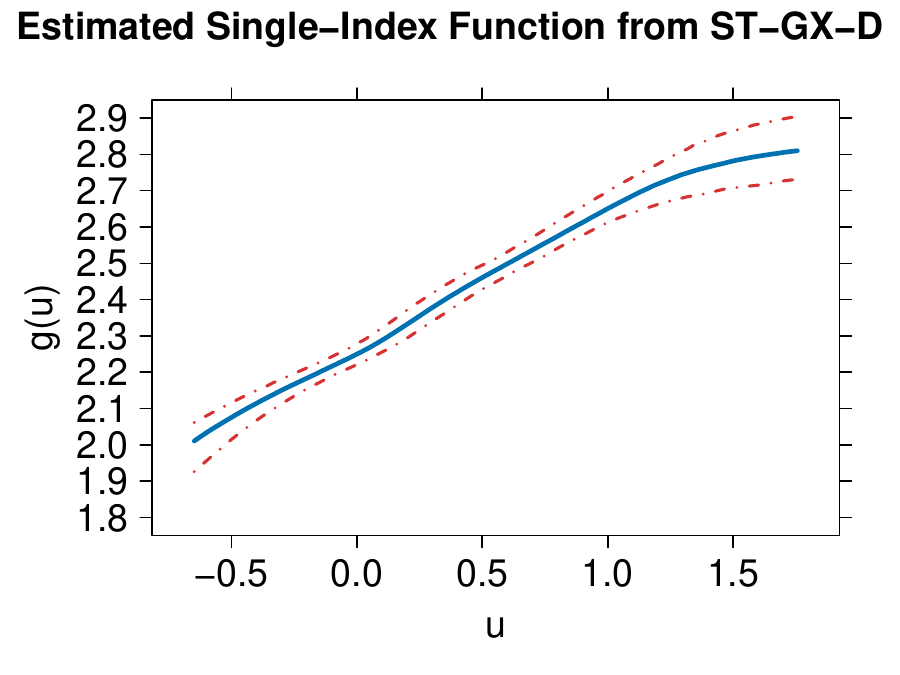}
\includegraphics[width=0.495\textwidth]{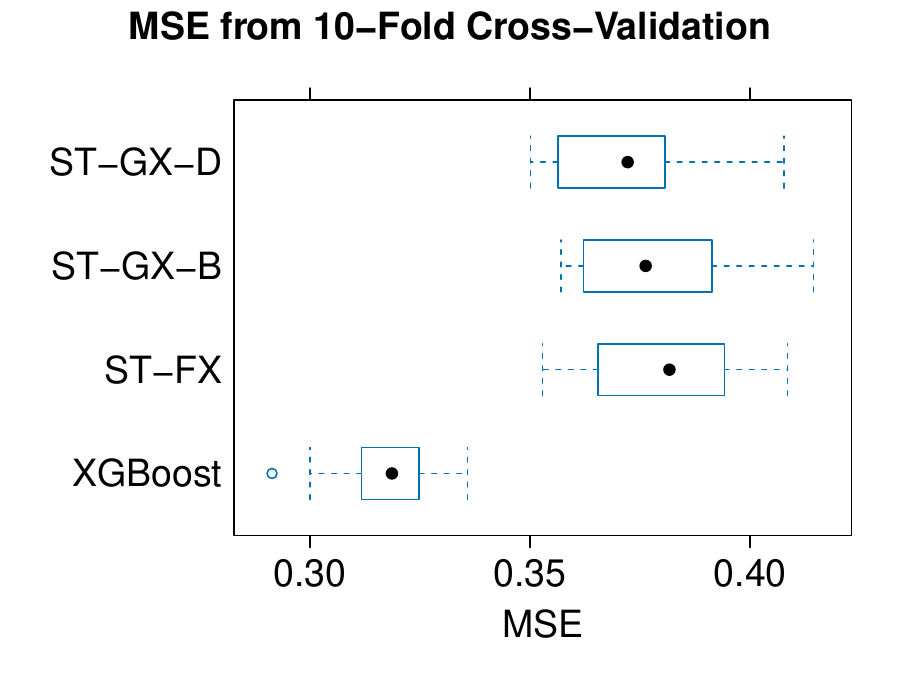}
\includegraphics[width=0.495\textwidth]{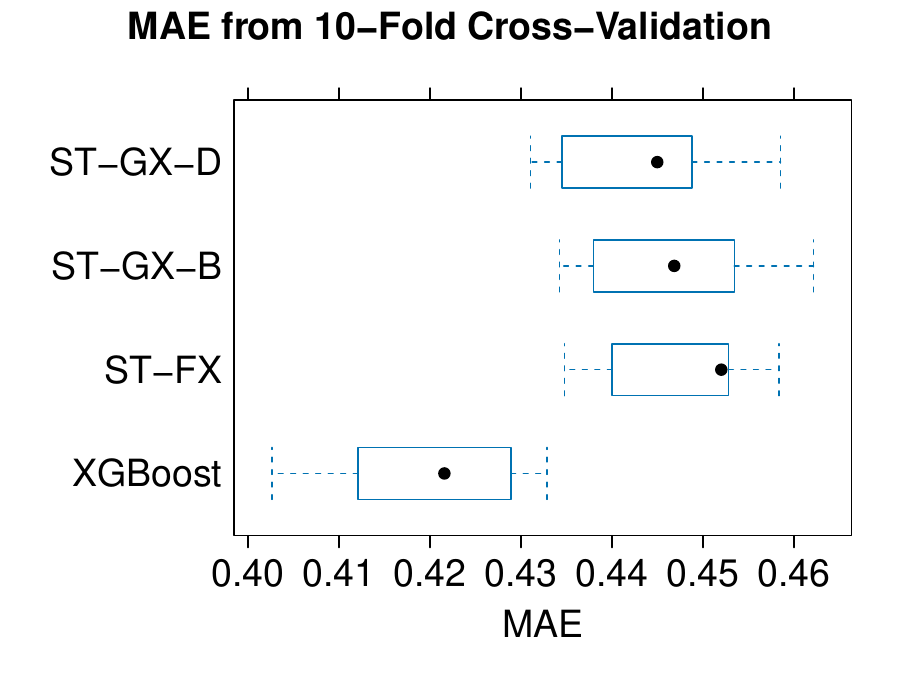}
\caption{\revone{Application to HP data: The residual diagnostic plot from fitting the ST-GX-D model (top left), the estimated single-index function with 95\% pointwise confidence bands from the ST-GX-D model (top right), and box-plots of the MSE (bottom left) and MAE (bottom right) from 10-fold cross-validation comparing the predictive performance of the ST-GX-D, ST-GX-B, ST-FX and XGBoost models.}} \label{fig:Real Data Application}
\end{figure}

\revone{Next, we used residual diagnostic plots to compare the fitted models and identify the error distribution or loss function best supported by the data; see Figure~\ref{fig:Real Data Application}. For each model, we compared the empirical residual histogram with the theoretical density curve constructed from the final parameter estimates. The diagnostics showed a clear hierarchy. First, the normal models (N-GX-D, N-GX-B, and N-FX) failed to capture both skewness and heavy-tailedness: their residual histograms were clearly left-skewed, rather than symmetric, and had much higher peaks than the theoretical normal densities. Second, the median regression models (MED-GX-D and MED-GX-B), which implicitly assume Laplace errors, captured heavy tails somewhat better but failed to accommodate the pronounced asymmetry. Third, the symmetric Student-$t$ models (T-GX-D and T-GX-PB) captured the heavy tails extremely well, but not the skewness. Conversely, the skew-normal models (SN-GX-D, SN-GX-B, and SN-FX) captured the skewness, but failed to account for the heavy tails, leading to misaligned empirical and theoretical peaks. Finally, the ST models (ST-GX-D, ST-GX-B, and ST-FX) provided the best fit, with residual histograms closely aligned with the theoretical density curves by simultaneously capturing skewness and heavy tails. Based on these findings, we excluded all non-ST models from further consideration. The top-left panel of Figure~\ref{fig:Real Data Application} displays the residual diagnostic plot for the proposed ST-GX-D model, with the theoretical ST density overlaid on the residual histogram. The remaining diagnostic plots are reported in S-Figures 5--7 of the Supplementary Material.}

\revone{Furthermore, we used 10-fold cross-validation (CV) to compare the out-of-sample predictive performance of the three final ST models (ST-GX-D, ST-GX-B, and ST-FX), with eXtreme Gradient Boosting (XGBoost) \citep{chen2016xgboost}. The XGBoost baseline was trained for up to 100 boosting rounds, with early stopping after 10 rounds to prevent overfitting, and was optimized using root mean squared error as the training evaluation metric. The bottom panels of Figure~\ref{fig:Real Data Application} display boxplots of out-of-sample MSE and MAE across the 10 CV test folds. ST-GX-B and ST-FX demonstrated broadly similar predictive behavior, while the proposed ST-GX-D model achieved competitive accuracy. Across all models, XGBoost had the lowest median out-of-sample MSE and MAE. Although this indicates superior raw predictive performance for XGBoost on this dataset, ST-GX-D remained highly accurate while offering a major interpretability advantage. In this PrD application, identifying and understanding clinical risk factors is arguably more important than maximizing raw prediction accuracy. Unlike XGBoost, whose complex tree ensemble obscures the precise nature of individual covariate effects, ST-GX-D provides a structured patient risk index with directional and relative interpretability. This makes it a practically useful model for assessing the comparative contributions of specific risk factors to disease severity.
}

\begin{table}
	\centering
	\begin{adjustbox}{max width=0.95\textwidth}
		\begin{tabular}[t]{lrrr}
			\toprule
			& Estimate & Lower Bound (2.5\%) & Upper Bound (97.5\%)\\
			\midrule
			Age (rescaled)  & 0.0504 & 0.0330 & 0.0697\\
			Gender: Male (Ref: Female) & 0.4458 & 0.3631 & 0.5061\\
			Race: Black or African American (Ref: Other Races) & 0.7219 & 0.6430 & 0.8226\\
			Race: White (Ref: Other Races) & -0.3137 & -0.3840 & -0.2411\\
			Diabetes: Yes (Ref: No Diabetes) & 0.1620 & 0.0998 & 0.2223\\
			Tobacco Usage: Yes (Ref: Non-User) & 0.3600 & 0.2920 & 0.4071\\
			Flossing Frequency: Daily (Ref: Less Than Daily) & -0.0561 & -0.0928 & -0.0209\\
			Insured (Ref: Uninsured) & -0.1107 & -0.1668 & -0.0614\\
			\addlinespace
			$w$ (skewness parameter) & 0.6423 & 0.6342 & 0.6507\\
			$\sigma$ (scale parameter) & 0.4207 & 0.4143 & 0.4273\\
			$\delta$ (degrees of freedom) & 5.2788 & 4.9647 & 5.6207\\
			\bottomrule
		\end{tabular}
	\end{adjustbox}
	\caption{Results from fitting the ST-GX-D model to the HP data. The table entries are the point estimate, and the corresponding lower and upper bounds of the 95\% confidence interval, for all model parameters. For the binary covariates, we list the reference group (Ref) in parentheses.}
	\label{tab:Real Data Application: Point Estimation}
\end{table}

Finally, we employed the parametric bootstrap procedure with 1000 repetitions to quantify the uncertainty of the point estimates for the ST-GX-D model. Using the 2.5\% and 97.5\% percentiles of the bootstrap estimates, we constructed 95\% confidence intervals for all model parameters. Table~\ref{tab:Real Data Application: Point Estimation} reports these point estimates and 95\% CIs, implying that the estimate of the skewness parameter $w$ was \revone{0.6423}, with a 95\% CI of \revone{(0.6342, 0.6507)}. Since both endpoints of the 95\% CI for $w$ were larger than 0.5, we conclude that the conditional density of PD given the covariates was indeed left-skewed. Additionally, the estimate of the degree of freedoms $\delta$ was \revone{5.2788}, with a 95\% CI of \revone{(4.9647, 5.6207)}. This suggests that the conditional density for PD was also heavy-tailed, further justifying the use of the ST distribution \eqref{eq:ST_Distribution_Density} for modeling the errors in our single-index model.

Our analysis also provides interpretable insights into the relationships between PrD and the explanatory variables. Table~\ref{tab:Real Data Application: Point Estimation} shows that none of the 95\% confidence intervals for the coefficients in $\boldsymbol{\beta}$ contained 0, indicating that all covariates were statistically significant. In particular, we found that age was positively associated with higher PD, i.e. the severity of PrD significantly increases with age. Male subjects exhibited significantly higher PrDs than females, underscoring the association between gender and periodontal health. Black or African-American individuals were also found to be more vulnerable to PrD compared to White individuals, highlighting the racial disparities in oral health outcomes. In addition, diabetes and tobacco usage were identified as being significantly associated with greater risk of PrD, whereas daily flossing was associated with significantly lower risk (or lower PrDs). Finally, individuals without dental insurance were at significantly higher risk of PrD than those who had insurance. These results suggest that there is a critical need to educate subjects about daily flossing and expand access to dental care in order to promote and maintain good oral health. Our findings are consistent with those in established studies in the PrD literature \citep{Marlow2011,Fleming2018}, confirming the appropriateness of our ST-GX-D model. Our single-index modal regression model was able to accurately capture the underlying relationships between the explanatory variables and PrD severity.

In the top right panel of Figure~\ref{fig:Real Data Application}, we plot the estimate of the single index function $g(u)$ as a function of the index $u$. The estimated function is the blue solid curve, and the 95\% pointwise confidence intervals are the dashed red curves. \revone{The top right panel of Figure~\ref{fig:Real Data Application} clearly shows that the estimated single-index function deviates from a straight line.  The curve increases at a relatively steady rate for values of $u < 1.0$, but for larger values of $u$ ($u \geq 1.0$), the curve begins to level off and increase much more slowly. This confirms that the relationship between the PD and the covariates is nonlinear, thus justifying the use of a flexible NN-based approach to model the single-index function.}

\textbf{Index interpretation}:  To demonstrate the practical application of our model, we provide the fitted equation $\hat{u} = \hat{\boldsymbol{\beta}}^\top \mathbf{x}$ below for calculating the indexes:
\begin{equation}\label{fitted-u}
\begin{aligned} 
\hat{u} =& ~\frac{\text{Age}-54.981}{15.107} \times 0.0504
+ \mathbb{I}\left(\text{Gender: Male}\right) \times 0.4458
+ \mathbb{I}\left(\text{Race: Black or African American}\right) \times 0.7219\\
&-\mathbb{I}\left(\text{Race: White}\right) \times 0.3137
+\mathbb{I}\left(\text{Diabetes: Yes}\right) \times 0.1620
+\mathbb{I}\left(\text{Tobacco Usage: Yes}\right) \times 0.3600\\
&-\mathbb{I}\left(\text{Flossing Frequency: Daily}\right) \times 0.0561
-\mathbb{I}\left(\text{Insured}\right) \times 0.1107. 
\end{aligned}
\end{equation}
Consider the following two hypothetical subjects: (1) a 60-year-old Black/African American male with diabetes, who is a tobacco user, who does not floss daily, and who has no dental insurance; and (2) a 30-year-old White female without diabetes, who does not use tobacco, who flosses daily, and who has dental insurance. Based on \eqref{fitted-u}, the first subject has an estimated index of \revone{$\hat{u} \approx 1.7064$}, while the second subject has an estimated index of \revone{$\hat{u} \approx -0.5638$}. Thus, the second subject has a lower predicted risk of developing PrD than the first subject.

\revone{As discussed before Section~\ref{sec: Parameter Estimation and Deep Neural Network Architecture}, the identifiability constraint allows the absolute ST-GX-D coefficient magnitudes to be interpreted as relative covariate importance. We therefore qualitatively compare them with the XGBoost Gain metric. Both models agree on the primary categorical risk factors, ranking Black or African American race, male gender, and tobacco usage among the most important variables. However, XGBoost assigns substantially higher importance to age, likely because it captures complex nonlinear age patterns to maximize accuracy, whereas ST-GX-D imposes a strictly linear additive age effect within the risk index. This contrast highlights a key clinical modeling trade-off: XGBoost gains accuracy through opaque nonlinear relationships, while ST-GX-D provides a structured composite risk index with clear directional and relative interpretability for practical decision making.}

\section{Simulation Studies} \label{sec:simulation studies}

Our simulation studies are comprised of four distinct schemes to study the finite sample properties of the ST-GX-D - our proposed single-index modal regression. The first scheme (Scheme I) evaluates the accuracy of point estimation and uncertainty quantification of ST-GX-D for a nonconvex, continuous single-index function. The second scheme (Scheme II) demonstrates the ST-GX-D model's empirical superiority over the ST-GX-B approach (i.e. the method using Bernstein polynomial bases) when the true single-index function has discontinuities. 
The third scheme (Scheme III) investigates ST-GX-D's robustness to misspecification of the residual error distribution $e_i$ in \eqref{eq:DNNSIM}. \revone{This scheme shows that ST-GX-D is able to maintain strong performance, even under model misspecification and outlier contamination, and demonstrates its competitive performance relative to XGBoost.} \revone{Finally, the fourth scheme (Scheme IV) compares ST-GX-D to both ST-FX and XGBoost, demonstrating that ST-GX-D achieves competitive out-of-sample predictive performance against both the more flexible deep learning model and the powerful tree-based ensemble method.} In all schemes, we generated three covariates $(x_1, x_2, x_3)$ as follows. The binary covariate $x_1 \sim \text{Bernoulli}(0.5)$, and a continuous covariate $x_2$ is drawn from $\text{Uniform}(-3, 0)$ if $x_1 = 0$, or $\text{Uniform}(0, 3)$ if $x_1=1$. Finally, we simulated $x_3 \sim \text{Uniform}(-3.5, 2.5)$. We fixed the coefficient vector $\boldsymbol{\beta} = (\beta_1, \beta_2, \beta_3)^\top = \frac{1}{\sqrt{3}}(1, 1, 1)^\top$ throughout.

\subsection{Scheme I}

In Scheme I, we considered a continuous function  for the true single-index function, 
\begin{equation} \label{eq:scheme1}
    g\left(u\right) = 10 \times \Phi\left(2.5 u \right),
\end{equation}  
where, $\Phi\left(\cdot \right)$ represents the cumulative distribution function for the standard normal distribution. The noise was generated from a $\text{ST}(0.6, 0, 1, 5)$  distribution characterized by left skewness ($w = 0.6$), location $\theta=0$, scale $\sigma=1$, and heavy tails ($\delta = 5$). It should be noted that \eqref{eq:scheme1} is a \emph{nonconvex} function. Therefore, using ReLU networks with strictly positive weights would not be appropriate.


We generated data with two sample sizes, $n=1000$ and $n=2000$, using \revone{300} Monte Carlo replicates. \revone{For each replication, we recorded the point estimates and computed the estimation biases for $\boldsymbol{\beta}$, $w$, $\sigma$, and $\delta$. The empirical bias distributions are summarized via boxplots in Figure~\ref{fig:bias_point_estimation}.}  \revone{For both sample sizes, the trained ST-GX-D model produced accurate point estimates, with the biases for $\boldsymbol{\beta}, w, \sigma$, and $\delta$ all concentrated around 0. The bias variability for $\delta$ was more pronounced than for the other parameters, consistent with existing literature indicating that estimation of the degrees-of-freedom parameter is inherently challenging \citep{hasannasab2021alternatives}. Furthermore, as the sample size increased from $n=1000$ to $n=2000$, both the average biases and the variability of the estimates visibly decreased, confirming that larger samples reduce estimation variance and improve precision, as theoretically expected. Table~\ref{tab:bootstrap_coverage} further reports the empirical coverage rates of the 95\% parametric bootstrap confidence intervals, constructed using 1000 bootstrap iterations. For the single-index coefficients $\boldsymbol{\beta}$, the skewness weight $w$, and the scale parameter $\sigma$, the coverage rates were highly satisfactory and close to the nominal 0.95, especially for $n=2000$. In contrast, the degrees-of-freedom parameter $\delta$ showed systematic over-coverage, with rates of 0.990 and 0.997 for $n=1000$ and $n=2000$, respectively. This likely reflects the pronounced estimation difficulty discussed previously, and empirical bias associated with $\delta$, which can cause the resulting confidence interval coverage to deviate from the nominal level. }

\revone{
\begin{figure}[ht]
\centering
\includegraphics[width=\linewidth]{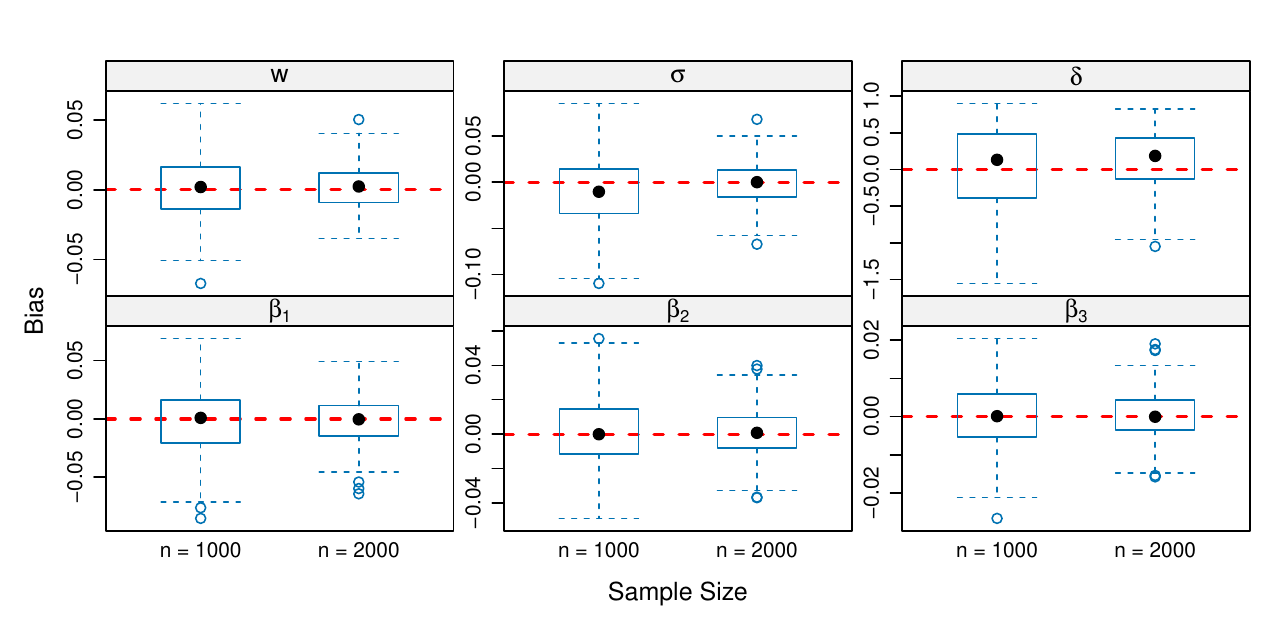}
\caption{\label{fig:bias_point_estimation}Boxplots of the estimation biases for the model parameters $\boldsymbol{\beta}$, $w$, $\sigma$, and $\delta$ under Scheme I, evaluated at sample sizes $n=1000$ and $n=2000$ over 300 Monte Carlo replications.}
\end{figure}

\begin{table}[ht]
    \centering
\begin{tabular}{cccccc}
\toprule
Sample Size ($n$) & Parameter & Nominal Level & Total Replicates & Successful Covers & Coverage Rate\\
\midrule
 & $\beta_1$ & 0.95 & 300 & 282 & 0.940\\

 & $\beta_2$ & 0.95 & 300 & 284 & 0.947\\

 & $\beta_3$ & 0.95 & 300 & 281 & 0.937\\

 & $w$ & 0.95 & 300 & 280 & 0.933\\

 & $\sigma$ & 0.95 & 300 & 270 & 0.900\\

\multirow{-6}{*}{\centering\arraybackslash 1000} & $\delta$ & 0.95 & 300 & 297 & 0.990\\
\cmidrule{1-6}
 & $\beta_1$ & 0.95 & 300 & 286 & 0.953\\

 & $\beta_2$ & 0.95 & 300 & 285 & 0.950\\

 & $\beta_3$ & 0.95 & 300 & 287 & 0.957\\

 & $w$ & 0.95 & 300 & 287 & 0.957\\

 & $\sigma$ & 0.95 & 300 & 288 & 0.960\\

\multirow{-6}{*}{\centering\arraybackslash 2000} & $\delta$ & 0.95 & 300 & 299 & 0.997\\
\bottomrule
\end{tabular}
\caption{\label{tab:bootstrap_coverage}Empirical 95\% coverage rates of the parametric bootstrap confidence intervals for the model parameters ($\boldsymbol{\beta}, w, \sigma, \delta$) at sample sizes $n=1000$ and $n=2000$, based on 300 Monte Carlo replications under Scheme I.}
\end{table}
}

We also validated the performance of ST-GX-D for estimating the true single-index function $g(u)$ in \eqref{eq:scheme1}. S-Figure 8 of the Supplementary Material shows the boxplot of the mean squared errors (MSEs) of our estimates for $g(u)$ from all replications. In general, the MSEs were close to 0, and the median of MSEs for $n = 2000$ is smaller than that for $n = 1000$. This demonstrates that ST-GX-D accurately estimated the true nonconvex function $g(u)$ in \eqref{eq:scheme1}, with estimation improving for larger $n$.

\subsection{Scheme II}

For Scheme II, we selected a more challenging function to estimate: a discontinuous function with abrupt changes in value, given as $g\left(u\right) = \lfloor u \rfloor$, where, $\lfloor u \rfloor$ denotes the floor function of $u$, or the greatest integer $\le u$. The noise term in the second scheme was the same as that in the first scheme. In this scheme, we generated $n=1000$ independent observations from the model \eqref{eq:DNNSIM} with $g(u)$ as the true single-index function. We then fit the ST-GX-D and ST-GX-B models to the simulated data. We repeated this for \revone{300} Monte Carlo replicates. The averaged simulation results are summarized in S-Table 1 (see Section 4.1 of the Supplementary Material) 
We observe that the ST-GX-D model yielded point estimates that were similarly accurate to those produced by ST-GX-B. For both models, most of the average biases fell close to 0, with the exception of the degrees of freedom parameter $\delta$ as previously observed in Scheme I. More importantly, ST-GX-D demonstrated superior performance in estimating the true discontinuous index function $g(u)$.  S-Figure 9 of the Supplementary Material displays boxplots of the MSEs for the 300 Monte Carlo estimates of $g(u)$ produced by both models. From this figure, it is clearly evident that ST-GX-D consistently achieved substantially smaller MSEs when estimating $g(u)$ compared to the ST-GX-B baseline. 

\subsection{Schemes III and IV: Central Findings }
In this subsection, we present only the key findings from Schemes III and IV, with complete details of the simulation designs and results relegated to Section 4 of the Supplementary Material. 

Scheme III demonstrates the ST-GX-D model's robustness under model misspecification. Namely, the true data generating mechanism for \eqref{eq:DNNSIM} had a mixture of an asymmetric Laplace distribution (see, S-Figure 10 in Supplementary Material) and a normal distribution centered at \revone{10} for the residual errors. This resulted in a left-skewed distribution with 1\% outliers in the upper tail. Based on \revone{300} Monte Carlo replicates, ST-GX-D achieved near-perfect median estimates of the true $\boldsymbol{\beta}$ coefficients (Supplementary Material, S-Figure 11) and the smallest median MSE for estimation of the true single-index function among all the fitted models (Supplementary Material, S-Figure~12). \revone{Furthermore, a 10-fold cross-validation evaluation showed that ST-GX-D successfully outperformed the state-of-the-art XGBoost algorithm in terms of the median out-of-sample MAE (Supplementary Material, S-Figure~13).}  These results highlight ST-GX-D's practical effectiveness and resilience to complex model misspecification.  In Scheme IV, we compared the predictive performance of the ST-GX-D model \eqref{eq:DNNSIM} to the fully flexible ST-FX model \eqref{eq:model FX} \revone{and the XGBoost model}. \revone{By employing a robust 10-fold cross-validation procedure}, ST-GX-D \revone{achieved the best overall predictive generalization, yielding a lower median out-of-sample MSE and MAE than both ST-FX and XGBoost} (Supplementary Material, S-Figure~14).  Notably, ST-GX-D achieved this superior predictive performance while maintaining relative interpretability through the parametric component $\boldsymbol{\beta}^\top \mathbf{x}$, which is a critical advantage over the more ``black box'' ST-FX \revone{and XGBoost models}.

\section{Discussion} \label{sec:discussion}

In this paper, we proposed a robust monotonic single-index model that is especially well-suited for skewed and heavy-tailed data. By employing the ST distribution \eqref{eq:ST_Distribution_Density} as the residual error distribution, we model the conditional \emph{mode} of the response given covariates (rather than the mean), thereby ensuring our method's robustness to outliers. We further use a NN to approximate the unknown monotonic single-index function. Our NN enforces monotonicity and flexibly captures various shapes of the monotone function (including nonconvex and discontinuous ones), while requiring minimal tuning of hyperparameters. Our method is implemented in the \textsf{R} package \href{https://doi.org/10.32614/CRAN.package.DNNSIM}{\textsc{DNNSIM}}, available on \textsc{CRAN} \citep{Liu2025}. Complete simulation code for reproducibility is available on \href{https://github.com/rh8liuqy/DNNSIM}{GitHub} at \url{https://github.com/rh8liuqy/DNNSIM}. While the real data application file (162 MB) exceeds GitHub's size limitations, the dataset and associated codes are available upon reasonable request to the corresponding author.

Our model offers a flexible yet interpretable framework for analyzing complex health data, particularly data arising in PrD research. The ST distribution captures the skewness and heavy tails  commonly encountered in PrD data, while the NN ensures accurate modeling of nonlinear associations between individual risk factors and a subject's most probable (i.e. the modal) value for pocket depth. Finally, due to the monotonicity constraint, practitioners can readily rank subjects' PD risk based on their index values. When applied to a real dataset (the HP data) derived from EHRs, our method provided interpretable insights into the covariate effects of risk factors such as age, diabetes status, tobacco use, flossing frequency, and insurance status, while also accurately quantifying PrD risk. In short, we have introduced a practical analytical tool for clinicians and researchers to study PrD.

Despite its robustness and flexibility, our current proposal treats all of the observations as independent, and does not incorporate subject-specific random effects (intercepts/slopes), which are common techniques in modeling hierarchical or longitudinal data. Future work would encompass extending our model to incorporate random effects, thus enhancing its applicability to more complex data structures \citep{Schumacher2021}. It may also be worthwhile to extend our method to handle discrete responses, e.g. categorical/ordinal levels of PrD among subjects, or zero-inflated counts for the number of teeth lost to gum diseases. These extensions would further broaden the applicability of our approach.

\revone{Finally, although Fisher consistency of the proposed estimator has been established in Proposition~\ref{prop:fisher_consistency}, full statistical consistency and contraction rates under the ST loss remain open. A natural direction is to develop these results within the sieve maximum likelihood framework of \citet{Shen2023}, who established existence, consistency, convergence rates, and a central limit theorem for single-layer neural-network sieves in nonparametric models. In our setting, this would require treating the monotone single-index architecture as a restricted NN sieve and extending the associated empirical process arguments to handle modal regression and heavy-tailed ST error distributions, rather than standard least-squares criteria. We leave this full asymptotic justification for future work.} 

\section*{Competing interests}
No competing interest is declared.

\section*{Author contributions statement}
Q.L and S.W. conceived the experiments; while S.W wrote the initial PyTorch code, the entirety of the code implementation, both in simulated and real data, was done by Q.L. Furthermore, Q.L, R.B and D.B developed the methodological framework, wrote, and reviewed the manuscript. 


\section*{Acknowledgements}
\revone{The authors thank the Associate Editor and the two reviewers for their constructive comments}. They gratefully acknowledge the HealthPartners Institute of Minnesota for providing both the motivating dataset and clinical context for this research. This work was partially supported by grants R21DE031879 and R01DE031134 from the United States National Institutes of Health. The majority of the paper was completed, when Qingyang Liu was a postdoctoral researcher at the Department of Statistics at Texas A\&M University (2023--2024), and the University of Wisconsin--Madison (2024--2026).

\bibliographystyle{abbrvnat}
\bibliography{reference}

@article{hasannasab2021alternatives,
  title={Alternatives to the {EM} algorithm for {ML} estimation of location, scatter matrix, and degree of freedom of the {S}tudent t distribution},
  author={Hasannasab, Marzieh and Hertrich, Johannes and Laus, Friederike and Steidl, Gabriele},
  journal={Numerical Algorithms},
  volume={87},
  number={1},
  pages={77--118},
  year={2021},
  publisher={Springer}
}

@book{caterini2018deep,
  title={{Deep Neural Networks in a Mathematical Framework}},
  author={Caterini, Anthony L and Chang, Dong Eui},
  year={2018},
  publisher={Springer}
}

@article{michalowicz2023periodontal,
  title={{Periodontal treatment and subsequent clinical outcomes and medical care costs: A retrospective cohort study}},
  author={Michalowicz, Bryan S and Anderson, Jeffrey P and Kottke, Thomas E and Dehmer, Steven P and Worley, Donald C and Kane, Sheryl and Basile, Sarah and Rindal, D Brad},
  journal={PloS one},
  volume={18},
  number={8},
  pages={e0290028},
  year={2023},
  publisher={Public Library of Science San Francisco, CA USA}
}

@article{BotelhoPeriodontology2022,
author = {Botelho, João and Machado, Vanessa and Leira, Yago and Proença, Luís and Chambrone, Leandro and Mendes, José João},
title = {Economic burden of periodontitis in the {U}nited {S}tates and {E}urope: An updated estimation},
journal = {Journal of Periodontology},
volume = {93},
number = {3},
pages = {373-379},
year = {2022}
}

@article{McKayCurtis2011,
  title = {A variable selection approach to monotonic regression with {B}ernstein polynomials},
  volume = {38},
  comment =  {1360-0532},
  comment = {http://dx.doi.org/10.1080/02664761003692423},
  comment = {10.1080/02664761003692423},
  number = {5},
  journal = {Journal of Applied Statistics},
  publisher = {Informa UK Limited},
  author = {McKay Curtis,  S. and Ghosh,  Sujit K.},
  year = {2011},
  pages = {961–976}
}

@article{yu2005three,
  title={A three-parameter asymmetric {L}aplace distribution and its extension},
  author={Yu, Keming and Zhang, Jin},
  journal={Communications in Statistics—Theory and Methods},
  volume={34},
  number={9-10},
  pages={1867--1879},
  year={2005},
  publisher={Taylor \& Francis}
}

@article{lin2007identifiability,
  title={Identifiability of single-index models and additive-index models},
  author={Lin, Wei and Kulasekera, KB},
  journal={Biometrika},
  volume={94},
  number={2},
  pages={496--501},
  year={2007},
  publisher={Oxford University Press}
}

@inproceedings{nair2010rectified,
  title={Rectified linear units improve restricted {B}oltzmann machines},
  author={Nair, Vinod and Hinton, Geoffrey E},
  booktitle={ICML '10: Proceedings of the 27th International Conference on Machine Learning},
  pages={807--814},
  location = {Haifa, Israel},
  year={2010}
}

@article{Rumelhart1986,
  title = {Learning representations by back-propagating errors},
  volume = {323},
  comment =  {1476-4687},
  comment = {http://dx.doi.org/10.1038/323533a0},
  comment = {10.1038/323533a0},
  number = {6088},
  journal = {Nature},
  publisher = {Springer Science and Business Media LLC},
  author = {Rumelhart,  David E. and Hinton,  Geoffrey E. and Williams,  Ronald J.},
  year = {1986},
  pages = {533–536}
}

@article{Liu2024,
  title = {Bayesian modal regression based on mixture distributions},
  volume = {199},
  comment =  {0167-9473},
  comment = {http://dx.doi.org/10.1016/j.csda.2024.108012},
  comment = {10.1016/j.csda.2024.108012},
  journal = {Computational Statistics and Data Analysis},
  publisher = {Elsevier BV},
  author = {Liu,  Qingyang and Huang,  Xianzheng and Bai,  Ray},
  year = {2024},
  pages = {108012}
}

@article{teicher1963identifiability,
  title={Identifiability of finite mixtures},
  author={Teicher, Henry},
  journal={The Annals of Mathematical Statistics},
  pages={1265--1269},
  year={1963},
  volume = {34},
  number = {4},
  publisher={JSTOR}
}

@article{daniels2010monotone,
  title={Monotone and partially monotone neural networks},
  author={Daniels, Hennie and Velikova, Marina},
  journal={IEEE Transactions on Neural Networks},
  volume={21},
  number={6},
  pages={906--917},
  year={2010},
  publisher={IEEE}
}

@inproceedings{runje2023constrained,
  title={Constrained monotonic neural networks},
  author={Runje, Davor and Shankaranarayana, Sharath M},
  booktitle={ICML'23: Proceedings of the 40th International Conference on Machine Learning},
  pages={29338--29353},
  year={2023},
  location = {Honolulu, Hawaii, USA}
}

@article{Liu2025,
  title = {{DNNSIM}: Single-Index Neural Network for Skewed Heavy-Tailed Data},
  comment = {http://dx.doi.org/10.32614/CRAN.package.DNNSIM},
  comment = {10.32614/cran.package.dnnsim},
  journal = {CRAN: Contributed Packages},
  publisher = {The R Foundation},
  author = {Liu,  Qingyang and Wang,  Shijie and Bai,  Ray and Bandyopadhyay,  Dipankar},
  year = {2025},
}

@article{Efron2012,
  title = {Bayesian inference and the parametric bootstrap},
  volume = {6},
  comment =  {1932-6157},
  comment = {http://dx.doi.org/10.1214/12-AOAS571},
  comment = {10.1214/12-aoas571},
  number = {4},
  journal = {The Annals of Applied Statistics},
  publisher = {Institute of Mathematical Statistics},
  author = {Efron,  Bradley},
  year = {2012},
  pages = {1971-1997}
}

@article{Villoria2024,
  title = {Periodontal disease: A systemic condition},
  volume = {96},
  comment =  {1600-0757},
  comment = {http://dx.doi.org/10.1111/prd.12616},
  comment = {10.1111/prd.12616},
  number = {1},
  journal = {Periodontology 2000},
  publisher = {Wiley},
  author = {Villoria,  German E. M. and Fischer,  Ricardo G. and Tinoco,  Eduardo M. B. and Meyle,  Joerg and Loos,  Bruno G.},
  year = {2024},
  pages = {7–19}
}

@article{Donos2017,
  title = {The periodontal pocket},
  volume = {76},
  comment =  {1600-0757},
  comment = {http://dx.doi.org/10.1111/prd.12203},
  comment = {10.1111/prd.12203},
  number = {1},
  journal = {Periodontology 2000},
  publisher = {Wiley},
  author = {Donos,  Nikolaos},
  year = {2017},
  pages = {7–15}
}

@article{Bandyopadhyay2010,
  title = {Linear mixed models for skew‐normal/independent bivariate responses with an application to periodontal disease},
  volume = {29},
  comment =  {1097-0258},
  comment = {http://dx.doi.org/10.1002/sim.4031},
  comment = {10.1002/sim.4031},
  number = {25},
  journal = {Statistics in Medicine},
  publisher = {Wiley},
  author = {Bandyopadhyay,  Dipankar and Lachos,  Victor H. and Abanto‐Valle,  Carlos A. and Ghosh,  Pulak},
  year = {2010},
  pages = {2643–2655}
}

@article{Ichimura1993,
  title = {Semiparametric least squares ({SLS}) and weighted {SLS} estimation of single-index models},
  volume = {58},
  comment =  {0304-4076},
  comment = {http://dx.doi.org/10.1016/0304-4076(93)90114-K},
  comment = {10.1016/0304-4076(93)90114-k},
  number = {1–2},
  journal = {Journal of Econometrics},
  publisher = {Elsevier BV},
  author = {Ichimura,  Hidehiko},
  year = {1993},
  pages = {71–120}
}

@article{Carroll1997,
  title = {Generalized Partially Linear Single-Index Models},
  volume = {92},
  comment =  {1537-274X},
  comment = {http://dx.doi.org/10.1080/01621459.1997.10474001},
  comment = {10.1080/01621459.1997.10474001},
  number = {438},
  journal = {Journal of the American Statistical Association},
  publisher = {Informa UK Limited},
  author = {Carroll,  R. J. and Fan,  Jianqing and Gijbels,  Irène and Wand,  M. P.},
  year = {1997},
  pages = {477–489}
}

@article{Yu2002,
  title = {Penalized Spline Estimation for Partially Linear Single-Index Models},
  volume = {97},
  comment =  {1537-274X},
  comment = {http://dx.doi.org/10.1198/016214502388618861},
  comment = {10.1198/016214502388618861},
  number = {460},
  journal = {Journal of the American Statistical Association},
  publisher = {Informa UK Limited},
  author = {Yu,  Yan and Ruppert,  David},
  year = {2002},
  pages = {1042–1054}
}

@article{wang2009spline,
  title={Spline estimation of single-index models},
  author={Wang, Li and Yang, Lijian},
  journal={Statistica Sinica},
  pages={765--783},
  year={2009},
  volume = {19},
  number = {2},
  publisher={JSTOR}
}

@article{Wu2010,
  title = {Single-index quantile regression},
  volume = {101},
  comment =  {0047-259X},
  comment = {http://dx.doi.org/10.1016/j.jmva.2010.02.003},
  comment = {10.1016/j.jmva.2010.02.003},
  number = {7},
  journal = {Journal of Multivariate Analysis},
  publisher = {Elsevier BV},
  author = {Wu,  Tracy Z. and Yu,  Keming and Yu,  Yan},
  year = {2010},
  pages = {1607–1621}
}

@article{Zhu2012,
  title = {Semiparametric quantile regression with high-dimensional covariates},
  comment =  {1017-0405},
  comment = {http://dx.doi.org/10.5705/ss.2010.199},
  comment = {10.5705/ss.2010.199},
  journal = {Statistica Sinica},
  publisher = {Statistica Sinica (Institute of Statistical Science)},
  author = {Zhu,  Liping and Huang,  Mian and Li,  Runze},
  year = {2012},
  volume = {22},
  number = {4},
  pages = {1379-1401}
}

@article{Ma2016,
  title = {Inference for single-index quantile regression models with profile optimization},
  volume = {44},
  comment =  {0090-5364},
  comment = {http://dx.doi.org/10.1214/15-AOS1404},
  comment = {10.1214/15-aos1404},
  number = {3},
  journal = {The Annals of Statistics},
  publisher = {Institute of Mathematical Statistics},
  author = {Ma,  Shujie and He,  Xuming},
  year = {2016},
  pages = {1234-1268}
}

@article{Gardes2017,
  title = {Tail dimension reduction for extreme quantile estimation},
  volume = {21},
  comment =  {1572-915X},
  comment = {http://dx.doi.org/10.1007/s10687-017-0300-x},
  comment = {10.1007/s10687-017-0300-x},
  number = {1},
  journal = {Extremes},
  publisher = {Springer Science and Business Media LLC},
  author = {Gardes,  Laurent},
  year = {2017},
  pages = {57–95}
}

@article{Xu2022,
  title = {Extreme Quantile Estimation Based on the Tail Single-index Model},
  comment =  {1017-0405},
  comment = {http://dx.doi.org/10.5705/ss.202020.0051},
  comment = {10.5705/ss.202020.0051},
  journal = {Statistica Sinica},
  publisher = {Statistica Sinica (Institute of Statistical Science)},
  author = {Xu,  Wen and Wang,  Huixia Judy and Li,  Deyuan},
  year = {2022},
  volume = {32},
  number = {2},
  pages = {893-914}
}

@book{hupf2020methods,
  title={Methods in Monotone Single-Index Models Using Functional and Scalar Covariates},
  author={Hupf, Bradley},
  year={2020},
  publisher={Doctoral dissertation, Florida State University}
}

@article{Acharyya2023,
  title = {A monotone single index model for missing-at-random longitudinal proportion data},
  volume = {51},
  comment =  {1360-0532},
  comment = {http://dx.doi.org/10.1080/02664763.2023.2173156},
  comment = {10.1080/02664763.2023.2173156},
  number = {6},
  journal = {Journal of Applied Statistics},
  publisher = {Informa UK Limited},
  author = {Acharyya,  Satwik and Pati,  Debdeep and Sun,  Shumei and Bandyopadhyay,  Dipankar},
  year = {2023},
  pages = {1023–1040}
}

@article{Balabdaoui2019,
  title = {Least squares estimation in the monotone single index model},
  volume = {25},
  comment =  {1350-7265},
  comment = {http://dx.doi.org/10.3150/18-BEJ1090},
  comment = {10.3150/18-bej1090},
  number = {4B},
  journal = {Bernoulli},
  publisher = {Bernoulli Society for Mathematical Statistics and Probability},
  author = {Balabdaoui,  Fadoua and Durot,  Cécile and Jankowski,  Hanna},
  year = {2019},
  pages = {3276-3310}
}

@article{WangJCGS2024,
  title = {Generative Quantile Regression with Variability Penalty},
  volume = {33},
  number = {4},
  journal = {Journal of Computational and Graphical Statistics},
  author = {Wang, Shijie and Shn, Minsuk and Bai, Ray},
  year = {2024},
  pages = {1202-1213}
}

@inproceedings{Hosseini2023ICLR,
  author       = {Alireza Mousavi Hosseini and
                  Sejun Park and
                  Manuela Girotti and
                  Ioannis Mitliagkas and
                  Murat A. Erdogdu},
  title        = {Neural Networks Efficiently Learn Low-Dimensional Representations
                  with {SGD}},
  booktitle    = {The Eleventh International Conference on Learning Representations,
                  {ICLR} 2023, Kigali, Rwanda, May 1-5, 2023},
  year         = {2023},
}

@article{Groeneboom2018,
  title = {Estimation in monotone single‐index models},
  volume = {73},
  comment =  {1467-9574},
  comment = {http://dx.doi.org/10.1111/stan.12138},
  comment = {10.1111/stan.12138},
  number = {1},
  journal = {Statistica Neerlandica},
  publisher = {Wiley},
  author = {Groeneboom,  Piet and Hendrickx,  Kim},
  year = {2018},
  pages = {78–99}
}

@article{archer1993application,
  title={Application of the back propagation neural network algorithm with monotonicity constraints for two-group classification problems},
  author={Archer, Norman P and Wang, Shouhong},
  journal={Decision Sciences},
  volume={24},
  number={1},
  pages={60--75},
  year={1993},
  publisher={Wiley Online Library}
}

@article{sill1997monotonic,
  title={Monotonic networks},
  author={Sill, Joseph},
  journal={NIPS '97:  Proceedings of the 11th International Conference on Neural Information Processing Systems},
  volume={10},
  year={1997},
  location = {Denver, CO},
  pages = {661–667}
}

@article{Sussmann1992,
  title = {Uniqueness of the weights for minimal feedforward nets with a given input-output map},
  volume = {5},
  comment =  {0893-6080},
  comment = {http://dx.doi.org/10.1016/S0893-6080(05)80037-1},
  comment = {10.1016/s0893-6080(05)80037-1},
  number = {4},
  journal = {Neural Networks},
  publisher = {Elsevier BV},
  author = {Sussmann,  Héctor J.},
  year = {1992},
  pages = {589–593}
}

@article{Marlow2011,
  title = {Health insurance status is associated with periodontal disease progression among {G}ullah {A}frican-{A}mericans with type 2 diabetes mellitus: Health insurance status association with periodontitis progression in a diabetic {G}ullah population},
  volume = {71},
  comment =  {0022-4006},
  comment = {http://dx.doi.org/10.1111/j.1752-7325.2011.00243.x},
  comment = {10.1111/j.1752-7325.2011.00243.x},
  number = {2},
  journal = {Journal of Public Health Dentistry},
  publisher = {Wiley},
  author = {Marlow,  Nicole M. and Slate,  Elizabeth H. and Bandyopadhyay,  Dipankar and Fernandes,  Jyotika K. and Leite,  Renata S.},
  year = {2011},
  pages = {143–151}
}

@article{Fleming2018,
  title = {Prevalence of daily flossing among adults by selected risk factors for periodontal disease—{U}nited {S}tates,  2011–2014},
  volume = {89},
  comment =  {1943-3670},
  comment = {http://dx.doi.org/10.1002/JPER.17-0572},
  comment = {10.1002/jper.17-0572},
  number = {8},
  journal = {Journal of Periodontology},
  publisher = {Wiley},
  author = {Fleming,  Eleanor B. and Nguyen,  Duong and Afful,  Joseph and Carroll,  Margaret D. and Woods,  Phillip D.},
  year = {2018},
  pages = {933–939}
}

@article{Schumacher2021,
  title = {Scale mixture of skew‐normal linear mixed models with within‐subject serial dependence},
  volume = {40},
  comment =  {1097-0258},
  comment = {http://dx.doi.org/10.1002/sim.8870},
  comment = {10.1002/sim.8870},
  number = {7},
  journal = {Statistics in Medicine},
  publisher = {Wiley},
  author = {Schumacher,  Fernanda L. and Lachos,  Victor H. and Matos,  Larissa A.},
  year = {2021},
  pages = {1790–1810}
}

@article{Rubio2015,
  title = {Bayesian modelling of skewness and kurtosis with Two-Piece Scale and shape distributions},
  volume = {9},
  comment =  {1935-7524},
  comment = {http://dx.doi.org/10.1214/15-EJS1060},
  comment = {10.1214/15-ejs1060},
  number = {2},
  journal = {Electronic Journal of Statistics},
  publisher = {Institute of Mathematical Statistics},
  author = {Rubio,  F. J. and Steel,  M. F. J.},
  year = {2015},
  pages = {1884-1912}
}

@book{Azzalini2013,
  title = {The Skew-Normal and Related Families},
  comment = {http://dx.doi.org/10.1017/CBO9781139248891},
  comment = {10.1017/cbo9781139248891},
  publisher = {Cambridge University Press},
  author = {Azzalini,  Adelchi and Capitanio,  Antonella},
  year = {2013},
}

@incollection{xiang2022modal,
  title={Modal Regression for Skewed, Truncated, or Contaminated Data with Outliers},
  author={Xiang, Sijia and Yao, Weixin},
  booktitle={Advances and Innovations in Statistics and Data Science},
  editor={Wenqing He and Liqun Wang and Jiahua Chen and Chunfang Devon Lin},
  pages={257--273},
  year={2022},
  publisher={Springer}
}

@article{Yao2014,
  title = {A New Regression Model: Modal Linear Regression},
  volume = {41},
  comment =  {1467-9469},
  comment = {http://dx.doi.org/10.1111/sjos.12054},
  comment = {10.1111/sjos.12054},
  number = {3},
  journal = {Scandinavian Journal of Statistics},
  publisher = {Wiley},
  author = {Yao,  Weixin and Li,  Longhai},
  year = {2014},
  pages = {656–671}
}

@article{chen2018modal,
  title={Modal regression using kernel density estimation: A review},
  author={Chen, Yen-Chi},
  journal={Wiley Interdisciplinary Reviews: Computational Statistics},
  volume={10},
  number={4},
  pages={e1431},
  year={2018},
  publisher={Wiley Online Library}
}

@article{feng2020statistical,
  title={A statistical learning approach to modal regression},
  author={Feng, Yunlong and Fan, Jun and Suykens, Johan AK},
  journal={Journal of Machine Learning Research},
  volume={21},
  number={2},
  pages={1--35},
  year={2020}
}

@article{Chen2016,
  title = {Nonparametric modal regression},
  volume = {44},
  comment =  {0090-5364},
  comment = {http://dx.doi.org/10.1214/15-AOS1373},
  comment = {10.1214/15-aos1373},
  number = {2},
  journal = {The Annals of Statistics},
  publisher = {Institute of Mathematical Statistics},
  author = {Chen,  Yen-Chi and Genovese,  Christopher R. and Tibshirani,  Ryan J. and Wasserman,  Larry},
  year = {2016},
  pages = {489-514}
}

@book{Schilling2009,
  title = {Bernstein Functions: Theory and Applications},
  comment = {http://dx.doi.org/10.1515/9783110215311},
  comment = {10.1515/9783110215311},
  publisher = {Walter de Gruyter},
  author = {Schilling,  René L. and Song,  Renming and Vondraček,  Zoran},
  year = {2009}
}

@article{deMelloeSilva2024,
  title = {Degree selection methods for curve estimation via {B}ernstein polynomials},
  volume = {40},
  number = {1},
  journal = {Computational Statistics},
  publisher = {Springer Science and Business Media LLC},
  author = {de Mello e Silva,  Juliana Freitas and Ghosh,  Sujit Kumar and Mayrink,  Vinícius Diniz},
  year = {2024},
  pages = {1–26}
}

@article{guan2020bayesian,
  title={Bayesian nonparametric policy search with application to periodontal recall intervals},
  author={Guan, Qian and Reich, Brian J and Laber, Eric B and Bandyopadhyay, Dipankar},
  journal={Journal of the American Statistical Association},
  volume={115},
  number={531},
  pages={1066--1078},
  year={2020},
  publisher={Taylor \& Francis}
}

@article{Shin2024,
  title = {Generative Multi-Purpose Sampler for Weighted {M}-estimation},
  volume = {33},
  ISSN = {1537-2715},
  comment = {http://dx.doi.org/10.1080/10618600.2023.2292668},
  comment = {10.1080/10618600.2023.2292668},
  number = {3},
  journal = {Journal of Computational and Graphical Statistics},
  publisher = {Informa UK Limited},
  author = {Shin,  Minsuk and Wang,  Shijie and Liu,  Jun S.},
  year = {2024},
  month = jan,
  pages = {1084–1097}
}

@article{ADAM,
  comment = {10.48550/ARXIV.1412.6980},
  comment = {https://arxiv.org/abs/1412.6980},
  author = {Kingma,  Diederik P. and Ba,  Jimmy},
  title = {Adam: A Method for Stochastic Optimization},
  journal = {arXiv},
  year = {2014},
  copyright = {arXiv.org perpetual,  non-exclusive license}
}

@inproceedings{chen2016xgboost,
  title={Xgboost: A scalable tree boosting system},
  author={Chen, Tianqi and Guestrin, Carlos},
  booktitle={Proceedings of the 22nd ACM SIGKDD International Conference on Knowledge Discovery and Data Mining},
  pages={785--794},
  year={2016}
}

@article{Shen2023,
  title = {Asymptotic properties of neural network sieve estimators},
  volume = {35},
  ISSN = {1029-0311},
  comment = {http://dx.doi.org/10.1080/10485252.2023.2209218},
  comment = {10.1080/10485252.2023.2209218},
  number = {4},
  journal = {Journal of Nonparametric Statistics},
  publisher = {Informa UK Limited},
  author = {Shen,  Xiaoxi and Jiang,  Chang and Sakhanenko,  Lyudmila and Lu,  Qing},
  year = {2023},
  month = May,
  pages = {839–868}
}

\end{document}


\maketitle

\section{Additional illustration of the ST density }

\begin{figure}[!htbp]
	\centering
	\includegraphics[width=0.8\textwidth]{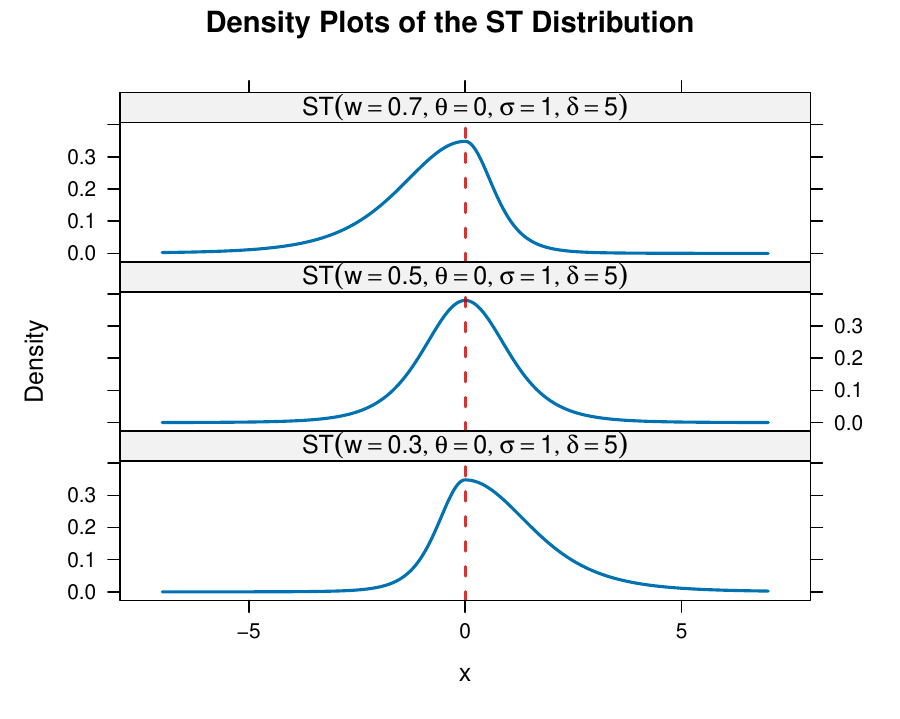}
	\caption{Density plots of the ST distribution with the skewness parameter $w = 0.3,0.5$ and $0.7$.}
	\label{fig:Density plots of the ST distribution} 
\end{figure}

The three panels of the Figure plot the densities of the ST distribution (introduced in Section 2 of the main paper) with various choices of the skewness parameter $w = 0.3, 0.5$ and $0.7$ respectively.

\section{Proofs of Theoretical Results}\label{appendix: Theorems and Proofs}

\setcounter{theorem}{0}
For the convenience of the reader, we restate the lemmas, propositions, and theorems from the main article and provide the associated proofs.

\begin{lemma}
	The ST distribution in equation (2) of the main article is identifiable.
\label{lemma:identifiable}
\end{lemma}
\begin{proof}[Proof of Lemma 1]
First, we show that $ \theta $ in the probability density function (PDF) $f_{\text{ST}}(x \mid w, \theta, \sigma, \delta)$ is identifiable, regardless of the values of $ w $, $ \sigma $, and $ \delta $. As shown in \citet{Liu2024}, the ST distribution is a two-component mixture distribution with a left-skewed density and a right-skewed density, both sharing a global mode at $ x = \theta $. This is because both $ f_{\mathrm{LT}}\left(x \mid \theta, \sigma \sqrt{w/(1-w)}, \delta\right) $ and $ f_{\mathrm{RT}}\left(x \mid \theta, \sigma \sqrt{(1-w)/w}, \delta\right) $ are respectively left-truncated and right-truncated Student-$ t$ densities, where the truncation point is at the mode $ x = \theta $. Suppose that $ f_{\mathrm{ST}}(x \mid w_1, \theta_1, \sigma_1, \delta_1) = f_{\mathrm{ST}}(x \mid w_2, \theta_2, \sigma_2, \delta_2) $ for all $ x \in \mathbb{R}^{1} $. Then, the global modes of both densities must coincide, i.e., $ \theta_1 = \theta_2 $. Thus, $ \theta $ is identifiable, regardless of the values of $w$, $\sigma$, and $\delta$.
	
Next, we show that $ w $, $ \sigma $, and $ \delta $ are identifiable. Without loss of generality, we assume $ \theta = 0 $.
By Theorem 1 of \citet{teicher1963identifiability}, it suffices to find two values $ x_1 $ and $ x_2 $ such that the following determinant is non-zero, i.e.
	$$
	\left|\begin{array}{ll}
		F_{\mathrm{LT}}\left(x_{1} \mid \theta = 0, \sigma \sqrt{\frac{w}{1-w}}, \delta\right) & F_{\mathrm{RT}}\left(x_{1} \mid \theta = 0, \sigma \sqrt{\frac{1-w}{w}}, \delta\right) \\
		F_{\mathrm{LT}}\left(x_{2} \mid \theta = 0, \sigma \sqrt{\frac{w}{1-w}}, \delta\right) & F_{\mathrm{RT}}\left(x_{2} \mid \theta = 0, \sigma \sqrt{\frac{1-w}{w}}, \delta\right)
	\end{array}\right| \neq 0,
	$$
where, $ F_{\mathrm{LT}} $ and $ F_{\mathrm{RT}} $ represent the cumulative distribution functions (CDFs) of the left- and right-truncated Student-$ t $ distributions respectively. Straightforward algebra shows that
	$$F_{\mathrm{RT}}\left(x ~\bigg|~ \theta = 0, \sigma \sqrt{\frac{1-w}{w}}, \delta\right) = \left\{2 F_{t}\left(x\sigma^{-1} \sqrt{\frac{w}{1-w}} ~\bigg|~ \delta\right) - 1\right\} \mathbb{I}(x \ge 0),
	$$
where, $ F_{t}\left(\cdot \mid \delta\right) $ represents the CDF of the Student-$ t $ distribution with degrees of freedom $ \delta $.
	
Let $ x_1 = \sigma \sqrt{\frac{1-w}{w}} $ and $ x_2 = 2 \sigma \sqrt{\frac{1-w}{w}} $. Then, the determinant becomes
	$$
	\left|\begin{array}{ll}
		1 & 2F_{t}(1 \mid \delta) \\
		1 & 2F_{t}(2 \mid \delta)
	\end{array}\right| = 2F_{t}(2 \mid \delta) - 2F_{t}(1 \mid \delta) > 0,
	$$
where, the inequality holds because $ F_{t}(\cdot \mid \delta) $ is the CDF of a continuous random variable and is therefore monotonically increasing. This establishes that $ w $, $ \sigma $, and $ \delta $ are identifiable. 
\end{proof}

\begin{theorem}
Let $ m(\mathbf{x}) = g(\boldsymbol{\beta}^\top \mathbf{x}) $ be a function with a vector input $ \mathbf{x}$ and a scalar-valued output. Suppose the following conditions hold:
\begin{enumerate}[(C{\arabic*})]
	\item The support of $ m(\cdot) $, denoted as $ S $, is a bounded convex set with at least one interior point.
	\item The single-index function $ g(\cdot) $ is a monotonic increasing function on its support.
	\item The $ L_2 $ norm of $ \boldsymbol{\beta} $ is one, i.e., $ \boldsymbol{\beta}^{\top} \boldsymbol{\beta} = 1 $.
\end{enumerate}
Then, the model in equation (1) of the main article is identifiable.
\end{theorem}

\begin{proof}[Proof of Theorem 2]
The main strategy of this proof closely follows \citet{lin2007identifiability}, though the assumptions in our setting differ. 
	
We first show that $ w $, $ \sigma $, and $ \delta $ are identifiable regardless of $ g(\boldsymbol{\beta}^{\top} \mathbf{x}) $. Suppose
	$$
	f_{\mathrm{ST}} \left( y \mid w_{1}, g_{1}(\boldsymbol{\beta}^{\top}_{1} \mathbf{x}), \sigma_{1}, \delta_{1} \right) = f_{\mathrm{ST}} \left( y \mid w_{2}, g_{2}(\boldsymbol{\beta}^{\top}_{2} \mathbf{x}), \sigma_{2}, \delta_{2} \right).
	$$
By Lemma \ref{lemma:identifiable}, it follows that $ w_{1} = w_{2} $, $ \sigma_{1} = \sigma_{2} $, $ \delta_{1} = \delta_{2} $, and $ g_{1}(\boldsymbol{\beta}^{\top}_{1} \mathbf{x}) = g_{2}(\boldsymbol{\beta}^{\top}_{2} \mathbf{x}) $. Thus, $ w $, $ \sigma $, and $ \delta $ are identifiable regardless of $ g(\boldsymbol{\beta}^{\top} \mathbf{x}) $.
	
Next, we show that $\boldsymbol{\beta} $ is identifiable, and therefore, so is $g(\boldsymbol{\beta}^\top \mathbf{x})$. Suppose $ \boldsymbol{\beta}_1 \neq \boldsymbol{\beta}_2 $. Recall, by Condition (C3) that $ \boldsymbol{\beta}_1^{\top} \boldsymbol{\beta}_1 = \boldsymbol{\beta}_2^{\top} \boldsymbol{\beta}_2 = 1 $. By the Cauchy-Schwarz inequality, $ |\boldsymbol{\beta}_{1}^{\top} \boldsymbol{\beta}_{2}| < 1 $. Under Condition (C1), there exists a sphere $ B = B(\mathbf{x}_{0}, r) \subset S $ for some $ \mathbf{x}_{0}$, such that $ \mathbf{x}_{0} + t \boldsymbol{\beta}_1 \in S $ and $ \mathbf{x}_{0} + t \boldsymbol{\beta}_2 \in S $ for all $ t \in (-r, r) $. By Condition (C3), we have
$$
	\begin{aligned}
		& g_1\left(\boldsymbol{\beta}_1^{\top} \mathbf{x}_0 + t\right) = g_1\left(\boldsymbol{\beta}_1^{\top}(\mathbf{x}_0 + t \boldsymbol{\beta}_1)\right) = g_2\left(\boldsymbol{\beta}_2^{\top}(\mathbf{x}_0 + t \boldsymbol{\beta}_1)\right) = g_2\left(\boldsymbol{\beta}_2^{\top} \mathbf{x}_0 + t \boldsymbol{\beta}_2^{\top} \boldsymbol{\beta}_1\right), 
        \end{aligned}
$$
        and
$$
        \begin{aligned}
		& g_2\left(\boldsymbol{\beta}_2^{\top} \mathbf{x}_0 + t\right) = g_2\left(\boldsymbol{\beta}_2^{\top}(\mathbf{x}_0 + t \boldsymbol{\beta}_2)\right) = g_1\left(\boldsymbol{\beta}_1^{\top}(\mathbf{x}_0 + t \boldsymbol{\beta}_2)\right) = g_1\left(\boldsymbol{\beta}_1^{\top} \mathbf{x}_0 + t \boldsymbol{\beta}_2^{\top} \boldsymbol{\beta}_1\right).
	\end{aligned}
$$
Since $ |\boldsymbol{\beta}_{1}^{\top} \boldsymbol{\beta}_{2}| < 1 $, we thus obtain
	$$
	\begin{aligned}
		g_1\left(\boldsymbol{\beta}_1^{\top} \mathbf{x}_0 + t\right) & = g_2\left(\boldsymbol{\beta}_2^{\top} \mathbf{x}_0 + t \boldsymbol{\beta}_2^{\top} \boldsymbol{\beta}_1\right) \\
		& = g_1\left(\boldsymbol{\beta}_1^{\top} \mathbf{x}_0 + t (\boldsymbol{\beta}_2^{\top} \boldsymbol{\beta}_1)^2\right) \\
		& = \cdots \\
		& = g_1\left(\boldsymbol{\beta}_1^{\top} \mathbf{x}_0 + t (\boldsymbol{\beta}_2^{\top} \boldsymbol{\beta}_1)^{2n}\right) \\
		& = \cdots \\
		& = g_1\left(\boldsymbol{\beta}_1^{\top} \mathbf{x}_0\right), \quad \text{for all } t \in (-r, r).
	\end{aligned}
	$$
Thus, $ g_1 (\boldsymbol{\beta}_1^{\top} \mathbf{x}_0 + t ) = g_1 (\boldsymbol{\beta}_1^{\top} \mathbf{x}_0 ) $ for all $ t \in (-r, r) $, which contradicts Condition (C2) that $ g_1(\cdot) $ is a monotonic increasing function. Therefore, $ \boldsymbol{\beta}_1 = \boldsymbol{\beta}_2 $ must hold. Finally, since $ \boldsymbol{\beta}_1 = \boldsymbol{\beta}_2 $, it follows directly that $ g_1( \boldsymbol{\beta}_1^\top \mathbf{x} ) = g_2( \boldsymbol{\beta}_2^\top \mathbf{x} ) $. This completes the proof.
\end{proof}

\revone{
\begin{proposition}
\label{prop:fisher_consistency}
Assume the data are generated according to Equations (1) and (2) in the main article with true parameters
$\boldsymbol{\varphi}_0 = (g_0,\boldsymbol{\beta}_0,w_0,\sigma_0,\delta_0)$
satisfying conditions (C1)--(C3). Additionally, assume that $\mathbf{x}$ is a random vector with distribution $p_{\mathbf{x}}$ whose support satisfies Condition (C1) for both $\boldsymbol{\beta}_0$ and any candidate $\boldsymbol{\beta}$ under consideration (C4). Let $\boldsymbol{\varphi} = (g,\boldsymbol{\beta},w,\sigma,\delta)$ be any candidate parameters satisfying the same regularity conditions, and define the population negative log-likelihood
\begin{equation}
\Psi(\boldsymbol{\varphi}) = \mathbb{E}_{(\mathbf{x},Y)}\Bigl[-\log f_{\mathrm{ST}}\bigl(Y \mid w,\; \theta = g(\boldsymbol{\beta}^\top \mathbf{x}),\; \sigma,\; \delta\bigr)\Bigr],
\label{eq:pop_loss}
\end{equation}
where the expectation is taken with respect to the true joint distribution of $(\mathbf{x},Y)$.
Then
$$
\Psi(\boldsymbol{\varphi}) \geq \Psi(\boldsymbol{\varphi}_0)
$$
for all $\boldsymbol{\varphi}$, with equality if and only if $\boldsymbol{\varphi} = \boldsymbol{\varphi}_0$. Consequently, the true parameter vector $\boldsymbol{\varphi}_0$ is the unique minimizer of the population loss functional $\Psi$.
\end{proposition}

\begin{proof}[Proof of Proposition 3]
We aim to show that the true parameter vector $\boldsymbol{\varphi}_0$ is the unique minimizer of the population negative log-likelihood $\Psi(\boldsymbol{\varphi})$ defined in~\eqref{eq:pop_loss}.

\paragraph{Step 1: Decompose the loss function.}
Consider the difference in population loss between an arbitrary parameter $\boldsymbol{\varphi}$ and the true parameter $\boldsymbol{\varphi}_0$. Let $p_{\boldsymbol{\varphi}}(Y \mid \mathbf{x})$ denote the conditional density of $Y$ given $\mathbf{x}$ under parameter $\boldsymbol{\varphi}$. By the definition of the expectation and the negative log-likelihood, we have
\begin{align}
\Psi(\boldsymbol{\varphi}) - \Psi(\boldsymbol{\varphi}_0) 
&= \mathbb{E}_{\mathbf{x} \sim p_{\mathbf{x}}} \Bigl[ \mathbb{E}_{Y \sim p_{\boldsymbol{\varphi}_0}(\cdot \mid \mathbf{x})} \bigl[ \log p_{\boldsymbol{\varphi}_0}(Y \mid \mathbf{x}) - \log p_{\boldsymbol{\varphi}}(Y \mid \mathbf{x}) \bigr] \Bigr] \notag\\
&= \mathbb{E}_{\mathbf{x} \sim p_{\mathbf{x}}} \Bigl[ \mathbb{E}_{Y \sim p_{\boldsymbol{\varphi}_0}(\cdot \mid \mathbf{x})} \Bigl[ \log \frac{p_{\boldsymbol{\varphi}_0}(Y \mid \mathbf{x})}{p_{\boldsymbol{\varphi}}(Y \mid \mathbf{x})} \Bigr] \Bigr].
\label{eq:loss_diff}
\end{align}
The inner expectation in~\eqref{eq:loss_diff} is exactly the Kullback--Leibler (KL) divergence from $p_{\boldsymbol{\varphi}}$ to $p_{\boldsymbol{\varphi}_0}$ conditional on $\mathbf{x}$, i.e.,
\begin{equation*}
D_{\mathrm{KL}}\bigl(p_{\boldsymbol{\varphi}_0}(\cdot \mid \mathbf{x}) \,\|\, p_{\boldsymbol{\varphi}}(\cdot \mid \mathbf{x})\bigr)
= \mathbb{E}_{Y \sim p_{\boldsymbol{\varphi}_0}(\cdot \mid \mathbf{x})} \Bigl[ \log \frac{p_{\boldsymbol{\varphi}_0}(Y \mid \mathbf{x})}{p_{\boldsymbol{\varphi}}(Y \mid \mathbf{x})} \Bigr].
\end{equation*}
Thus,
\begin{equation}
\Psi(\boldsymbol{\varphi}) - \Psi(\boldsymbol{\varphi}_0) = \mathbb{E}_{\mathbf{x} \sim p_{\mathbf{x}}} \Bigl[ D_{\mathrm{KL}}\bigl(p_{\boldsymbol{\varphi}_0}(\cdot \mid \mathbf{x}) \,\|\, p_{\boldsymbol{\varphi}}(\cdot \mid \mathbf{x})\bigr) \Bigr].
\label{eq:loss_kl}
\end{equation}
It is well-known that the KL divergence is non-negative for every $\mathbf{x}$, i.e.,
\begin{equation*}
D_{\mathrm{KL}}\bigl(p_{\boldsymbol{\varphi}_0}(\cdot \mid \mathbf{x}) \,\|\, p_{\boldsymbol{\varphi}}(\cdot \mid \mathbf{x})\bigr) \ge 0,
\end{equation*}
and the KL divergence equals zero if and only if $p_{\boldsymbol{\varphi}}(y \mid \mathbf{x}) = p_{\boldsymbol{\varphi}_0}(y \mid \mathbf{x})$ for almost all $y$. Since the expectation of a non-negative random variable is non-negative, we obtain
\begin{equation*}
\Psi(\boldsymbol{\varphi}) - \Psi(\boldsymbol{\varphi}_0) \ge 0 \quad \Longrightarrow \quad \Psi(\boldsymbol{\varphi}) \ge \Psi(\boldsymbol{\varphi}_0),
\end{equation*}
which establishes that $\boldsymbol{\varphi}_0$ is a global minimizer of $\Psi$.

\paragraph{Step 2: Uniqueness of the minimizer.}
It remains to show that if $\Psi(\boldsymbol{\varphi}) = \Psi(\boldsymbol{\varphi}_0)$, then $\boldsymbol{\varphi} = \boldsymbol{\varphi}_0$. From~\eqref{eq:loss_kl}, $\Psi(\boldsymbol{\varphi}) = \Psi(\boldsymbol{\varphi}_0)$ implies
\begin{equation*}
\mathbb{E}_{\mathbf{x} \sim p_{\mathbf{x}}} \Bigl[ D_{\mathrm{KL}}\bigl(p_{\boldsymbol{\varphi}_0}(\cdot \mid \mathbf{x}) \,\|\, p_{\boldsymbol{\varphi}}(\cdot \mid \mathbf{x})\bigr) \Bigr] = 0.
\end{equation*}
Since the integrand is non-negative, this forces
\begin{equation*}
D_{\mathrm{KL}}\bigl(p_{\boldsymbol{\varphi}_0}(\cdot \mid \mathbf{x}) \,\|\, p_{\boldsymbol{\varphi}}(\cdot \mid \mathbf{x})\bigr) = 0
\end{equation*}
for almost all $\mathbf{x}$ with respect to the measure induced by $p_{\mathbf{x}}$. Consequently, for almost all $\mathbf{x}$ and all $y$,
\begin{equation}
f_{\mathrm{ST}}\bigl(y \mid w_0,\; \theta_0 = g_0(\boldsymbol{\beta}_0^\top \mathbf{x}),\; \sigma_0,\; \delta_0\bigr)
= f_{\mathrm{ST}}\bigl(y \mid w,\; \theta = g(\boldsymbol{\beta}^\top \mathbf{x}),\; \sigma,\; \delta\bigr).
\label{eq:density_equality}
\end{equation}
By Lemma 1, the ST distribution is identifiable. Hence, from~\eqref{eq:density_equality} we deduce that
\begin{equation}
\begin{aligned}
w &= w_0, \quad \sigma = \sigma_0, \quad \delta = \delta_0, \label{eq:param_equal}\\
g(\boldsymbol{\beta}^\top \mathbf{x}) &= g_0(\boldsymbol{\beta}_0^\top \mathbf{x}) \quad \text{for almost all } \mathbf{x}.
\end{aligned}
\end{equation}
By Condition~(C4) that $\mathbf{x}$ is a random vector whose distribution $p_{\mathbf{x}}$ is compatible with Condition~(C1), we can apply Theorem~2 to conclude that
\begin{equation*}
\boldsymbol{\beta} = \boldsymbol{\beta}_0 \quad \text{and} \quad g = g_0.
\end{equation*}
Collecting these results with~\eqref{eq:param_equal}, we conclude $\boldsymbol{\varphi} = \boldsymbol{\varphi}_0$. Thus, equality of the population loss functions implies that $\boldsymbol{\varphi} = \boldsymbol{\varphi}_0$, proving that $\boldsymbol{\varphi}_0$ is the \emph{unique} minimizer of $\Psi$. This establishes the Fisher consistency for the proposed estimation framework.
\end{proof}
}

\begin{theorem}
	If $\boldsymbol{A}^{(k)} > 0$ for $k = 1,\dots, K+1$, then the DNN $g(u)$ in equation (4) of the main article is monotonically increasing. Here, $\boldsymbol{A}^{(k)} > 0$ denotes that all elements of $\boldsymbol{A}^{(k)}$ are greater than zero.
\end{theorem}

\begin{proof}[Proof of Theorem \revone{4}]
	Let $\operatorname{sech}(x) = 2 \exp{(x)}/(\exp{(2 x)}+1)$ represent the hyperbolic secant function, which is the first derivative of the hyperbolic tangent function. 
	It is well known that 
	\begin{equation}
		0 < \operatorname{sech}(x) \le 1, \quad \text{for } x \in \mathbb{R}^{1}.
		\label{eq:the derivative tanh}
	\end{equation}
	The first derivative of $g(u)$ in equation (4) of the main article is given by
	$$
	G^{\prime}(u) = \left\{\boldsymbol{A}^{(K+1)} \boldsymbol{A}^{(K)} \cdots \boldsymbol{A}^{(1)}\right\}  \operatorname{sech} \circ \cdots \circ \operatorname{sech}\left(u\right) \quad \text{for } u \in \mathbb{R}^{1}.
	$$
	By \eqref{eq:the derivative tanh} and the assumption that $\boldsymbol{A}^{(k)} > 0$ for $k = 1,\dots,K+1$, the first derivative of $g(u)$ is strictly positive, i.e., $G^{\prime}(u) > 0$. Therefore, $g(u)$ is monotonically increasing.
\end{proof}


\begin{theorem}
	For any univariate continuous, monotonic increasing function $m(\cdot): \Omega \mapsto \mathbb{R}^{1}$, where $\Omega$ is a compact subset of $\mathbb{R}^{1}$, there exists a DNN $G(\cdot)$ in equation (4) of the main article with at most $k$ hidden layers and strictly positive weights such that, for any $u \in \Omega$ and $\epsilon > 0$,
    $$|m(u) - g(u)| < \epsilon.$$
\label{thm:universal approximation theory}
\end{theorem}

\begin{proof}[Proof of Theorem \revone{5}]
The proof of Theorem \ref{thm:universal approximation theory} is based on a straightforward modification of the proof of Theorem 3.1 in \citet{daniels2010monotone}. Briefly, Theorem 3.1 of \citet{daniels2010monotone} establishes the universal approximation theorem for DNN models with positive weights under the sigmoid activation function. The proof of Theorem 3.1 in \citet{daniels2010monotone} relies on the fact that the Heaviside function $ H(x) = \mathbb{I}(x \ge 0) $ can be approximated by the sigmoid function as $ \lim_{a \rightarrow \infty} 1/(1 + \exp(-a x)) = H(x) $ \citep{runje2023constrained}.

Instead of the sigmoid activation function, we employ the hyperbolic tangent function. It is straightforward to show that the Heaviside function can also be approximated by the hyperbolic tangent function after an affine transformation as $ \lim_{a \rightarrow \infty} (\tanh(a x) + 1)/2 = H(x)$. Therefore, the proof of Theorem 4 follows directly from the proof of Theorem 3.1 in \citet{daniels2010monotone} with the sigmoid activation function replaced by the hyperbolic tangent activation function.  
\end{proof}

\section{Application to HP Data: Additional Results}

\begin{figure}[!htbp]
	\centering
	\includegraphics[width=0.49\textwidth]{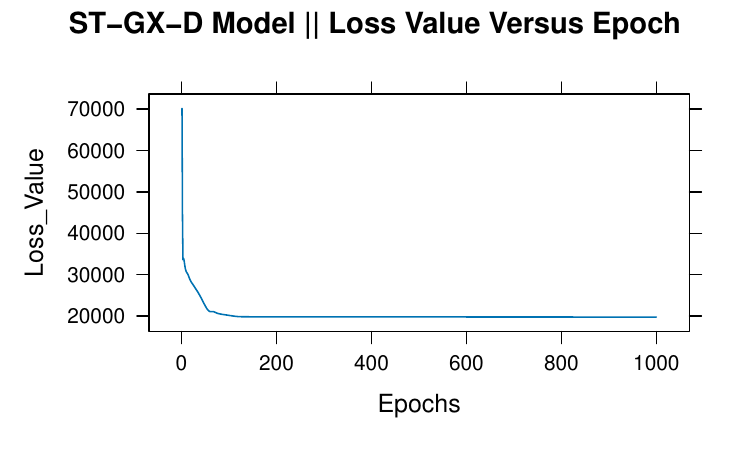}
	\includegraphics[width=0.49\textwidth]{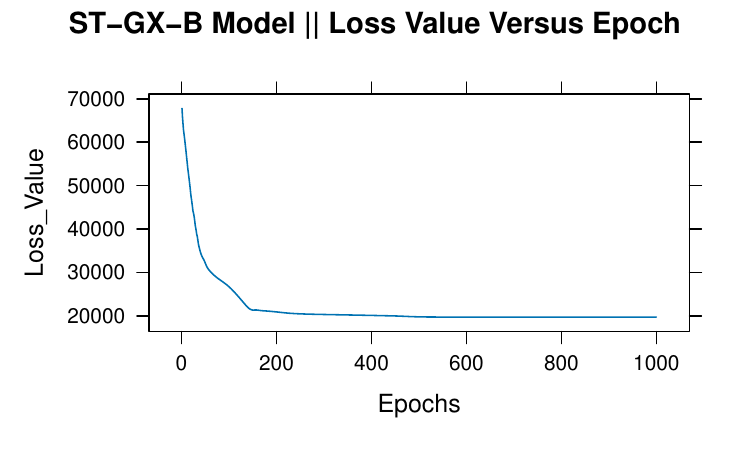}
	\includegraphics[width=0.49\textwidth]{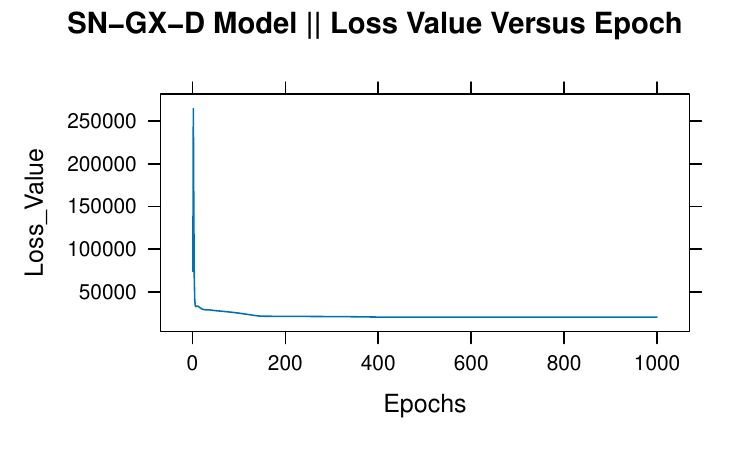}
	\includegraphics[width=0.49\textwidth]{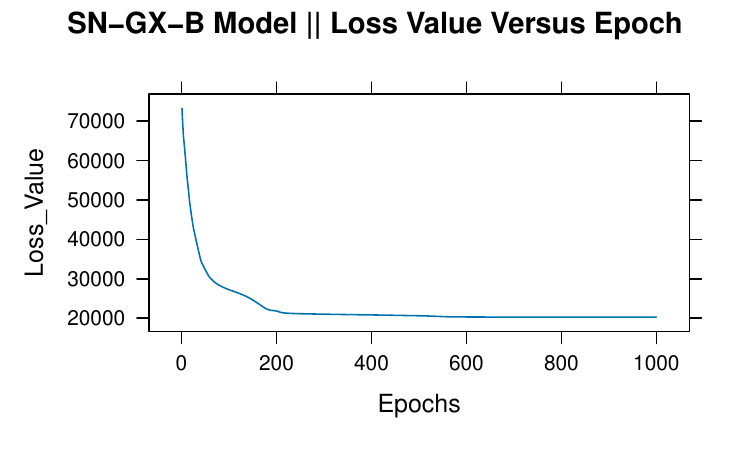}
	\includegraphics[width=0.49\textwidth]{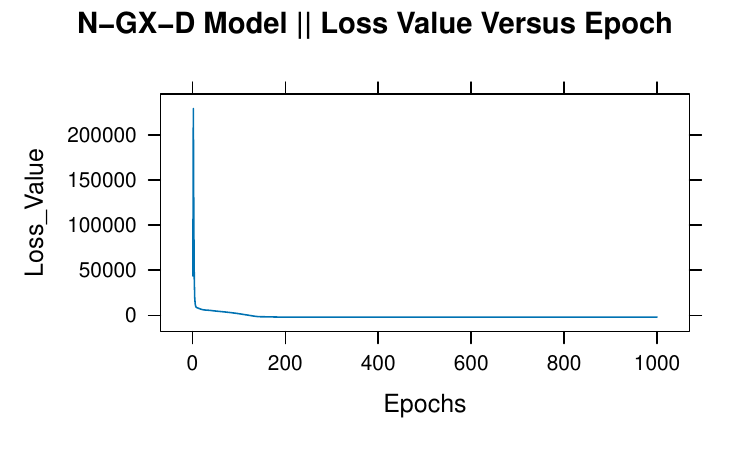}
	\includegraphics[width=0.49\textwidth]{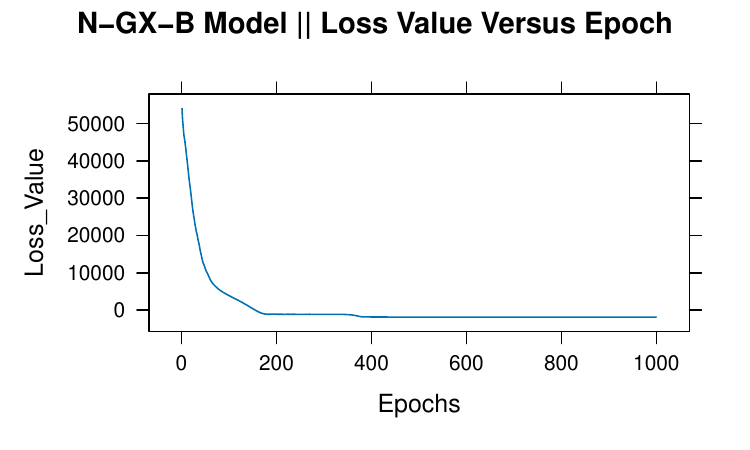}
	\caption{\label{fig:Real Data Application: Loss Value Versus Epoch 1}\revone{Plots of the loss value at the end of each epoch versus epoch number, obtained from fitting the likelihood-based single-index models to the HP dataset.  This figure displays the convergence of the neural network architectures (ST-GX-D, SN-GX-D, and N-GX-D) alongside their Bernstein polynomial counterparts (ST-GX-B, SN-GX-B, and N-GX-B).}}
\end{figure}

\begin{figure}[!htbp]
	\includegraphics[width=0.49\textwidth]{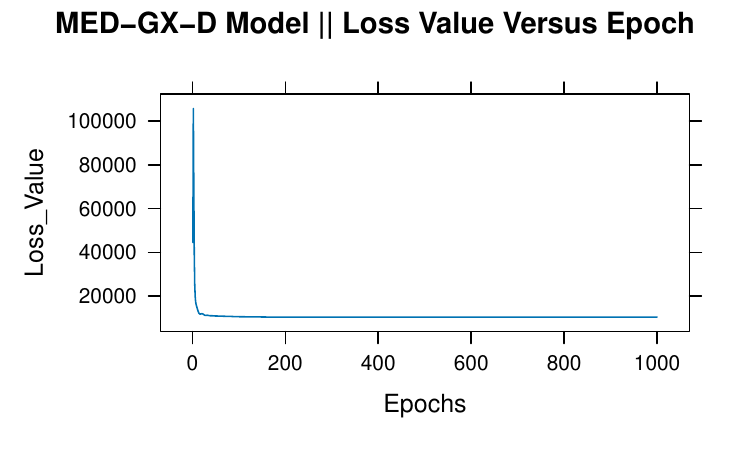}
	\includegraphics[width=0.49\textwidth]{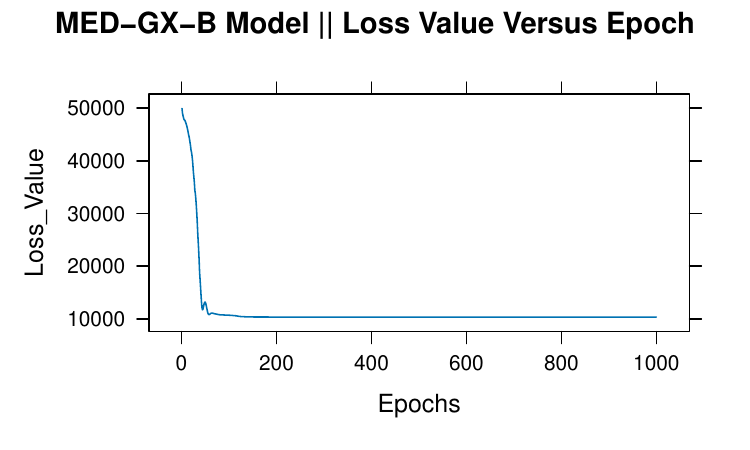}
	\includegraphics[width=0.49\textwidth]{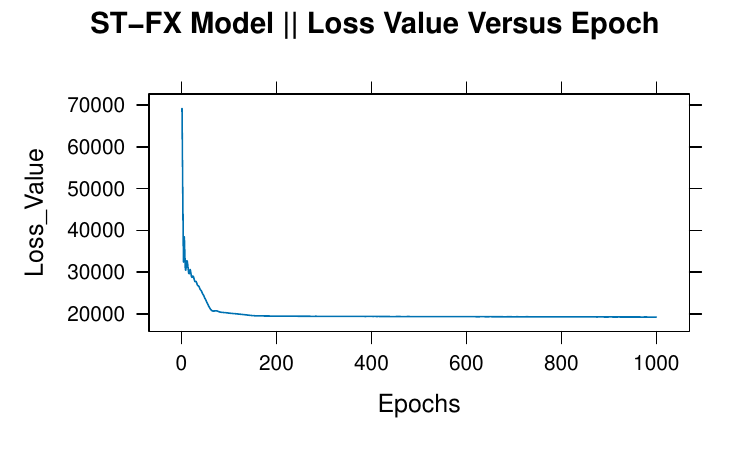}
	\includegraphics[width=0.49\textwidth]{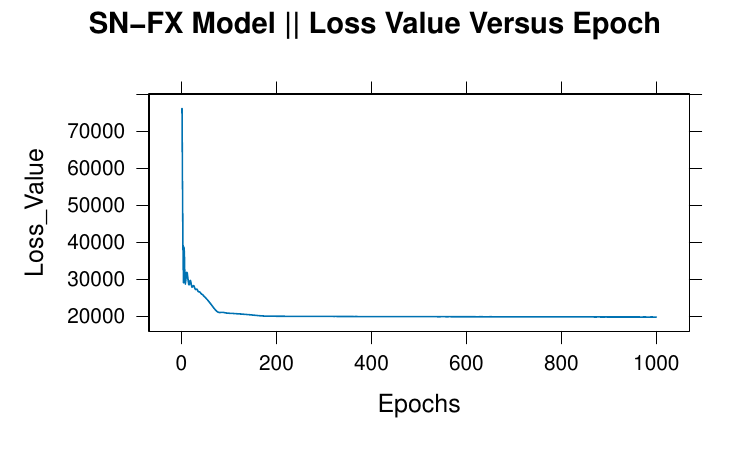}
	\includegraphics[width=0.49\textwidth]{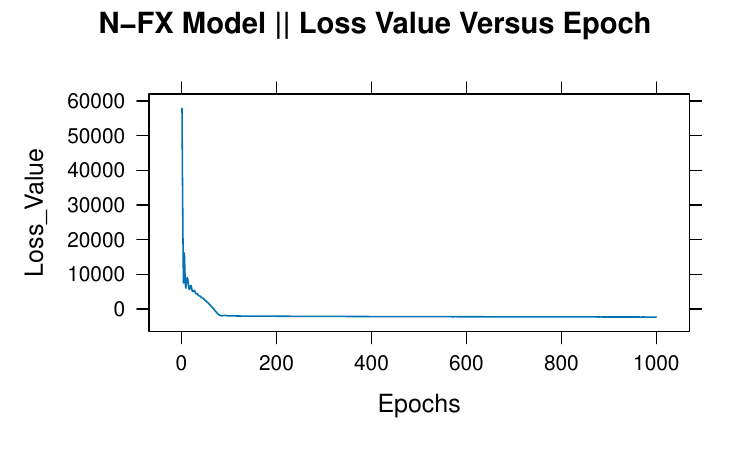}
	\caption{\label{fig:Real Data Application: Loss Value Versus Epoch 2}\revone{Plots of the loss value at the end of each epoch versus epoch number for the median regression and fully flexible models fitted to the HP dataset. This figure illustrates the convergence for the median single-index models (MED-GX-D and MED-GX-B) as well as the models that directly approximate the central location without a single-index constraint (ST-FX, SN-FX, and N-FX).}}
\end{figure}

\begin{figure}[!htbp]
	\includegraphics[width=0.49\textwidth]{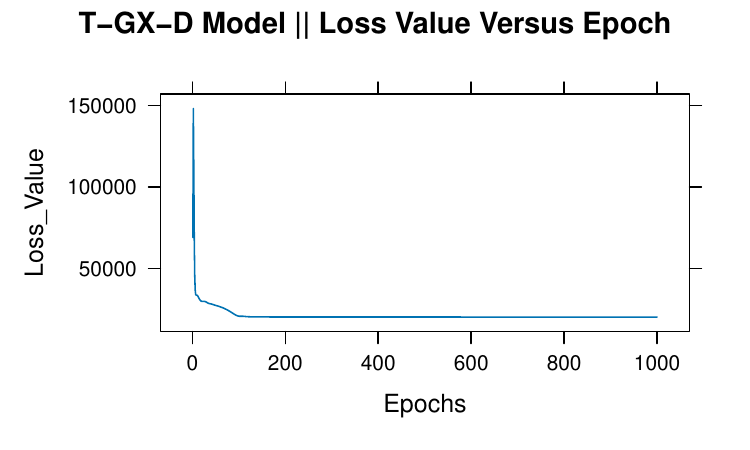}
	\includegraphics[width=0.49\textwidth]{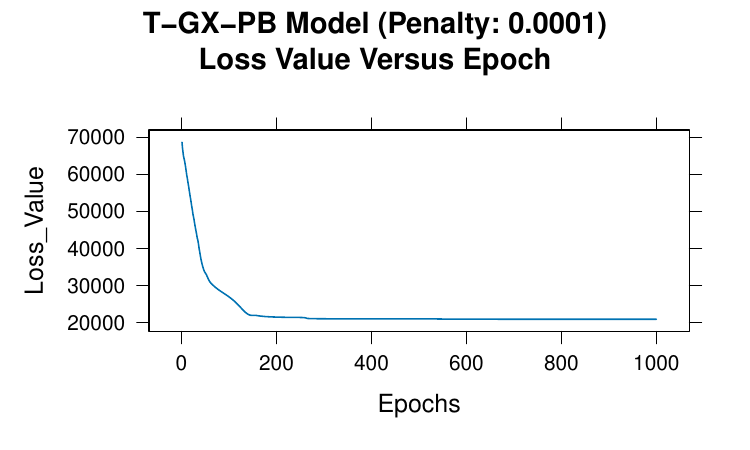}
	\includegraphics[width=0.49\textwidth]{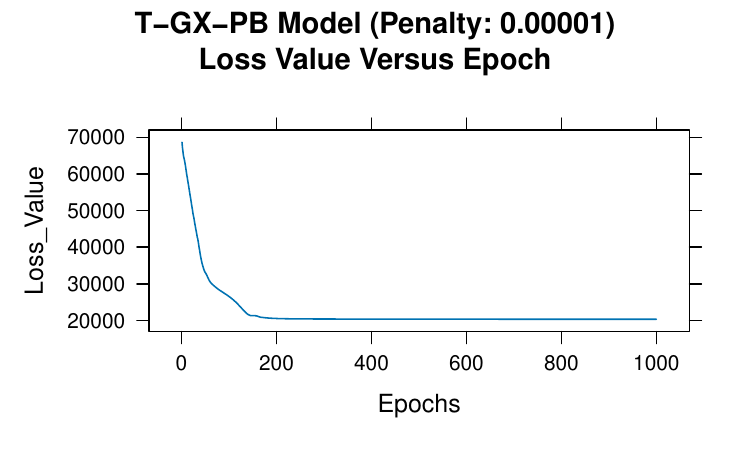}
	\includegraphics[width=0.49\textwidth]{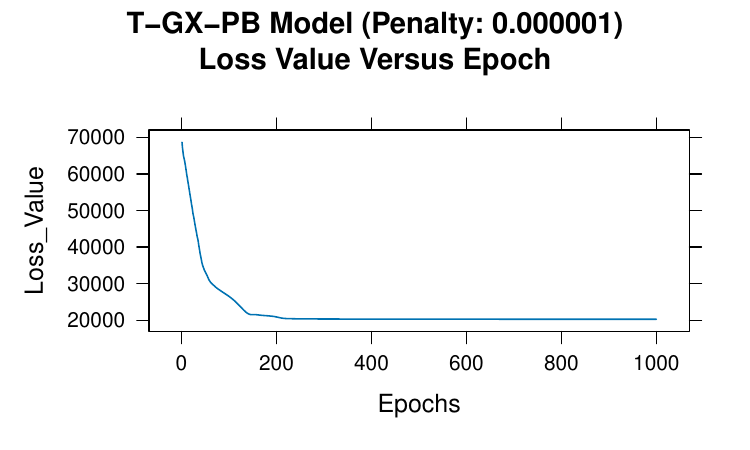}
	\includegraphics[width=0.49\textwidth]{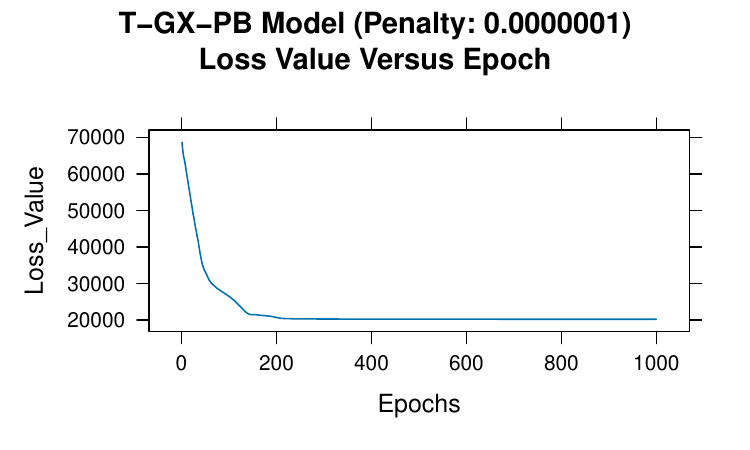}
	\caption{\label{fig:Real Data Application: Loss Value Versus Epoch 3}\revone{Plots of the loss value at the end of each epoch versus epoch number for the Student-$t$ single-index models fitted to the HP dataset.  This figure shows the optimization convergence of the NN approach (T-GX-D) and the penalized Bernstein polynomial models (T-GX-PB) evaluated across various $L_2$ penalty parameters.}}
\end{figure}

\revone{S-Figures~\ref{fig:Real Data Application: Loss Value Versus Epoch 1}, \ref{fig:Real Data Application: Loss Value Versus Epoch 2}, and \ref{fig:Real Data Application: Loss Value Versus Epoch 3} present the loss curves for all the models fitted to the HP dataset in Section~3 of the main paper. On the horizontal axis, we plot the epoch number, and on the vertical axis, we plot the loss value at the end of each epoch.  These figures collectively demonstrate that all the evaluated models successfully converged to local minima.}

\begin{figure}[!htbp]
\includegraphics[width=0.49\textwidth]{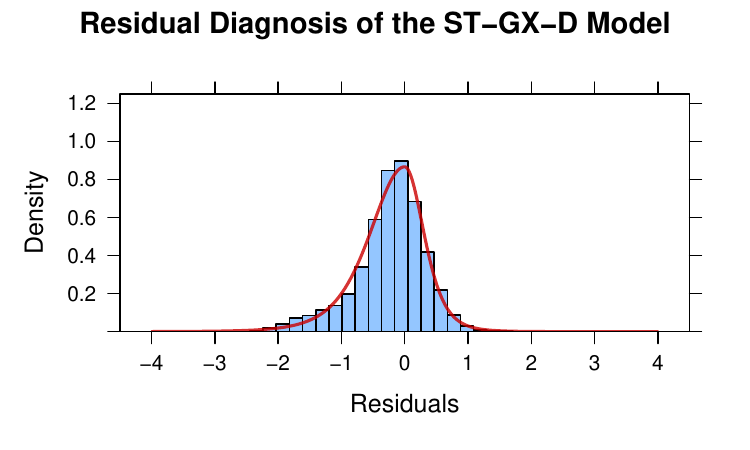}
\includegraphics[width=0.49\textwidth]{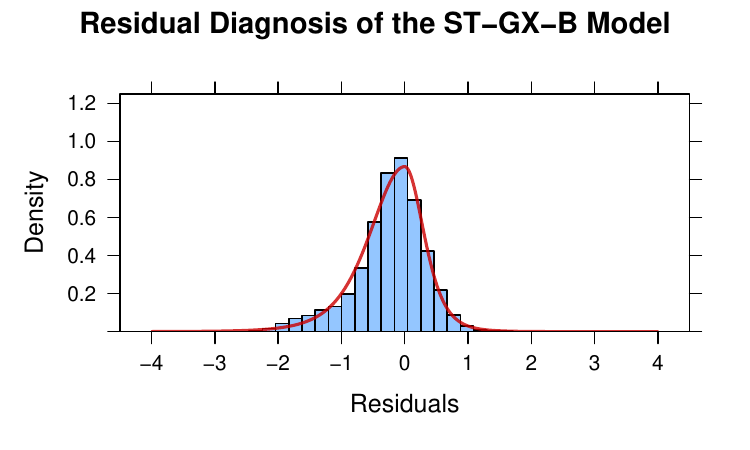}
\includegraphics[width=0.49\textwidth]{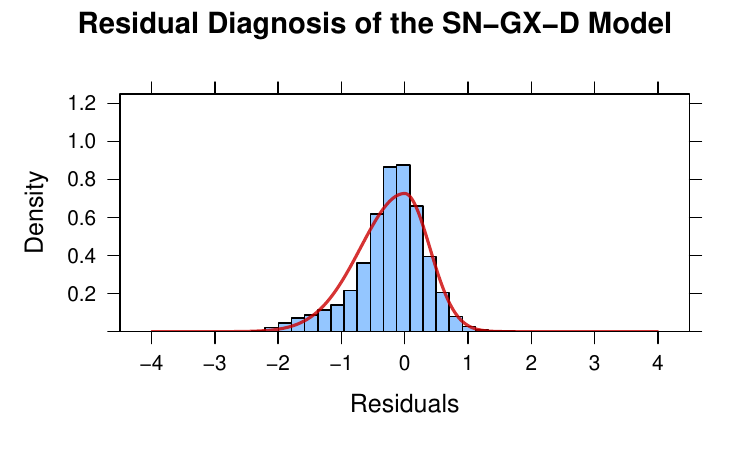}
\includegraphics[width=0.49\textwidth]{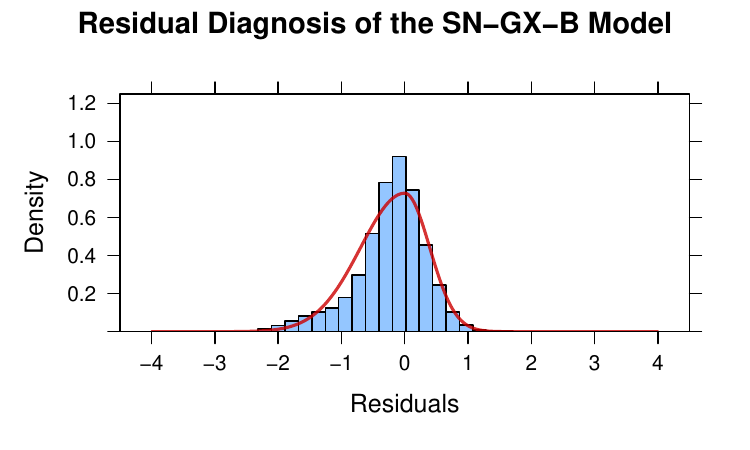}
\includegraphics[width=0.49\textwidth]{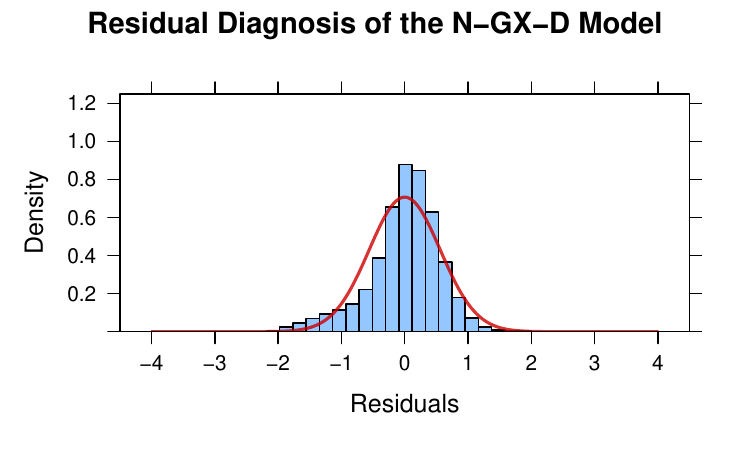}
\includegraphics[width=0.49\textwidth]{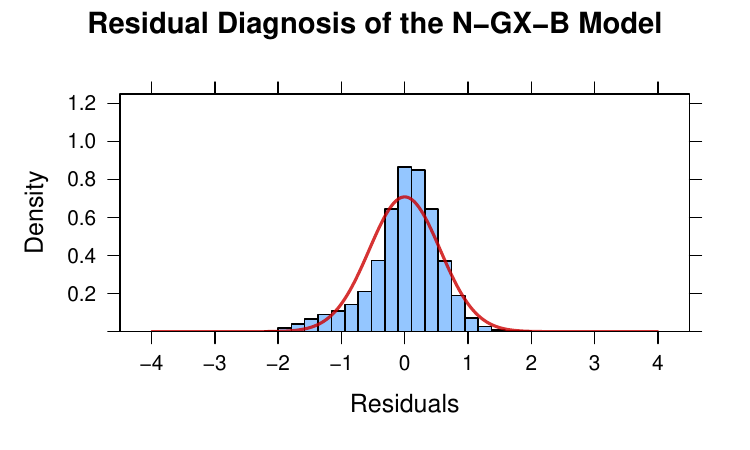}
\includegraphics[width=0.49\textwidth]{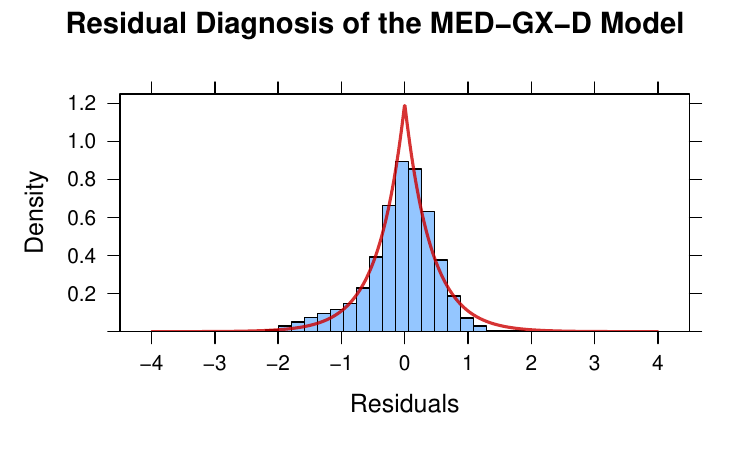}
\includegraphics[width=0.49\textwidth]{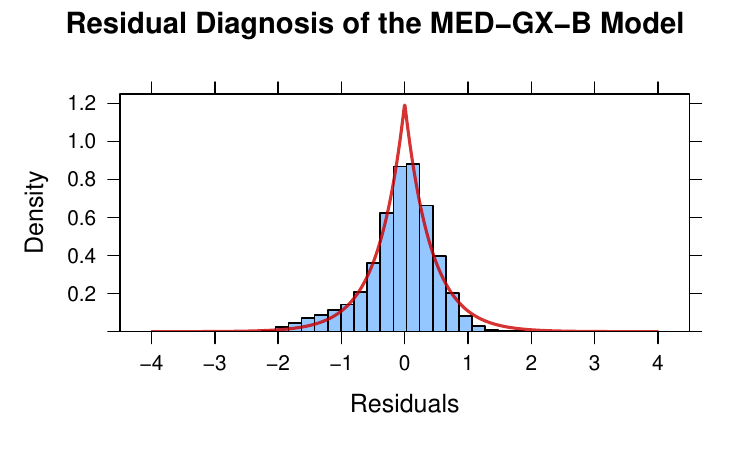}
\caption{\label{fig:Real Data Diagnostic Plots 1}\revone{Residual diagnostic plots obtained from fitting the primary single-index models to the HP dataset. This figure compares the neural network architectures (ST-GX-D, SN-GX-D, N-GX-D, and MED-GX-D) against their Bernstein polynomial counterparts (ST-GX-B, SN-GX-B, N-GX-B, and MED-GX-B).}}
\end{figure}

\begin{figure}[!htbp]
\centering
\includegraphics[width=0.49\textwidth]{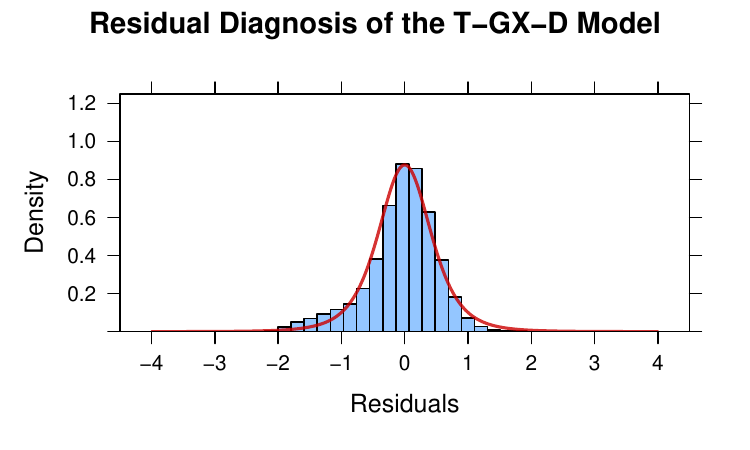}
\includegraphics[width=0.49\textwidth]{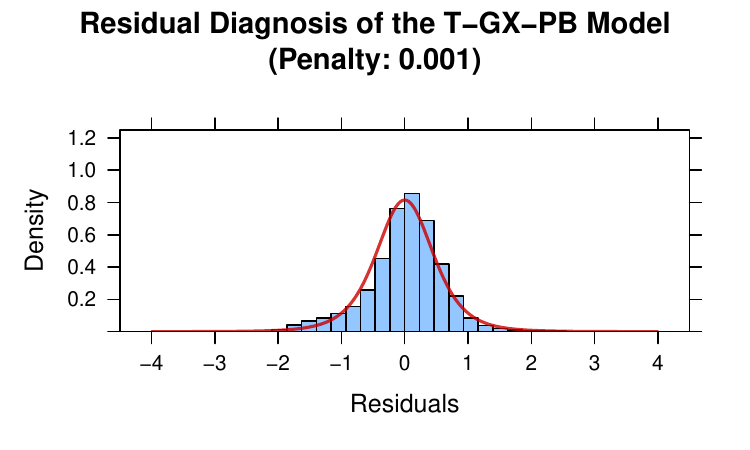}
\includegraphics[width=0.49\textwidth]{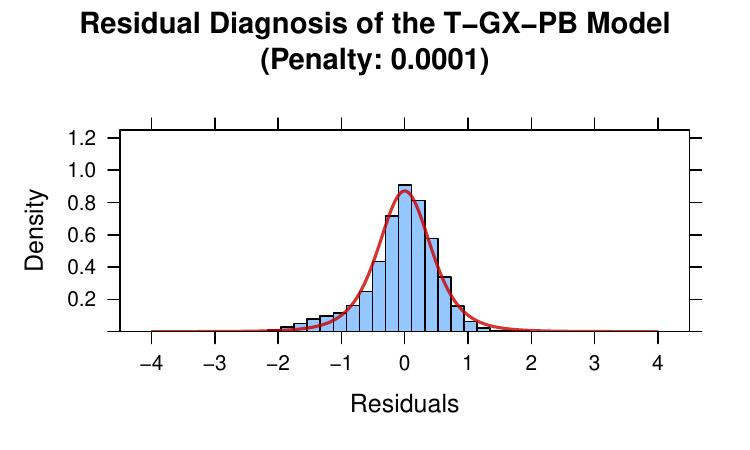}
\includegraphics[width=0.49\textwidth]{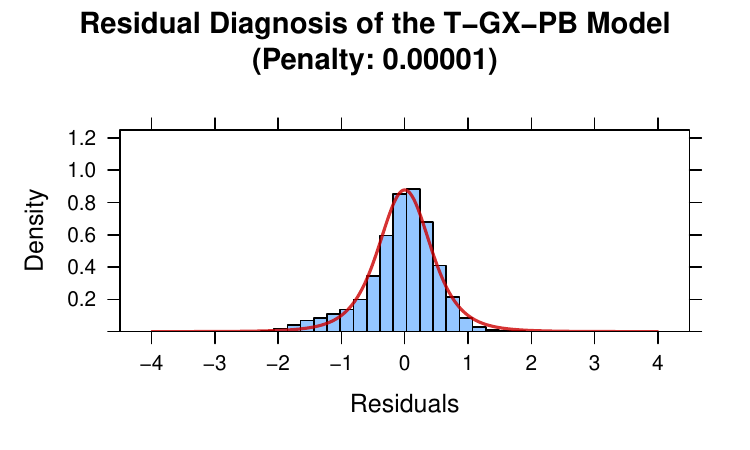}
\includegraphics[width=0.49\textwidth]{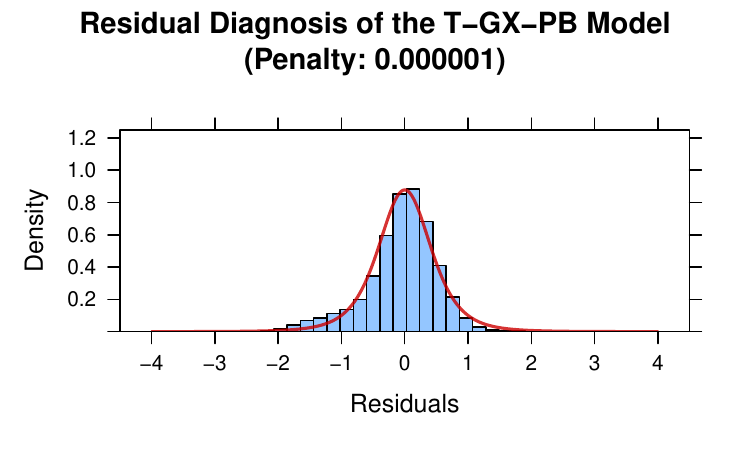}
\includegraphics[width=0.49\textwidth]{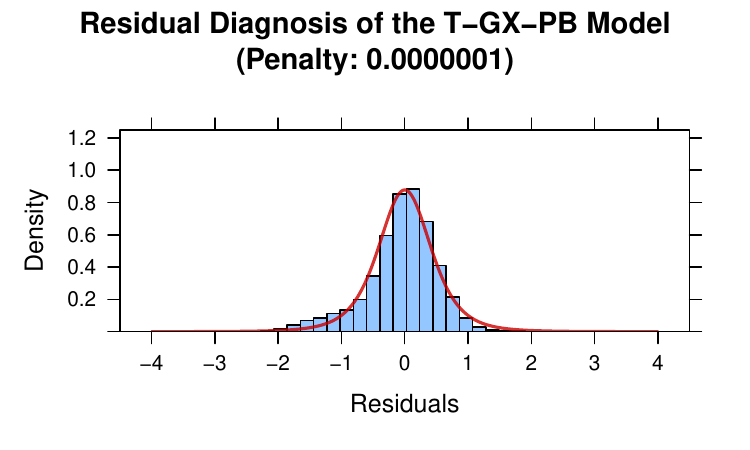}
\caption{\label{fig:Real Data Diagnostic Plots 2}\revone{Residual diagnostic plots obtained from fitting the Student-$t$ single-index models to the HP dataset. This figure illustrates the NN approach (T-GX-D) alongside the penalized Bernstein polynomial models (T-GX-PB) evaluated across various $L_2$ penalty parameters.}}
\end{figure}

\begin{figure}[!htbp]

\includegraphics[width=0.49\textwidth]{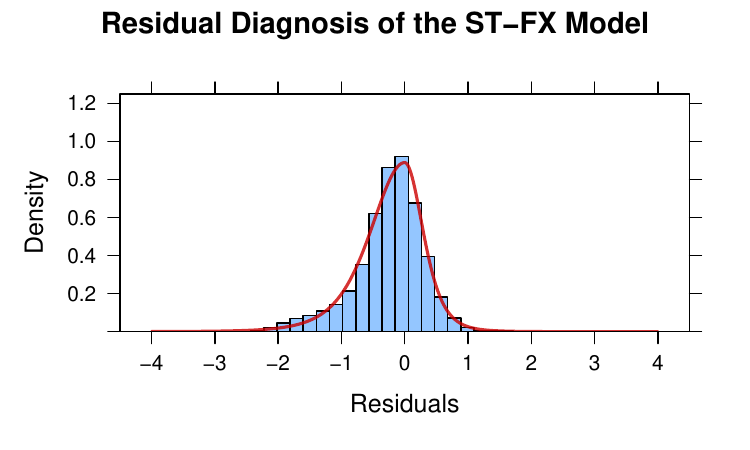}
\includegraphics[width=0.49\textwidth]{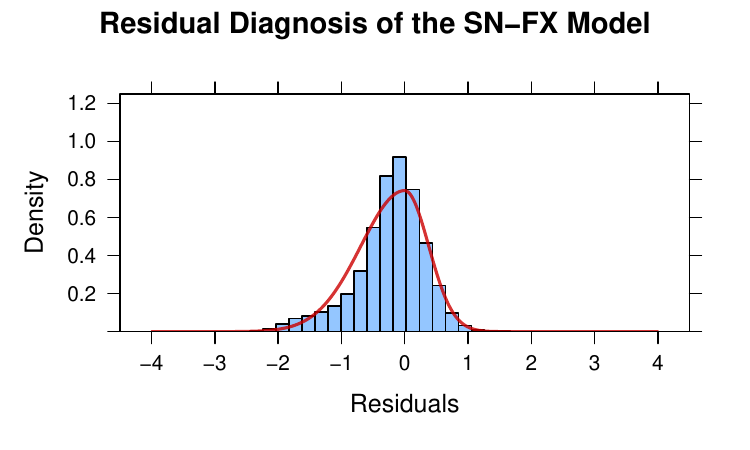}
\includegraphics[width=0.49\textwidth]{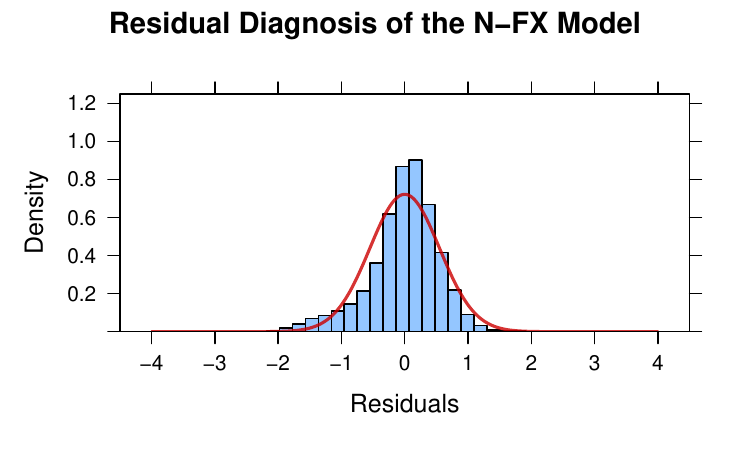}
\caption{\label{fig:Real Data Diagnostic Plots 3}\revone{Residual diagnostic plots obtained from fitting the fully flexible NN models to the HP dataset.  This figure displays the residuals for the models that directly approximate the central location without a single-index constraint under different error distributions (ST-FX, SN-FX, and N-FX).}}
\end{figure}

\revone{S-Figures~\ref{fig:Real Data Diagnostic Plots 1}, \ref{fig:Real Data Diagnostic Plots 2}, and \ref{fig:Real Data Diagnostic Plots 3} display the residual diagnostic plots for all the models fitted to this dataset. In each of these plots, the solid red line represents the estimated theoretical probability density function. We created this reference red line by evaluating the respective theoretical distribution's density equation (such as the ST, Laplace, Student-$t$, or standard normal density) using the final maximum likelihood parameter estimates, including the scale, skewness, and degrees of freedom yielded by the trained model. By comparing the empirical residual histograms to these theoretical red reference lines, it is evident that the models assuming the ST distribution (i.e., ST-GX-D, ST-GX-B, and ST-FX) were the most suitable, accurately aligning with the empirical residuals by capturing both the skewness and the heavy-tailedness simultaneously. Models assuming the SN distribution successfully captured the asymmetry of the residuals but failed to appropriately model the heavy tails.  Conversely, models assuming the Student-$t$ distribution captured the heavy tails extremely well but inherently could not account for the asymmetry.  Similarly, the median regression models, which structurally correspond to a Laplace distribution error assumption, captured the heavy tails reasonably well (though not quite as effectively as the Student-$t$ models) but also failed to capture the asymmetry.}

\newpage

\section{Additional Simulation Results}

\subsection{Schemes I and II: Additional Figures/Tables}

\begin{figure}[!htbp]
\centering
\includegraphics[width=0.76\textwidth]{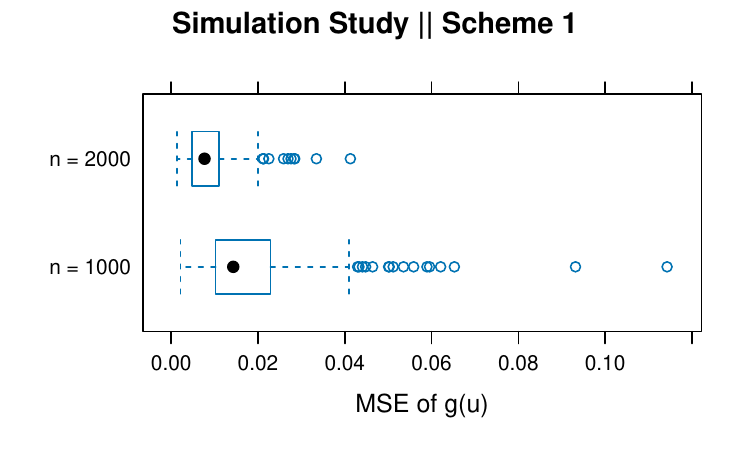}
\caption{Boxplots of the MSEs corresponding to the estimation of the true single-function $g(u)$ from fitting the ST-GX-D model to data generated from Scheme I. The estimates are based on 300 Monte Carlo replicates.}
\label{fig:Simulation study: boxplot of the MSE of the single-index function in the first scheme}
\end{figure}

\begin{figure}[!htbp]
	\centering
\includegraphics[width=0.76\textwidth]{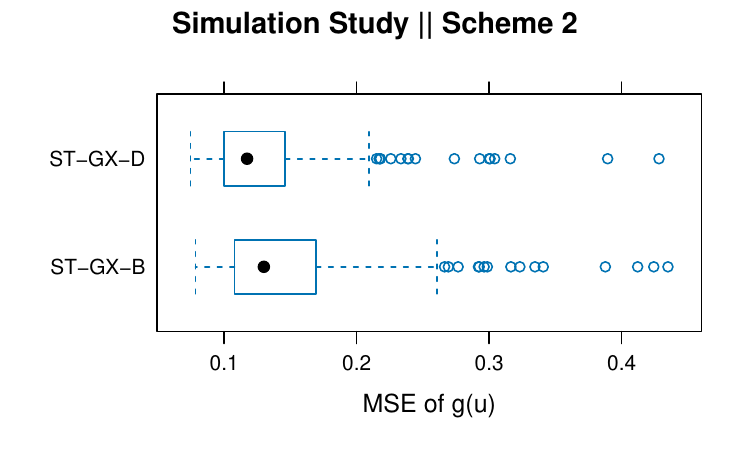}
\caption{Boxplots of the MSEs corresponding to the estimation of the true single-function $g(u)$ from fitting the ST-GX-D and ST-GX-B models to data generated from Scheme II. The estimates are based on 300 Monte Carlo replicates.}
\label{fig:Simulation study: boxplot of the MSE of single-index function in the second scheme}
\end{figure}

In the following, S-Figures~\ref{fig:Simulation study: boxplot of the MSE of the single-index function in the first scheme} and ~\ref{fig:Simulation study: boxplot of the MSE of single-index function in the second scheme} are the additional figures associated with Schemes I and II from Sections 4.1 and 4.2, respectively, of the main article

\revone{
\begin{table}[ht]
\centering
\begin{tabular}{lrrrr}
\toprule
 & \multicolumn{2}{c}{ST-GX-D} & \multicolumn{2}{c}{ST-GX-B} \\
\cmidrule(lr){2-3} \cmidrule(lr){4-5}
Parameter & APE & ABPE & APE & ABPE \\
\midrule
$\beta_1$ & 0.5681 & -0.0092 & 0.5592 & -0.0182 \\
$\beta_2$ & 0.5750 & -0.0024 & 0.5796 &  0.0022 \\
$\beta_3$ & 0.5717 & -0.0057 & 0.5723 & -0.0050 \\
$w$       & 0.5961 & -0.0039 & 0.5961 & -0.0039 \\
$\sigma$  & 1.0338 &  0.0338 & 1.0357 &  0.0357 \\
$\delta$  & 5.4801 &  0.4801 & 5.4962 &  0.4962 \\
\bottomrule
\end{tabular}
\caption{Simulation results from Scheme II. Here, APE = average of the point estimates, and ABPE = average bias of the point estimates.}
\label{tab:Simulation study: average point estimation from the second scheme}
\end{table}
}

\newpage 

\subsection{Scheme III: Complete Details and Results}

In Scheme III, we introduced a mixture noise model. The true single-index function was the same as that in Scheme II, i.e. $g(u) = \lfloor u \rfloor$. However, $99\%$ of the noise was generated from the asymmetric Laplace distribution $\text{ALD}(0, 0.5, 0.7)$, where $\text{ALD}(\mu, \sigma, p)$ denotes an asymmetric Laplace distribution with location parameter $\mu$, scale parameter $\sigma$, and skewness parameter $p$ \citep{yu2005three}. The probability density function of the $\text{ALD}(\mu, \sigma, p)$ distribution is
\begin{equation}
	f(x \mid \mu, \sigma, p)=\frac{p(1-p)}{\sigma} \exp -\rho_p\left(\frac{y-\mu}{\sigma}\right),
	\label{eq:density function of ALD}
\end{equation}
where, $x,\mu \in \mathbb{R}$, $\sigma > 0$, $p \in (0,1)$ and $\rho_p(u)=u\left(p-\mathbb{I}_{u<0}\right)$ is the check function.

\begin{figure}[!htbp]
	\centering
	\includegraphics[width=0.8\textwidth]{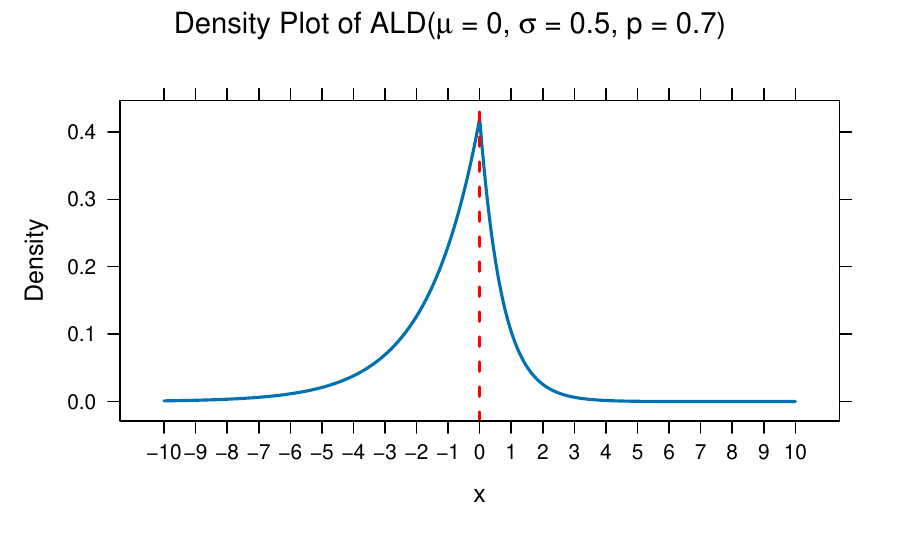}
	\caption{Density plot of the ALD$(0, 0.5, 0.7)$ distribution. The blue solid line represents the density curve, while the red dashed line indicates the location of the mode of the density function.}
	\label{fig:density plot of the ALD distribution}
\end{figure}

The density plot of the $\text{ALD}(0, 0.5, 0.7)$ distribution is shown in S-Figure \ref{fig:density plot of the ALD distribution}. It is evident from this figure that the $\text{ALD}(0, 0.5, 0.7)$ density is left-skewed. 
The remaining $1\%$ of the noise was generated from a normal distribution with mean 10 and standard deviation 0.1, which introduces outliers in the upper tail of the simulated dataset. The combination of $\text{ALD}(0, 0.5, 0.7)$ noise and normal noise centered around 10 ensured that the overall error distribution was left-skewed with $1\%$ large outliers.

\revone{We fit the ST-GX-D, SN-GX-D, N-GX-D, MED-GX-D, T-GX-D, ST-GX-B, SN-GX-B, N-GX-B, MED-GX-B, T-GX-PB, and eXtreme Gradient Boosting (XGBoost)  \citep{chen2016xgboost} models to 300 simulated datasets. To identify these competing models, we follow the same naming rule as in the main article. Specifically, MED-GX-D represents the single-index model that employs a NN to estimate the unknown single-index function while utilizing the mean absolute error as the loss function for median regression. Similarly, MED-GX-B represents the corresponding median regression model that relies on a Bernstein polynomial basis to estimate the single-index function. Furthermore, T-GX-D denotes the single-index model utilizing a NN paired with the negative log-likelihood of a symmetric Student-$t$ distribution as the loss function. Finally, T-GX-PB designates the symmetric Student-$t$ single-index model where the single-index function is estimated using a Bernstein polynomial augmented with an $L_2$ penalty applied directly to the polynomial coefficients to regularize the estimation process.} 

\begin{figure}[!htbp]
	\centering
	\includegraphics[width=0.8\textwidth]{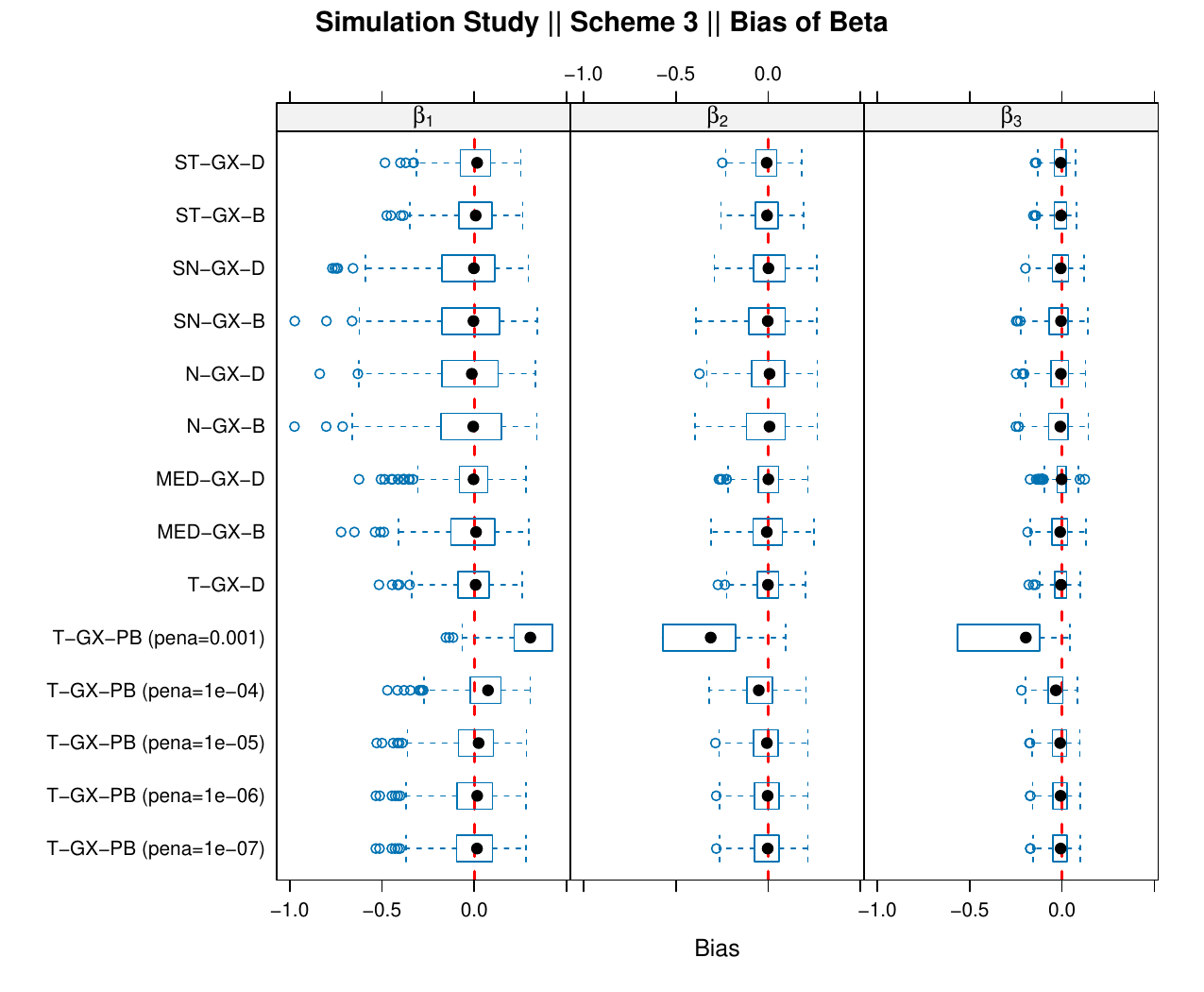}
	\caption{Boxplots of the point estimates for $\beta_1$, $\beta_2$, and $\beta_3$ with data generated under Scheme III, and based on 300 Monte Carlo replications. The vertical dashed red lines represent the true values $\beta_1=1/\sqrt{3}$, $\beta_2 = 1/\sqrt{3}$, and $\beta_3 = 1/\sqrt{3}$.}
	\label{fig:simulation study, boxplot of beta, third scheme}
\end{figure}

\revone{S-Figure~\ref{fig:simulation study, boxplot of beta, third scheme} presents boxplots of the estimation biases for the regression coefficients $\beta_1$, $\beta_2$, and $\beta_3$ obtained from fitting the competing models.  In this figure, the vertical dashed line indicates a bias of exactly zero. We observe that ST-GX-D achieved highly accurate and stable point estimates for $\boldsymbol{\beta}$. Specifically, the mean bias of the 300 point estimates for $\beta_1$ from the ST-GX-D model was practically zero, and its overall estimates exhibited very tight interquartile ranges. However, to be completely transparent with the results, several other robust formulations such as T-GX-D and ST-GX-B also produced highly competitive coefficient estimates with minimal bias. The most critical takeaway from these coefficient estimates is that for any given likelihood or loss function, the neural network (NN) architectures (denoted by the D suffix) systematically matched or outperformed their Bernstein polynomial counterparts (denoted by B or PB). Note that the results for $w$, $\sigma$, and $\delta$ in the ST distribution are not included because under model misspecification, the true values and biases for these parameters are not well-defined. S-Figure~\ref{fig:simulation study, boxplot of MSE of g(u), third scheme} presents the boxplots of the MSEs for the estimation of the true index function $g(u)$ using 300 Monte Carlo replicates.  Among all the evaluated models, we observe that the ST-GX-D model decisively achieved the smallest median MSE and the narrowest interquartile range. Most importantly, a consistent pattern emerges when comparing the functional estimation strategies: for the exact same likelihood assumption, the NN approach is universally better than the Bernstein polynomial approach. Specifically, ST-GX-D, SN-GX-D, N-GX-D, MED-GX-D, and T-GX-D consistently achieved smaller median MSEs than ST-GX-B, SN-GX-B, N-GX-B, MED-GX-B, and T-GX-PB respectively. This provides strong empirical evidence regarding the superiority of NNs over Bernstein polynomial bases for flexibly estimating the single-index function, especially when the true underlying function is discontinuous. Finally, models employing the highly flexible ST error assumption generally exhibited smaller median MSEs for the estimation of $g(u)$ compared to all other competing approaches, including those assuming skewed normal or normal errors, as well as the robust symmetric Student-$t$ and median regression formulations.  This underscores the exceptional capability of the ST distribution to comprehensively accommodate both skewness and heavy-tailedness, thereby providing the high degree of robustness against complex model misspecification and outlier contamination.}

\begin{figure}[!htbp]
	\centering
\includegraphics[width=0.8\textwidth]{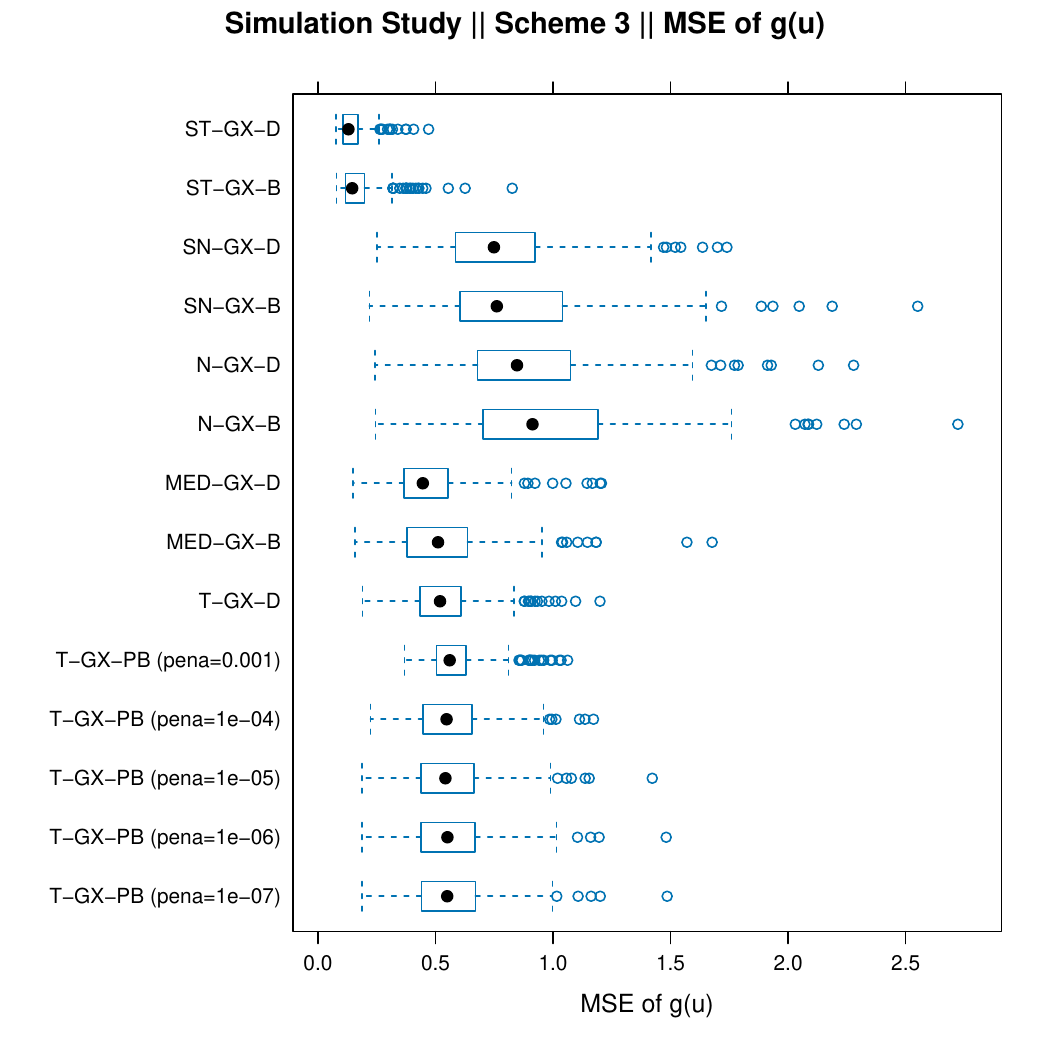}
\caption{Boxplots of MSEs from estimation of the true single-index function $g(u)$ using the 6 competing models, with data generated from Scheme III. For the estimation, 300 Monte Carlo replicates were used.}
\label{fig:simulation study, boxplot of MSE of g(u), third scheme}
\end{figure}

\begin{figure}[!htbp]
	\centering
\includegraphics[width=0.85\textwidth]{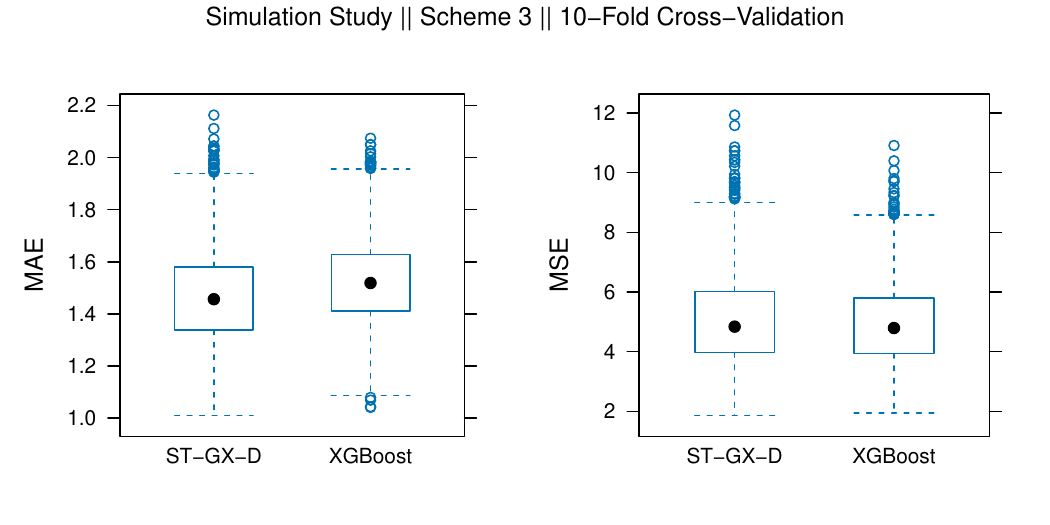}
\caption{Boxplots of the out-of-sample MSE and MAE for the proposed ST-GX-D model vs. XGBoost under the Scheme III simulation setting. Results were obtained from a 10-fold cross-validation procedure. }
\label{fig:xgboost_cv_comparison}
\end{figure}

\revone{To further evaluate the predictive performance of our proposed method against state-of-the-art machine learning algorithms, we compared ST-GX-D with XGBoost using a 10-fold cross-validation procedure under the Scheme III simulation setting.  The XGBoost model was trained over 100 rounds with early stopping and was explicitly optimized using the root mean squared error (RMSE) as its evaluation metric. S-Figure~\ref{fig:xgboost_cv_comparison} presents the resulting boxplots of the out-of-sample MSE and Mean Absolute Error (MAE) across the cross-validation folds. The summary statistics reveal that XGBoost achieved a marginally lower median MSE (4.79) compared to ST-GX-D (4.84). However, it is crucial to recognize that XGBoost was directly optimized to minimize RMSE, which provides it with an inherent and somewhat unfair structural advantage when evaluated on the MSE metric. In contrast, ST-GX-D is optimized by minimizing the negative log-likelihood of the ST distribution to estimate the conditional mode and accommodate heavy tails. Despite this objective function mismatch, ST-GX-D successfully outperformed XGBoost in terms of the median MAE, achieving a median absolute error of 1.46 compared to 1.52 for XGBoost. Because MAE is fundamentally less sensitive to large predictive deviations than MSE, this robust performance firmly highlights the practical effectiveness of ST-GX-D when handling complex skewed and heavy-tailed noise.}

\subsection{Scheme IV: Complete Details and Results}

In Scheme IV, we generated data from equation (6) of the main article, with noise drawn from the same ST distribution as described in Schemes I and II. The ground truth for $f(x_1, x_2, x_3)$ was defined as  
\begin{equation} \label{eq:scheme-4}
f(x_1, x_2, x_3) = x_1 + x_2 + x_3.    
\end{equation}

Note that this is \emph{not} the same functional form as the model in equation (1) of the main manuscript, which 
requires the regression coefficients $\boldsymbol{\beta}$ to have unit norm, i.e. $\boldsymbol{\beta}^\top \boldsymbol{\beta} = 1$. Here, in \eqref{eq:scheme-4}, $\boldsymbol{\beta} = (1,1,1)^\top$, with an L2 norm of $3$.

\begin{figure}[!htbp]
	\centering
	\includegraphics[width=0.85\textwidth]{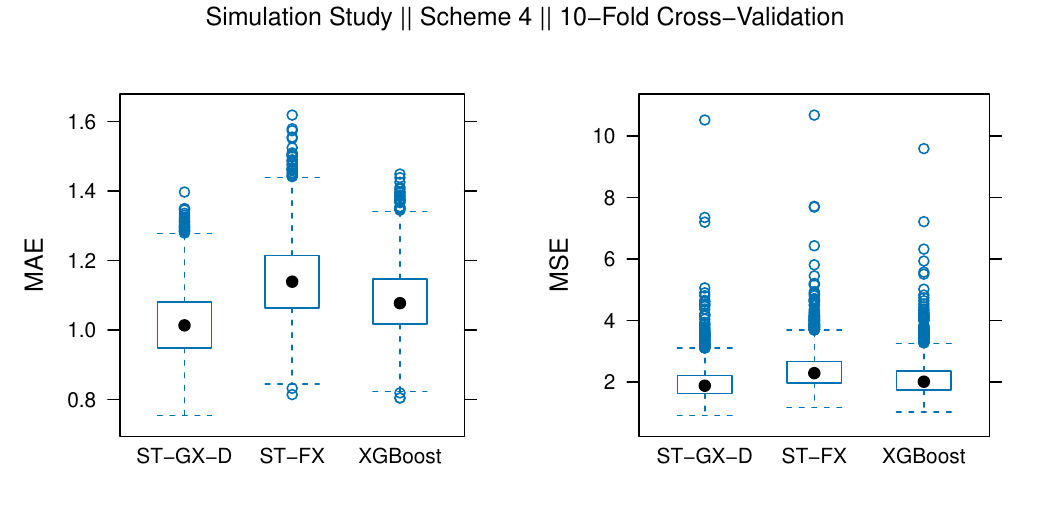}
	\caption{Boxplots of the out-of-sample MSE and MAE 
 for the ST-GX-D, ST-FX, and XGBoost models under the Scheme
IV simulation setting. Results were obtained from a 10-fold cross-validation procedure.}
	\label{fig: CV scheme4}
\end{figure}

Under the model (6) from the main article with $f(\mathbf{x})$ defined as in \eqref{eq:scheme-4}, we simulated 300 datasets with $n = 1000$. \revone{For each dataset, we fit the proposed ST-GX-D, ST-FX, and the XGBoost model. The out-of-sample prediction accuracy of these models was again evaluated based on the MSE and MAE computed from the test sets during a 10-fold cross-validation procedure.  S-Figure \ref{fig: CV scheme4} illustrates the boxplots of these cross-validated metrics across the Monte Carlo replicates. The summary statistics reveal that ST-GX-D achieved the best overall predictive performance, yielding the lowest median MSE of 1.88 and a median MAE of 1.01. In comparison, the XGBoost model yielded a median MSE of 2.01 and a median MAE of 1.08, while the ST-FX model produced the highest median MSE of 2.29 and median MAE of 1.14. This reliably demonstrates that the ST-GX-D model consistently achieves better out-of-sample generalization than ST-FX and XGBoost.}

\revone{\section{Neural Network Training Details and Sensitivity Analysis}

To thoroughly investigate the sensitivity of ST-GX-D to varying hyperparameter configurations, we conducted an extensive grid search evaluating the MSE of the estimated single-index function across different network depths, widths, learning rates, and training epochs over 300 independent Monte-Carlo replications, following the exact simulation setting of Scheme~1 in the main article with a sample size of $n = 1000$. Throughout all simulation settings, a consistent decay rate of 0.999 was applied to the learning process to ensure stable weight updates. As visualized in the boxplots of S-Figure~\ref{fig:sensitivity MSE of g}, the performance distributions clearly demonstrate that excessively deep networks or poorly scaled learning rates introduce severe numerical instability and heavily inflated MSEs. This conclusion is further supported by the estimation biases of index parameters $\beta_1, \beta_2,$ and $\beta_3$ shown in S-Figures~\ref{fig:sensitivity bias of beta1} to \ref{fig:sensitivity bias of beta3}. Specifically, deeper architectures featuring two or three hidden layers or models trained with a comparatively large learning rate of 0.01 frequently struggled to converge and produced erratic estimates in numerous simulation scenarios. Conversely, shallower network designs paired with a moderate learning rate of 0.001 and trained for at least 1000 epochs consistently delivered robust and superior performance characterized by low medians and compact interquartile ranges. Among all evaluated configurations, the architecture denoted as D:1 W:256 (meaning exactly one hidden layer of width 256) achieved exceptional stability and optimal accuracy. While a narrower network like D:1 W:128 is also shallow, our results indicate that it remains notably more sensitive to variations in learning rates and training epochs. The D:1 W:256 architecture strikes an ideal balance because it provides ample flexibility to approximate complex nonlinear index functions reliably across different hyperparameter settings. Furthermore, it is computationally more convenient compared to more complicated models and successfully avoids the destructive overfitting and divergence risks inherent to deeper networks. Therefore, we explicitly utilize D:1 W:256 trained for at least 1000 epochs with a learning rate of 0.001 as our default architecture.

Regarding the estimation of the single-index function $g(\cdot)$, the associated NN takes the one-dimensional estimated linear predictor $\boldsymbol{\beta}^\top \mathbf{x}$ as input. The network processes this scalar through an initial linear layer, a dynamic number of hidden layers (configured as one hidden layer of width 256 in our default setup), and a final linear layer to produce a one-dimensional output. The hyperbolic tangent activation function is applied after the input layer and each hidden layer. For the random initialization of this specific network, we rely on the default initialization strategy provided by the \texttt{PyTorch} framework. This approach automatically draws initial weights and biases from a bounded uniform distribution scaled by the inverse square root of the layer width.  This standard initialization provides a highly stable starting point for learning the continuous function $g(\cdot)$ and does not require manual adjustments. 

Conversely, direct optimization of the remaining model parameters $\{\boldsymbol{\beta}, w, \sigma, \delta\}$ using stochastic gradient-based methods frequently leads to numerical instability. To address this issue for the single-index coefficients $\boldsymbol{\beta}$, we adopt a similar overparameterization strategy inspired by \citet{Shin2024}. Instead of estimating the $p$-dimensional vector $\boldsymbol{\beta}$ directly, we formulate its estimation as the output of a custom NN. The input to this network is a fixed $64 \times 64$ identity matrix. The architecture consists of three fully connected linear layers. The first two layers maintain a hidden dimension of 64 and utilize the ReLU activation function. The final linear layer maps the hidden representations to the target dimension $p$. We initialize the biases of this final output layer to exactly zero and apply a Xavier uniform initialization to the weights using a highly restricted gain of 0.01. Because the weights are drawn from a symmetric uniform distribution centered at zero and the biases are zero, the expected value of the unnormalized $\boldsymbol{\beta}$ vector at initialization is precisely zero. During the forward propagation, the network processes the identity matrix to produce a $64 \times p$ output matrix. We compute the column-wise average of this matrix to collapse it into a single vector of length $p$. Finally, we divide this resulting vector by its $L_2$ norm to strictly enforce the identifiability constraint $||\boldsymbol{\beta}||_2 = 1$. We employ a parallel NN formulation to estimate the remaining scalar parameters $w$, $\sigma$, and $\delta$. Each parameter is modeled using a dedicated three-layer feedforward NN that shares the same architectural principles as the network used for the single-index coefficients. Specifically, the input is the same fixed $64 \times 64$ identity matrix and the first two hidden layers contain 64 nodes with ReLU activation functions. The final linear layer of each dedicated network maps the hidden representation to a single numerical value. We then compute the overall average of this output to obtain a preliminary unconstrained scalar.

To satisfy the specific parametric space constraints of $w$, $\sigma$, and $\delta$, we apply deterministic mathematical transformations to these preliminary scalars and carefully initialize the network biases. The parameter $w$ controls the skewness weight and must be bounded strictly between 0 and 1. We achieve this by applying the logistic sigmoid function to the preliminary scalar. By initializing the final layer bias to exactly zero, the expected initial value of $w$ prior to training is set to a balanced 0.5. The scale parameter $\sigma$ must be strictly positive. We enforce this positivity constraint by applying an exponential transformation. Initializing the final layer bias to zero ensures the expected initial scale estimate is precisely 1.0. Finally, the degrees of freedom parameter $\delta$ requires a strict lower bound of 2 to ensure a finite variance and numerical stability during the estimation process. We apply a scaled and shifted sigmoid function to restrict $\delta$ to the continuous interval between 2 and 200. To provide a robust starting point for modeling the heavy-tailed distribution, we initialize the final layer bias to exactly $-3.88$. Following the transformation, this carefully chosen bias guarantees that the initial degrees of freedom parameter is positioned near 6.0 prior to any gradient updates. While our proposed initialization strategy generally provides stable parameter estimation, the optimization of NNs is generally non-convex. If convergence to a local optimum remains a practical concern, we encourage practitioners to experiment with multiple different initial values. By running the estimation process with several distinct random seeds or initialization biases, one can systematically evaluate the resulting models and select the configuration that achieves the highest maximized log-likelihood.

\begin{figure}[!htbp]
\centering
\includegraphics[width=\linewidth]{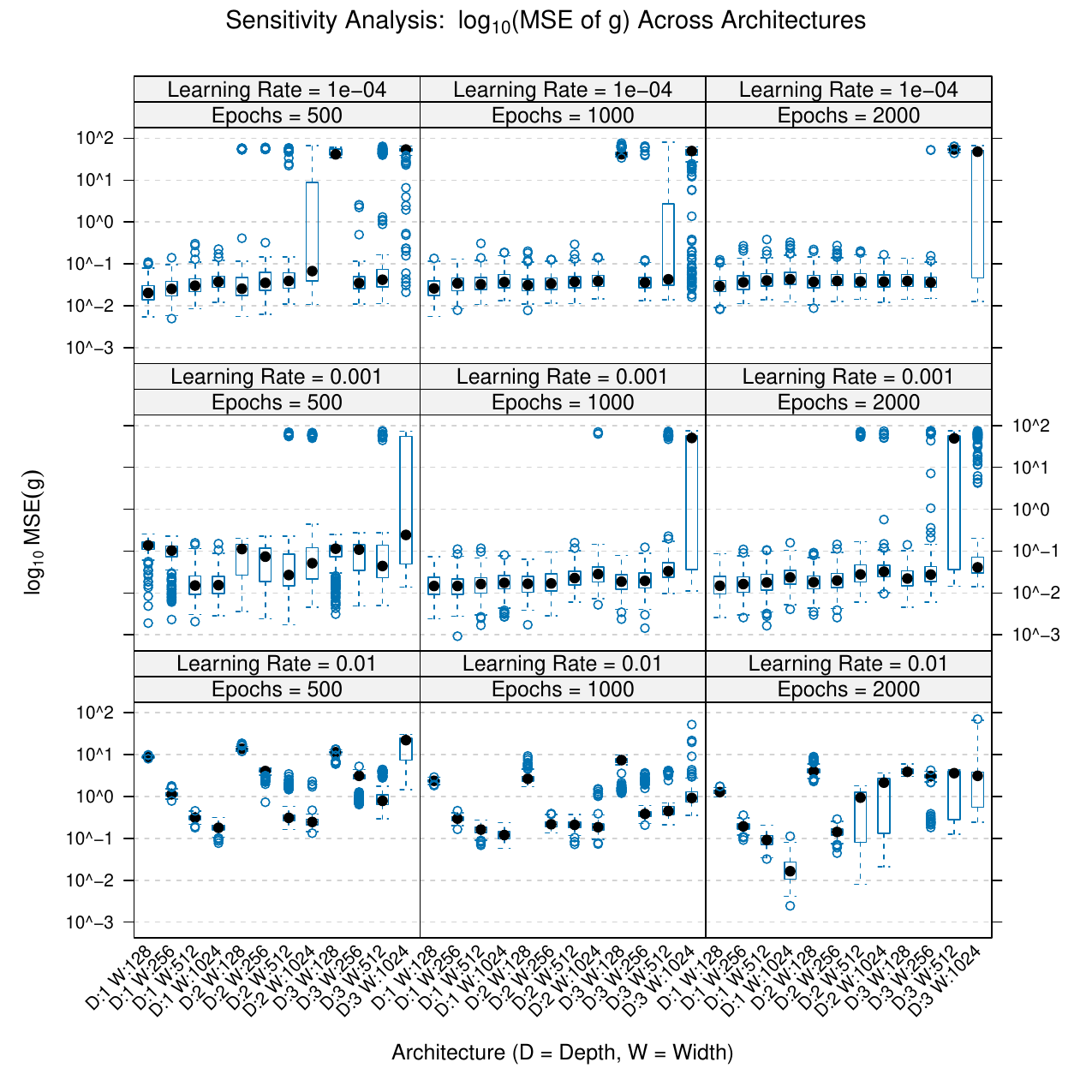}
\caption{\label{fig:sensitivity MSE of g}Sensitivity of the estimated single-index function $g(u)$ to NN hyperparameter configurations. }
\end{figure}

\begin{figure}[!htbp]
\centering
\includegraphics[width=\linewidth]{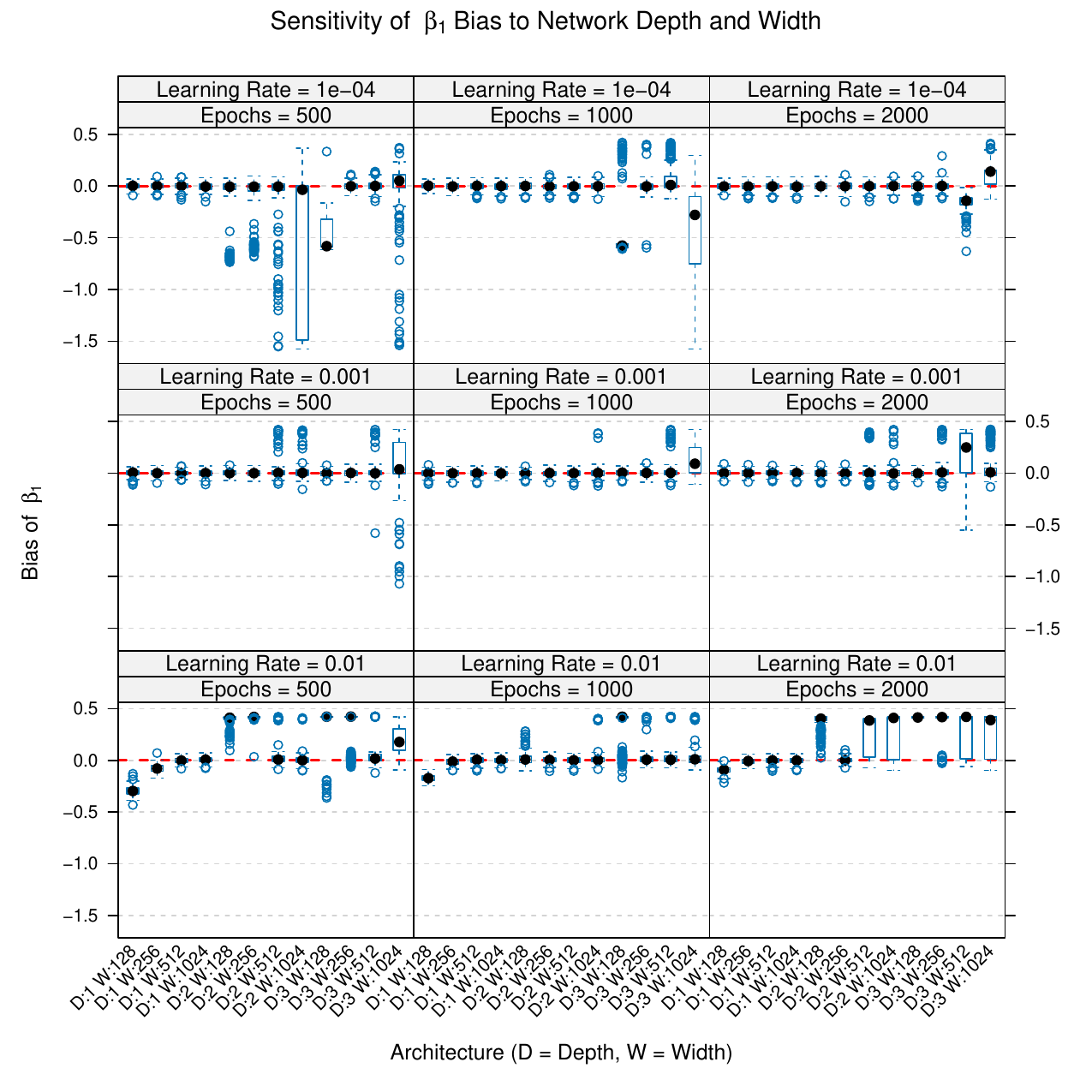}
\caption{\label{fig:sensitivity bias of beta1}Sensitivity of the estimated single-index function $g(u)$ to NN hyperparameter configurations. }
\end{figure}

\begin{figure}[!htbp]
\centering
\includegraphics[width=\linewidth]{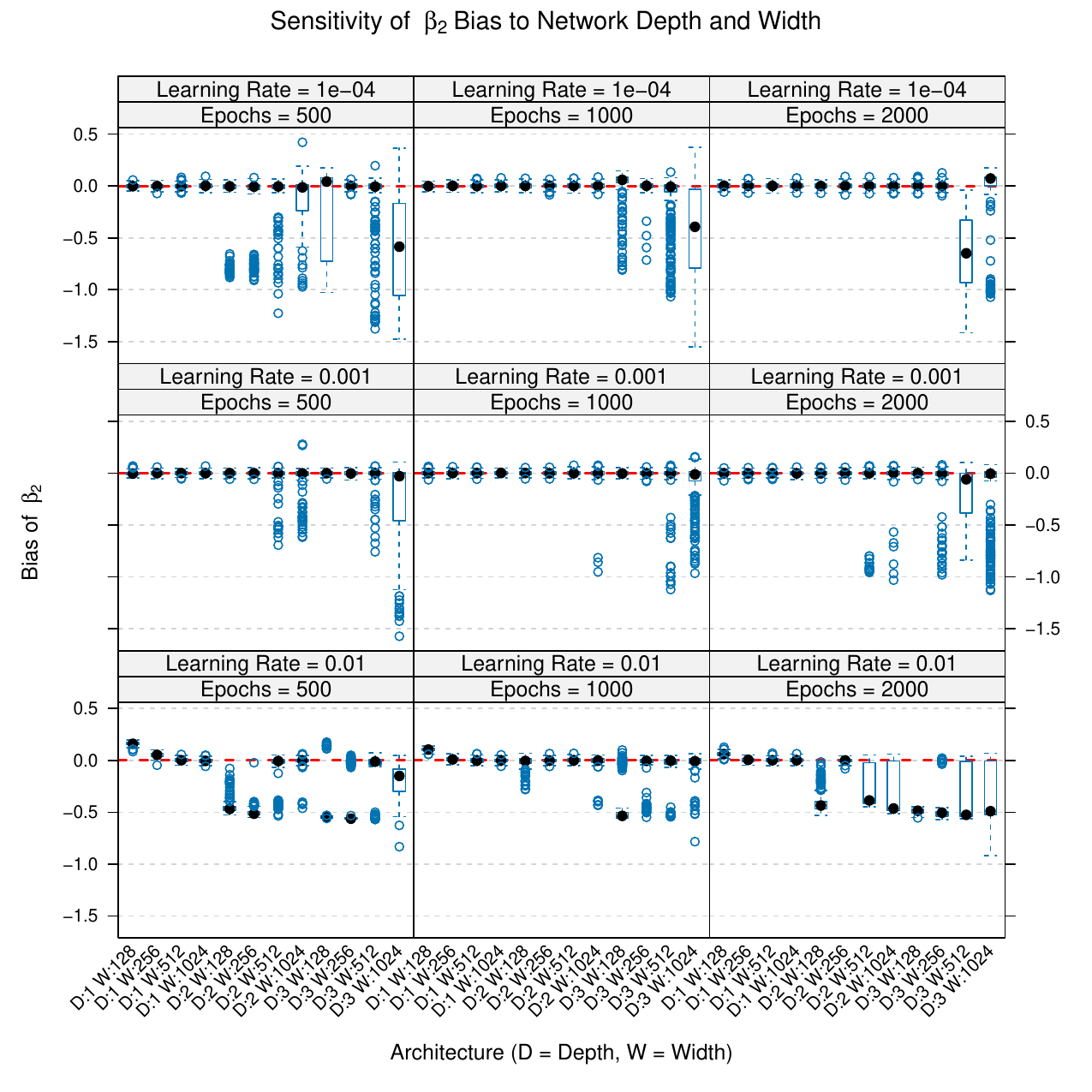}
\caption{\label{fig:sensitivity bias of beta2}Sensitivity of the estimated single-index function $g(u)$ to NN hyperparameter configurations. }
\end{figure}

\begin{figure}[!htbp]
\centering
\includegraphics[width=\linewidth]{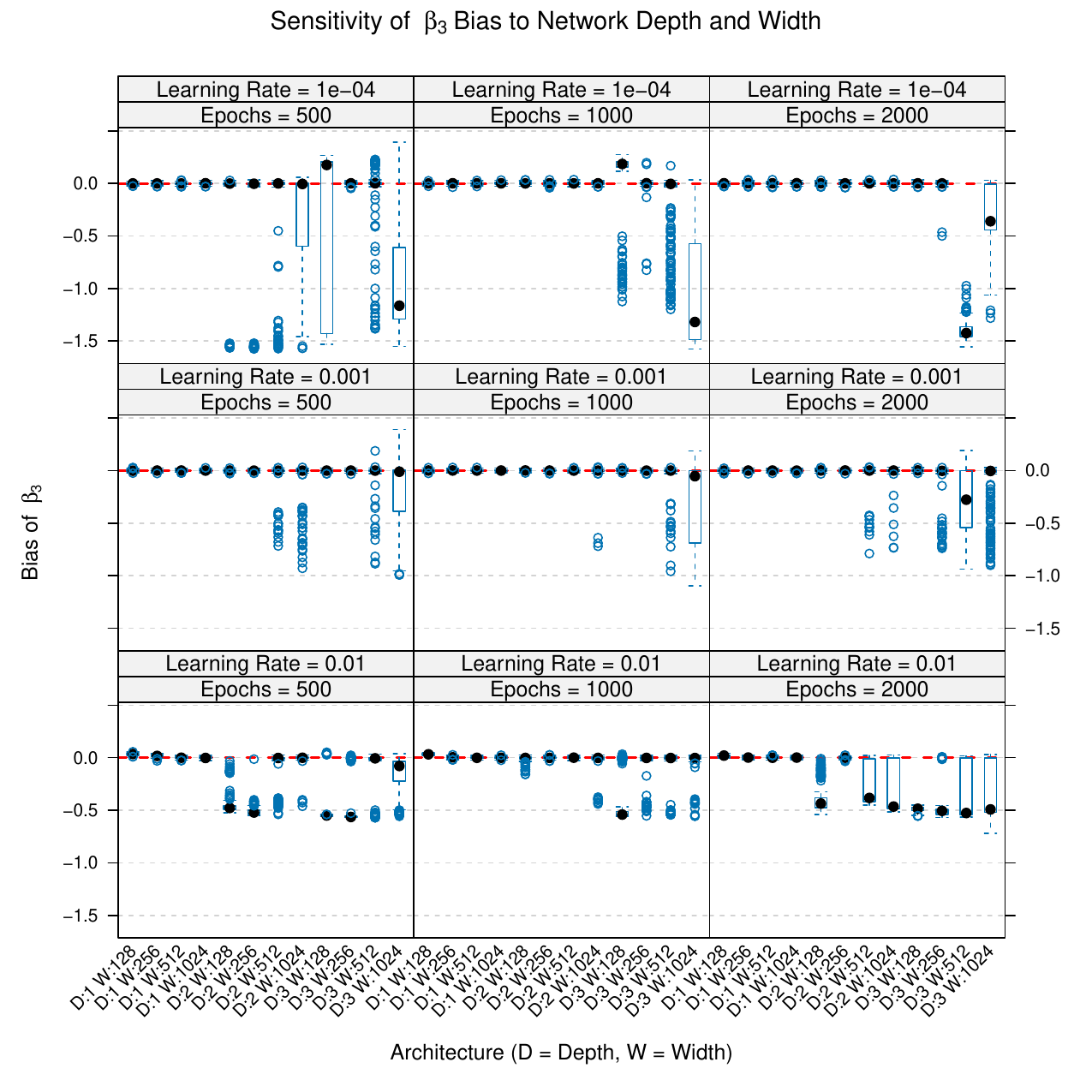}
\caption{\label{fig:sensitivity bias of beta3}Sensitivity of the estimated single-index function $g(u)$ to NN hyperparameter configurations. }
\end{figure}

}

\bibliographystyle{apalike}
\bibliography{reference}